\documentclass[iop]{emulateapj}

\shortauthors{Matthews}
\shorttitle{Radio Stars 2 Workshop Summary}

\makeatletter
\usepackage{hyperref,cleveref,etoolbox}
\patchcmd\H@refstepcounter{\protected@edef}{\protected@xdef}{}{}
\makeatother

\begin{document}
\newcommand{\ang}{\rm \AA}
\newcommand{\msun}{M$_\odot$}
\newcommand{\lsun}{L$_\odot$}
\newcommand{\days}{$d$}
\newcommand{\degree}{$^\circ$}
\newcommand{\ud}{{\rm d}}
\newcommand{\as}[2]{$#1''\,\hspace{-1.7mm}.\hspace{.0mm}#2$}
\newcommand{\am}[2]{$#1'\,\hspace{-1.7mm}.\hspace{.0mm}#2$}
\newcommand{\ad}[2]{$#1^{\circ}\,\hspace{-1.7mm}.\hspace{.0mm}#2$}
\newcommand{\lsim}{~\rlap{$<$}{\lower 1.0ex\hbox{$\sim$}}}
\newcommand{\gsim}{~\rlap{$>$}{\lower 1.0ex\hbox{$\sim$}}}
\newcommand{\HA}{H$\alpha$}
\newcommand{\HII}{\mbox{H\,{\sc ii}}}
\newcommand{\kms}{\mbox{km s$^{-1}$}}
\newcommand{\HI}{\mbox{H\,{\sc i}}}
\newcommand{\KI}{\mbox{K\,{\sc i}}}
\newcommand{\nan}{Nan\c{c}ay}
\newcommand{\jks}{Jy~km~s$^{-1}$}
\slugcomment{Accepted to PASP}

\title{Radio Stars: from kHz to THz}

\author{Lynn D. Matthews\altaffilmark{1}}
\altaffiltext{a}{MIT Haystack Observatory, 99 Millstone Road, Westford, MA
  01886 USA. lmatthew@haystack.mit.edu}

\begin{abstract}

Advances in technology and instrumentation have now opened up 
virtually the entire radio spectrum  to the study of 
stars. An international
workshop  {\it Radio Stars: from kHz to THz} was held at the Massachusetts Institute of Technology
Haystack Observatory on 2017 November 1--3 to enable the discussion of 
progress in solar and stellar astrophysics enabled by  radio
wavelength observations.  Topics covered 
included the Sun as a radio star, radio emission from hot and cool stars (from the pre- to
post-main-sequence), ultracool dwarfs, stellar activity, stellar winds and mass loss,
planetary nebulae, cataclysmic variables, classical novae, and the
role of radio stars in understanding the Milky Way. 
This article summarizes meeting highlights along with some contextual
background information.
\end{abstract}


\section{Background and Motivation for the Workshop}
Detectable radio emission\footnote{Throughout 
this article, the term ``radio'' is used
to refer to emission spanning kHz to THz frequencies
(i.e., wavelengths from meter to submillimeter).}  is ubiquitous among stars spanning
virtually every temperature, mass, and evolutionary stage.
This emission can be either thermal or nonthermal and may stem from
any of several  mechanisms, including bremsstrahlung
(free-free), 
gyromagnetic radiation, spectral lines (radio recombination lines
(RRLs); rotational transitions of molecules; hyperfine atomic transitions),
plasma radiation, and the electron cyclotron maser  (ECM) mechanism. While 
radio emission typically comprises only a small fraction of the total
stellar luminosity, radio wavelength studies provide
insight into a broad range of
stellar phenomena that cannot be studied by any other
means 
(Dulk 1985; Hjellming 1988; Linsky 1996;
G\"udel 2002; Paredes 2005). 

Nonthermal stellar radio
emission is associated with processes involving shocks or magnetic fields, including
magnetospheric activity, stellar pulsations, wind-wind
interactions, and jets. 
The detection of thermal stellar emission is
most common from objects with relatively large emitting surfaces (e.g.,
luminous stars; \HII\ regions; disks; stellar outflows and ejecta).
While the various types of stellar radio emission
manifest themselves over several decades in
frequency (kHz--THz), they are united by fundamental commonalities in
telescope design,
detector technologies, and data processing techniques, and they are
often complementary in
probing the different regions of the atmosphere, wind, and circumstellar
environment within a given star or stellar system.

Recent technological advances have led to dramatic improvements in
sensitivity and achievable angular, temporal, and spectral resolution
for observing stellar radio emission. As a result, stars are now
being routinely observed over essentially the entire radio spectrum. 
To discuss the fruits of these advances,  an international workshop ``Radio
Stars: from kHz to THz'' (hereafter RS-2) was convened at the
Massachusetts Institute of Technology (MIT) Haystack Observatory in Westford,
Massachusetts on
2017 November 1-3. This was a follow-on to the first Haystack Radio
Stars workshop (hereafter RS-1) held five years prior (Matthews 2013).
RS-2 had 51 registered participants 
representing 10 countries.  
A primary motivation was to 
bring together stellar astrophysicists (including observers, theorists, 
and modelers) from a variety of
sub-disciplines to explore common themes in the study of stars across
the Hertzsprung-Russell (H-R) diagram, particularly those that exploit the unique
potential of the latest generation of radio instruments.

\section{Scientific Sessions}
The science program of RS-2 comprised 
oral presentations (12
invited reviews and 24 contributed talks),
posters, and a moderated closing discussion session. The complete program and
presentation abstracts can be viewed on the meeting 
web site\footnote{{\url{http://www.haystack.mit.edu/workshop/Radio-Stars2017/}}}.
Slides and audio recordings from many of the
presentation are also available
at this site.

The workshop was structured around 
eleven scientific sessions: (I) An
Overview of Stellar Radio Astronomy; (II) Radio Emission from Young
Stars; (III) Radio Emission from Hot Stars; (IV) The Sun as a Radio
Star; (V) Active Stars; (VI) Cool and Ultracool Dwarfs (UCDs); (VII) Radio
Stars as Denizens of the Galaxy; (VIII) Radio Astrophysics across the
H-R Diagram; (IX) Radio Emission from Cool, Evolved Stars; (X) Radio
Emission beyond the AGB; (XI) The Future of Stellar Radio
Science. 
In this review, I summarize some highlights from these various
sessions, which illustrate the myriad areas of stellar astrophysics that are being
advanced by radio observations. My summary is roughly organized
around the session topics, although I sometimes deviate from this in
an effort to synthesize information from different sessions related to
a common theme. I also draw attention to unsolved
problems and puzzles that were highlighted at the meeting and 
where future radio wavelength observations are
expected to play a key role in advancing our understanding.

\section{An Overview of Stellar Radio Astronomy} 
A framework for the entire meeting was supplied by the opening 
presentation of
J. Linsky (University of Colorado/National Institute of Standards and
Technology). His talk underscored  
that we are in the midst an exceptionally productive era for stellar radio
astrophysics, and he highlighted
a wide variety of timely topics, many of which were discussed further
by subsequent speakers.

Linsky identified four key areas 
where stellar radio astronomy has unique potential to advance our understanding of
stellar astrophysics. To paraphrase: (1) What are the physical processes through
which stars control the evolution and habitability of planets? (2) How
do young stars and their planets evolve in their common disk
environment? (3) How is energy from the stellar magnetic field
converted into heat, particle acceleration, and mechanical energy (e.g.,
mass loss)? (4) What physical processes control mass loss across the
H-R diagram, and how does the mass loss affect the different stages of
stellar evolution?  Linsky also emphasized
the growing need for models with predictive power---both at radio
wavelengths and across the entire electromagnetic spectrum.

As summarized  below, numerous insights into Linsky's key
questions were put forward during the workshop. But at the same
time---underscoring the vitality of the field---new questions have 
begun to emerge as the result of better data
and modeling. 

\section{Radio Emission from Pre-Main Sequence Stars}
\protect\label{sec:YSOs}
\subsection{Radio Emission Processes in Young Stars}
\protect\label{sec:overYSOs}
L. Loinard (Universidad Nacional Aut\'onoma de M\'exico) provided an overview of the merits of
multi-frequency radio wavelength observations for
understanding fundamental physical processes (accretion,
ejection, magnetism)  in the evolution of 
low-mass young stellar objects (YSOs), from the protostellar (ages $\sim10^{5}$~yr) through
the pre-main sequence (PMS) stages ($\sim50\times10^{6}$~yr). 
For YSOs, different radio
emitting mechanisms and wavebands are relevant for studying material on different
physical scales (e.g., Feigelson \& Montmerle 1999; G\"udel 2002): thermal dust emission 
from envelopes/disks ($\sim$200-10000~AU scales);
thermal bremsstrahlung from the base of partially ionized
jets ($\sim$200~AU scales);
and gyro-emission from
magnetic structures linking the YSOs to their disks 
($\lsim$0.5~AU scales).
In at least one
confirmed case,
nonthermal (synchrotron) emission is also seen from jets on
pc scales (as indicated by linearly polarized
emission; Carrasco-Gonz\'alez et
al. 2010), but its presence is strongly suspected in additional
sources (e.g., Rodr\'\i
guez-Kamenetzky et al. 2016; see also Section~\ref{sec:YSOjets}).

The typical radio spectrum of a low-mass YSO is dominated by thermal
bremsstrahlung at frequencies below $\sim50$~GHz and thermal dust
emission at higher frequencies. However, 
Loinard drew attention to IRAS
16293$-$2422, whose `B' component has a spectral energy distribution (SED) that can be fit
with a single steep power law ($\alpha\approx2.3$) all the
way from 2.5--690~GHz. The source size increases at lower
frequencies, implying the emission dominated by dust even at cm wavelengths
(for free-free emission, the opposite trend would be expected),
implying this YSO is in a very early accretion stage. 
Data from the Atacama Large Millimeter/submillimeter Array (ALMA) 
were  able to resolve for the first time an inverse P~Cygni line profile,
implying strong infall and providing a rare glimpse into how the material is
infalling at different positions along the envelope (Hernandez et al. in
prep.).

Recently, the ALMA Partnership (2015) resolved the disk of the
young, solar-type star HL~Tau at mm and sub-mm wavelengths
into a spectacular series of rings and dust structures that have been
interpreted as evidence for ongoing planet formation. However, Loinard noted that
the region inside 10~AU (where terrestrial planets are expected to form)
is optically thick at ALMA bands. But  the disk becomes
optically thin at 7~mm, which allowed Carrasco-Gonz\'alez et al. (2016) to
undertake a comprehensive investigation of the inner
disk using the Karl G. Jansky Very Large Array (VLA) at an angular
resolution comparable to ALMA's. This underscores the complementarity of
longer wavelength facilities for the study of planet formation. Loinard
noted that a next-generation VLA (ngVLA; Murphy et al. 2018) 
would enable even more detailed studies of individual gas
clumps and other features of the terrestrial planet forming region of
such sources. 

\subsection{New Constraints on YSO Masses}
\begin{figure} 
\centering
\scalebox{0.7}{\rotatebox{0}{\includegraphics{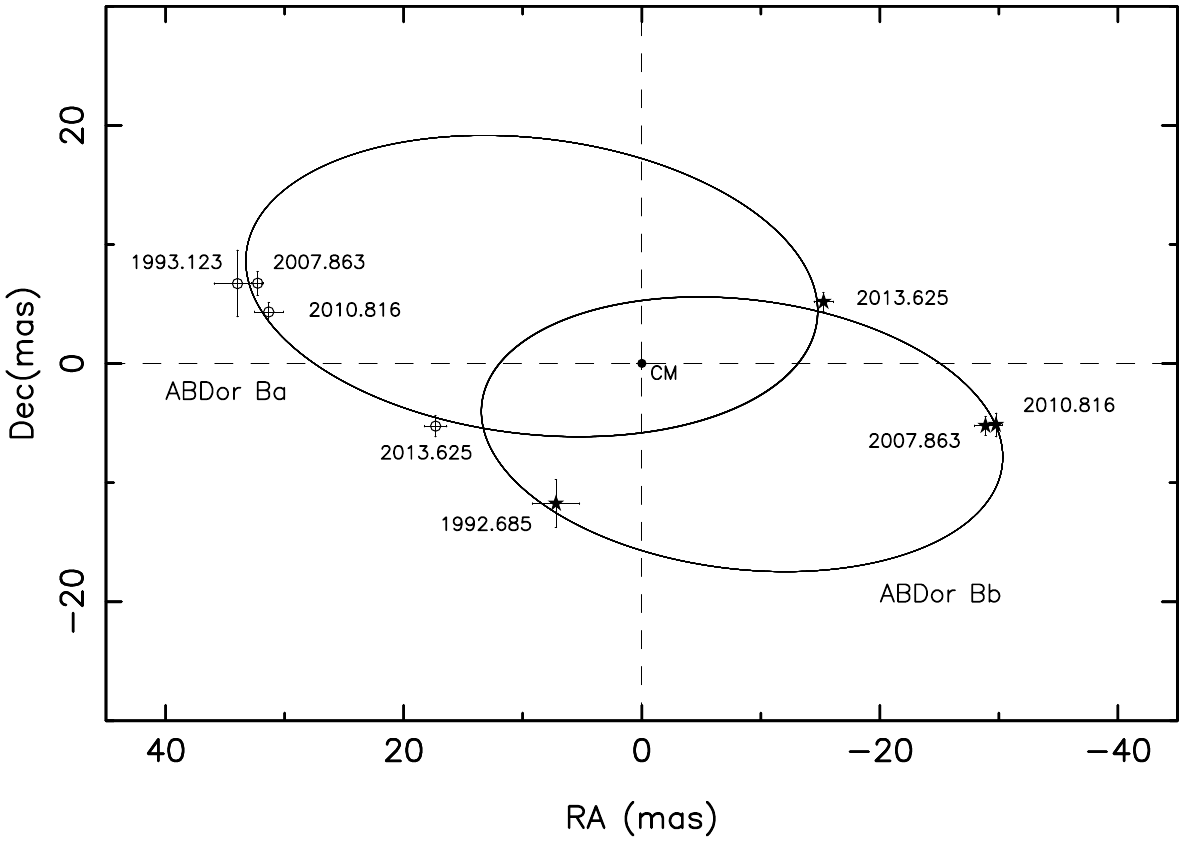}}}
\caption{Absolute orbits of components `a' and `b' of the PMS binary
  system AB~Dor~B, obtained from  IR data (circles) and VLBI astrometry spanning roughly six
  years (star symbols). The resulting mass determinations of the individual stars support the
  existence of a still unexplained
discrepancy between empirical and theoretical masses
 for low-mass PMS stars. From Azulay et al. (2015); reproduced with permission from Astronomy \&
  Astrophysics (A\&A), \copyright ESO.}
\label{fig:azulay-orbit}
\end{figure}

A longstanding problem with current models for low-mass
PMS stars is a discrepancy between the predicted
and empirically derived masses. 
Theoretical values are systematically too low by $\sim$10--30\%
(e.g., Hillenbrand \& White 2004), and the disagreement becomes worse
for lower mass objects.
R. Azulay (Max Planck Institut f\"ur Radioastronomie) described
an effort to obtain an empirical calibration for mass models by using
multi-epoch astrometry from
very long baseline interferometry (VLBI) to derive
dynamical masses for binaries within the AB~Dor moving group.
Observations of the binary AB~Dor~B were conducted with the
Australian VLBI network at 8.4~GHz during 3 epochs spanning roughly 6
years (Azulay et al. 2015; Figure~\ref{fig:azulay-orbit}).  A second AB~Dor binary,
HD~160934, was observed  at 5~GHz using the European VLBI Network (EVN)
during 3 epochs spanning 18 months (Azulay et al. 2017b). 
In both
cases, radio (synchrotron) emission  was detected from the two binary components and the
 dynamical mass estimates derived from orbital fitting 
are systematically higher by 10--40\% than predicted by  theoretical
H-R diagrams for PMS stars, consistent with
previous trends. Azulay suggested that the most likely explanation is
inadequate treatment of  convection in the models. 
Further astrometric studies by Azulay and collaborators 
are targeting additional stars and brown dwarfs (Azulay
et al. 2017a; Guirado et al. 2018).  

\subsection{Radio Emission from the Jets of YSOs}
\protect\label{sec:YSOjets}
Historically, the 
study of free-free emission from thermal jets has been one of the most important applications
of radio observations to the study of YSOs (see review by Anglada,
Rodr\'\i guez, \& Carrasco-Gonz\'alez 2018). 
Typically this emission arises from the base ($\lsim$100~AU) 
of large-scale optical jets. Most theoretical models predict that these jets are
accelerated within the central AU, and there is a clear
connection between accretion and ejection in the sense that the mass
traced by the free-free jet emission should be proportional
to the mass being accreted (modulo an efficiency factor of
$\sim$10\%). While this is largely born out by observations, Loinard noted
that an important goal for future studies will be to systematically relate the knots 
frequently observed in YSO jets (assumed to arise from
episodic ejection) with the  accretion process in a given star.

Until now, YSO jets had been previously little explored at longer radio
wavelengths and had not been detected below 
$\sim$1~GHz. However,  R. Ainsworth (Jodrell
Bank Centre for Astrophysics), has begun exploiting newly available
low-frequency
radio telescopes as a tool to more accurately
constrain the physical properties of these jets and facilitate the detection of nonthermal emission, if
present. Ainsworth presented results from a pilot study performed with the
Giant Metrewave Telescope (GMRT), as well as follow-up observations
from the Low Frequency Array (LOFAR).

The GMRT observations by Ainsworth's team targeted three bright,
well-studied protostars in Taurus.
All three were detected at both 325 and 610~MHz, and
the SEDs  are consistent with an
extrapolation from higher frequencies where the emission is partially
optically thin and characterized by a single power law (Ainsworth et
al. 2016). However, for one source (DG~Tau),
nonthermal emission was also detected from a second radio component
that appears to trace the shock front associated with a
jet impacting the ambient medium.  While synchrotron emission has 
been detected previously from at least one high-mass YSO jet (see
Section~\ref{sec:overYSOs}), Ainsworth's finding is evidence that 
the jets of low-mass YSOs are also capable of accelerating material to
relativistic velocities, despite being relatively low-powered.  Ainsworth noted that 
the  acceleration most likely stems from the Fermi mechanism
(diffusive acceleration), whereby
charged particles systematically gain energy through repeated
reflections (e.g., Padovani et al. 2015). 
Taking the measured radio luminosity from DG~Tau, 
equipartition implies that a minimum magnetic field strength
of $\approx$0.11~mG and a minimum
energy of $\approx4\times10^{40}$~erg are required (Ainsworth et
al. 2014). 
Ainsworth pointed out that jets of this type will be a source of
Galactic cosmic rays, as the associated electron energies are
estimated to be $\gsim$1~GeV.

Ainsworth reported that T~Tau was also
detected in follow-up observations at 150~MHz with
LOFAR, marking the lowest frequency at which a YSO
has ever been detected  (Coughlan et al. 2017). 
The object was partially
resolved, with hints of extension in the direction of a known
large-scale outflow. A turnover of
the SED from the higher-frequency
power law was detected for the first time, thus breaking the degeneracy
between emission measure and electron density and
enabling unique determinations of these parameters, along with the ionized gas mass. 
Ainsworth and her collaborators are continuing
efforts to characterize a larger sample of low-mass YSOs at low
frequencies using both LOFAR and the GMRT, and these
efforts are expected to benefit from recent GMRT sensitivity improvements.

\subsection{Do YSOs Obey the G\"udel-Benz Relation?}
S. Wolk (Harvard-Smithsonian Center for Astrophysics; hereafter CfA) provided an
update on a  joint VLA-{\it
  Chandra} program  on
which he first reported 
at RS-1. This effort, 
led by J. Forbrich (University of Hertfordshire), was  begun as a means of testing
whether YSOs adhere to the
G\"udel-Benz (GB) relation---a tight
power law correlation between the radio (gyrosynchrotron) emission and
X-ray luminosity spanning
over 10 orders of magnitude that holds for wide variety
of active stars (G\"udel \& Benz 1993; Benz \& G\"udel 1994).
Whether or not it holds for YSOs can provide
insights into high-energy processes in young stars,
including the early, intense irradiation of protoplanetary disks. 

The project described by Wolk has been taking
advantage of the sensitivity of the upgraded VLA,
combined with {\it Chandra}, to test whether the GB
relation holds for hundreds of young stars in clusters. 
Early findings showed almost none of the YSOs being
detected in the radio adhered to
the GB relation; instead, the majority of  stars detected in
both bands were brighter in the radio than predicted  (Forbrich, Osten,
\& Wolk 2011; Forbrich \&
Wolk 2013). To
insure that these results were not biased because of sensitivity
limitations or the lack of contemporaneous data, Wolk and Forbrich's team began a joint {\it
  Chandra}/VLA survey of the Orion Nebula Cluster comprising $\sim$30~h of coordinated
radio/X-ray observations. 
In total, they detected 556 compact sources---a factor of 7 increase 
 in the number of radio sources in the region compared with past works
 (Forbrich et al. 2016b). 
Dual frequency measurements (4.7 and 7.3~GHz) permitted
radio SED determinations. Only 112 of
the sources are detected in the near-infrared (NIR) 
as well as in radio and X-rays, while 171 of the radio
sources have no NIR or X-ray counterpart. 
In addition, 13 radio sources with
extreme variability were identified, exhibiting more than an order of
magnitude variation on timescales of less than 30 minutes; all are
X-ray sources, but only a subset shows X-ray flaring on
comparable timescales (Forbrich et al. 2017). 

Obtaining a comprehensive understanding of these results remains an
ongoing challenge. 
As noted by J. Linsky, the GB relation likely depends on a spatial
correlation between two different parts of the electron energy
distribution: thermal electrons
(sampled by the X-rays), and nonthermal electrons 
(sampled by gyrosynchrotron radio emission). However, in 
the  data shown by Wolk, the H$\alpha$, NIR,
radio, X-ray emission could be coming from different parts of a given object (jets,
wind, proplyds, etc.). Some new insights are expected to come from
follow-up VLBI observations presently being carried out for all 
556 of the detected Orion sources (Forbrich et al., in prep.).

\section{Radio Emission from Cool Dwarfs}
\protect\label{sec:cool}
\subsection{Winds from Solar-like Stars}
\protect\label{sec:coolwinds}
While the ``coronal'' winds from low-mass  
main-sequence stars typically produce only low rates of mass loss
($\sim10^{-14}M_{\odot}$ yr$^{-1}$; Section~\ref{sec:winds}), J. Linsky
reminded us that these winds
nonetheless  affect the star's evolution by
slowing rotation and altering angular momentum---and as a
consequence, by changing the dynamo and magnetic field. 
However, a problem for studying coronal winds in the radio
is the difficulty in distinguishing
photospheric and chromospheric emission from  wind emission.
The
frequency at which wind becomes optically thick depends on mass-loss rate
and the assumed wind temperature and speed, but typically
gyrosynchrotron emission  will dominate over the wind at lower
frequencies (cm-wavelengths), while in the mm
domain, the
chromosphere dominates. Linsky emphasized that unambiguously detecting and
disentangling these components poses a significant
challenge. As evidence, he pointed to recent VLA
observations of  solar mass stars  where either the stars were
undetected (Fichtinger et al. 2017), or where chromospheric emission
was seen, but the data could not constrain the possibility of an additional
coronal contribution to the radio flux
(Villadsen et al. 2014).
Linsky argued that new approaches will be crucial, such as
observations with wider wavelength coverage, including
mm-wavelengths where there is no nonthermal emission
component and where the chromospheric brightness temperature becomes
weaker. 

Another unknown for  solar-type stars
is the degree of constancy of their winds.
Linsky pointed to work by Osten \& Wolk (2015) who computed the expected
time-averaged transient mass loss in coronal mass ejections (CMEs)  for stars, taking
into account an equipartition between the kinetic energy in the CMEs
and the bolometric flux from a
stellar flare. These results predict high
mass-loss rates for stars like the young Sun or active M dwarfs---much
larger than the current solar mass-loss rate. However, testing
these predictions will require unambiguous detections of
CMEs from other stars---something that has not yet been done
(Section~\ref{sec:active}).

\subsection{Chromospheres}
\protect\label{sec:chromospheres}
Linsky noted that the study of chromospheres 
in the ultraviolet (UV) or X-rays requires the observation of emission lines
formed in non-local thermodynamic equilibrium (non-LTE) conditions 
over a range of temperatures. Further, different portions of the emission line
form in different parts
of the atmosphere, making it necessary to solve the radiative transfer
equation. In comparison,
radio wavelengths offer a number of  advantages for studying the thermal structure of the
same star. The dominant emission process is free-free continuum for
which LTE conditions hold (e.g., Wedemeyer et al. 2016).  Opacity in the radio is 
simple and well-understood (electron-ion and
H$^{-}$ free-free) and heating rates can be  derived empirically.
Radio brightness temperatures can be derived from a
simple  integral of electron temperature over height, whereas
temperatures in the UV require a more complex integration over a source function with
height, which is not related to the Planck function (as in the radio), but rather to the
statistical equilibrium distribution of the Planck function times the
departure coefficient. In the radio, the free-free emission from
longer to shorter wavelengths successively probes the photosphere to
upper chromosphere, whereas  in UV lines, the
photosphere contributes to the emission line wings, the chromosphere
to the peak, and the transition
region to the line core. 
 Finally, circular polarization of the mm
continuum can trace variations in electron
densities 
(e.g., White et al. 2017; Loukitcheva et al. 2017; see also
Section~\ref{sec:mmSun}) and provides one of the best available
means to measure chromospheric B-fields.

Linsky additionally pointed out that radio observations are likely the
best means to resolve the
long-standing question of whether A- and B-type stars (which lack convective
zones) have chromospheres. Chromospheres are difficult to measure in
the UV for
warmer stars because of the increasing contamination from
the photosphere, but radio measurements may provide some new insights.

\subsection{Activity on Cool Dwarfs} 
\protect\label{sec:active}
The study of stellar flaring at radio wavelengths 
has long been relevant for understanding the magnetic properties
of stars and uncovering the similarities and differences between their
magnetic activity and the Sun's. More
recently, this topic has seen a dramatic  surge in interest owing to
its relevance to assessing exoplanet habitability (see
Section~\ref{sec:exoplanets}). Not surprisingly, studies of stellar
activity received proportionally far greater attention at RS-2 than
RS-1  (cf. Matthews 2013). The present section focuses on
meeting content related to active stars
and binaries earlier than spectral type M7; activity on UCDs
(Section~\ref{sec:ultracool})  and  the Sun 
(Section~\ref{sec:sun-as-star}) are described
separately. However, there was 
naturally thematic overlap, and one outstanding
question is whether  the type of  large-scale
magnetospheric currents now inferred in UCDs may also be present in
some other  classes of active stars (see Section~\ref{sec:ultracool}). 

As discussed by S. White (Air Force Research Laboratory) and R. Mutel
(University of Iowa), the bulk of the radio emission
observed from active stars at cm wavelengths 
is  nonthermal gyrosynchrotron radiation arising
from a power law distributions of highly relativistic electrons in the
corona, thought to be produced from flare acceleration and/or
magnetic reconnection events.
Evidence for this interpretation comes from the
polarization properties of the emission, as well as 
VLBI observations of active stars where $T_{B}\sim10^{9}$~K is
observed. Unfortunately, the
harmonics are highly broadened in this regime, thus  as pointed out
by White, it is not possible
to identify a specific magnetic
field strength with a specific emission frequency. 
Mutel and White therefore discussed, respectively, the feasibility of gleaning
insights through the study other flavors of radio emission from active
stars, namely thermal gyrosynchrotron and gyroresonance. 

\subsubsection{A First Detection of  Thermal Gyrosynchrotron Emission
  Beyond the Sun} 
\protect\label{sec:firstthermal}
In hot plasmas displaying nonthermal gyrosynchrotron emission,
{\em thermal}
gyrosynchrotron emission may also arise from  the Maxwellian tail of a thermal electron
density distribution (e.g., Dulk 1985).  
Predicted observational signatures are deviations in the SED from a
pure power law, along with very high
fractional circular polarization ($\sim$30-100\%).

R. Mutel recounted his recent efforts to search for thermal gyrosynchrotron
from eight radio-loud stars selected from different parts of the H-R diagram.
The optical depth of thermal gyrosynchrotron, $\tau_{\nu}$, is strongly dependent on both the
magnetic field strength $B$ and the temperature $T$ 
($\tau_{\nu}\propto T^{7}B^{9}$). Given these strong dependencies, 
this emission is not expected to be detectable unless $T\sim10^{8}$~K
and $B\sim$1~kG (i.e., the plasma is very hot  with a very strong magnetic
field).  
Mutel's target sample therefore included three chromospherically active binaries,  two PMS
(weak-line T~Tauri) stars, and three
M dwarfs. Contemporaneous flux density measurements were obtained
for each star at
several frequencies between 15--45~GHz using the VLA. 

Mutel found that the
SEDs of the three active binaries showed no evidence for thermal
gyrosynchrotron, appearing instead as power law sources consistent with
nonthermal gyrosynchrotron. However, for the two PMS stars and the
M-dwarf binary UV~Ceti, 
compelling evidence for a thermal
gyrosynchrotron component was identified based on the shape of the SED and
the presence of enhanced circularly polarized emission above
30~GHz. Not surprisingly, these are the targets with the highest plasma
temperatures. Mutel reported that this is the first time that
evidence for thermal gyrosynchrotron emission has been seen from a star other than the
Sun.   

\subsubsection{Gyroresonant Emission from Active Stars}
\protect\label{sec:activity}
In analogy with previous work on the Sun, 
a potential means  of probing coronal magnetic field strengths in
non-relativistic plasmas somewhat cooler ($\sim10^{6}$~K) than
those targeted by Mutel (Section~\ref{sec:firstthermal}) is  gyroresonance emission (White
2004).  
In particular,
S. White evaluated the prospect of linear polarization
measurements of such emission as a tool for
understanding the magnetic properties of active stars. He
concluded that  in most cases, the combined signal
from the two stellar hemispheres will dramatically reduce the overall
amount of polarization, making it difficult to identify gyroresonance
emission. An exception may occur in the case of
stars with large polar spots viewed nearly pole-on.
However, even in these cases, there may be another complication uncovered by earlier
work, namely  that the presence of
plasma emission at lower frequencies can confuse the interpretation of
the gyroresonant emission, leading to an apparent polarization reversal
between low and high frequencies (White \& Franciosini 1995).

\subsubsection{Time Domain Studies of Active Stars}
Taking advantage of the frequency ranges now accessible with the
Murchison Widefield Array (MWA), 
C. Lynch (University of Sydney) described an ongoing program to
characterize the little-explored radio flares from M dwarfs that
occur at frequencies of a few hundred MHz. This topic has received
only scant attention since a few pioneering studies 
more than 30 years ago where m-wave
flaring was detected from 11 M dwarfs  (Spangler \& Moffett 1976; Davis et al. 1978; Kundu
et al. 1988). 
Lynch noted that although this particular emission was recognized as being coherent,
there has been an unresolved controversy as to whether the underlying
mechanism is plasma or ECM emission. 

\begin{figure} 
\centering
\scalebox{1.0}{\rotatebox{0}{\includegraphics{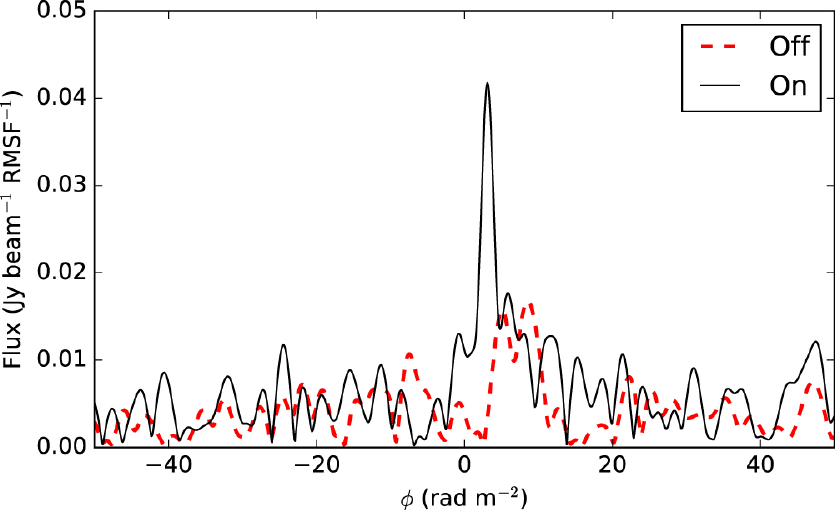}}}
\caption{Faraday dispersion function during a bright flare on
  the active star UV~Ceti (black curve), derived via rotation
  measure synthesis from 154~MHz MWA data. The implied linear
polarization fraction is $>$18\%, suggesting that the emission
arises via an ECM mechanism.  No linearly polarized emission was detected
during an adjacent non-flaring interval (red dashed line). From Lynch
et al. (2017); reproduced by permission of the American Astronomical
Society (AAS). }
\label{fig:elliptical-pol}
\end{figure}

Previous statistics on the strength and frequency of ``MHz''
flares, coupled with the known presence of $>$2000 M dwarfs in the
Southern Hemisphere within 25~pc  (Winters et al. 2015) led to the prediction that M-dwarf
flares should be ubiquitous in
transient surveys conducted by modern low-frequency instruments like
the MWA. But surprisingly, {\em no} flaring M dwarfs
were detected in two recent widefield MWA surveys (Tingay et al. 2016;
Rowlinson et al. 2016). To better understand why, Lynch and her
colleagues began deeper targeted observations of M dwarfs to
characterize their behavior at meter wavelengths. The observations
emphasized Stokes~V (circular polarization) data products to overcome the confusion
limits that otherwise plague deep integrations at MWA frequencies,
effectively
boosting the achievable sensitivity by a factor of $\sim$80.

Using this approach, Lynch et al. (2017) reported a detection of flaring from UV~Ceti in
each of four days when the star was observed with the MWA, marking the
first detection of meter-wavelength flaring from an M dwarf with
current-generation radio facilities.  The emission is coherent, and a
signature of linear polarization was observed from the brightest flare, indicating that the
emission is elliptically polarized and hence
is likely to be ECM (Figure~\ref{fig:elliptical-pol}). This finding permitted
an estimation of the magnetic field strength at the location
of the event ($\sim$28~G).  In addition, evidence for a periodicity of
5.45~hours was seen, echoing a behavior recently seen at higher
frequencies (see below).
However, follow-up observations 
by Lynch and colleagues in late 2017 did not detect this signal, indicating that it is not
persistent. Two other targeted M dwarfs, YZ~CMi and
CN~Leo, were undetected at the time of the meeting, but MWA observations
remain ongoing. Lynch stressed that information on flare rates and
luminosities at m-wavelengths remains  extremely limited and that
significant amounts of dedicated observing time with low-frequency telescopes will be
needed to improve these statistics. She  noted that unfortunately computational
limitations currently preclude the routine mining of data from other MWA
experiments for M dwarf flare events.

Another recent campaign aimed at studying activity on M dwarfs, but at
somewhat higher frequencies, is being led by J. Villadsen (National
Radio Astronomy Observatory; hereafter NRAO). A primary underlying motivation is
understanding the
extrasolar space weather associated with these stars (see
also Section~\ref{sec:weather}), and in particular whether they
are subject to CMEs. 
Villadsen emphasized that while  flares are observed frequently
from M dwarfs, CMEs and energetic protons are not---they are only
predicted.

The detection of CMEs from stars other than the Sun has
been a longstanding quest for astronomers, but so far no such 
events have been unambiguously identified (see also Section~\ref{sec:weather}).
While coherent bursts of emission from M dwarfs sometimes
exhibit frequency drifts reminiscent of solar CMEs, as with the
m-wave emission discussed by Lynch (see above), it has so far remained
impossible to differentiate between coherent plasma
emission (as occurs with solar CMEs) and
ECM emission.  Furthermore, as noted by Osten, it has so far proved
difficult to identify  an observational signature that can be ascribed
uniquely to CMEs and not to
flares.

In hopes of making progress in this quest, Villadsen and her colleagues
used the VLA to study a sample of five M dwarfs over
22 epochs in bands covering 0.22--0.48~GHz and
1--4~GHz (or 1--6~GHz). Their goal is to identify events with a
time-frequency evolution analogous to solar CMEs. Having multiple
frequency bands allows tracing plasma at different distances from the
star, 
with the higher frequencies probing the low corona and the
lower frequency band probing $\sim1.5-3R_{\star}$. This arises
from the dependence of the radio emitting frequency on the electron
density and magnetic field strength at a given height (e.g., Dulk 1985). In parallel,
Villadsen's team used the Very Long Baseline Array (VLBA) to
attempt to resolve the radio coronae before and after flares. While it
is known from previous VLBI experiments that the
radio corona of M dwarfs can become extremely extended during 
periods of activity (e.g., Benz, Conway, \& G\"udel 1998),
Villadsen's study is the first
attempt to connect high-resolution VLBI imaging with wide
bandwidth dynamic spectroscopy (Villadsen et al. 2017). 

The VLA data obtained by Villadsen and collaborators helped to
confirm that coherent bursts in M dwarfs are extremely common,
occurring in 13 of 22 observed epochs and on a variety of timescales, from 
seconds or minutes to events lasting
an hour or longer. The latter require some type of
persistent electron acceleration process which Villadsen 
suggested may be emerging magnetic flux driving
magnetic reconnection over extended intervals, analogous to
what happens on the Sun. 

However, while tantalizing hints of CME-{\em like} phenomena were seen in some objects,
unambiguous signatures of a CME remain elusive.
For example, in AD~Leo, an off-limb flare was observed with the VLBA,
nearly a full stellar diameter from the star, but the
dynamical spectrum taken with the VLA at the same time shows no
evidence of outward motion and no
obvious connection with this event. 
In the case of UV~Ceti, dynamic spectra from five different epochs all
show very similar structures, suggesting some kind of periodic
phenomenon that could be related to the periodicity seen at meter wavelengths (see above).
Some features in the UV~Ceti data exhibit a frequency drift with time, 
but Villadsen  argued that the emission is more
likely analogous to the periodic aurorae seen in
UCDs and planets (see Section~\ref{sec:ultracool}) and therefore
attributable to an ECM mechanism---not plasma emission from a CME. 
In the ECM scenario, frequency drift
results from geometric modulation rather than motion of blobs
of material.

Campaigns reported by other meeting participants had similar
outcomes. For example,
R. Osten (Space Telescope Science Institute) 
described recent VLA measurements of EQ~Peg at 230--470~MHz,
performed with simultaneous near-UV monitoring. In 20 hours of data, several
moderate flares were seen, but no ``type~II'' events (see also
Section~\ref{sec:lowfreqSun})
of the kind expected
to occur along with CMEs (Crosley \& Osten
2018). 

The confluence of these results naturally leads to the question: why
do we not see CMEs from other stars? One possibility raised by Osten is
that perhaps CMEs are not super-Alvf\'enic, hence no
shocks are produced. Two additional possibilities suggested by B. Chen
(New Jersey Institute of Technology)
are that either the ejected material is very clumpy or else that the brightness
temperature of the type~II radio burst  cannot become sufficiently 
high before their electromagnetic energy is converted back into
Langmuir waves.
Osten also noted that a related question that merits further study is
whether the large magnetic field strengths of M dwarfs might prevent
the magnetic breakout needed for the ejection of material from the
stellar surface.  This question is particularly relevant for the
question of exoplanet habitability (Section~\ref{sec:weather}).

\subsubsection{Insights into Solar and Stellar Activity from Theory and Modeling}
As the quality of radio observations of active phenomena on
the Sun and other stars grows, so does
the need for increasingly sophisticated theoretical models to aid in
the physical interpretation of the data. 
S.-P. Moschou (CfA) described
a new state-of-the-art tool that  is capable of producing realistic
radio synthetic images of both quiescent and transient phenomena 
over a wide range of radio wavelengths. The models 
incorporate a variety of important physics,  including
acceleration, scattering, and shocks. Because radio waves from a star
like the Sun are highly refracted, the curvature of the ray
paths must also be treated to properly compute the emergent radio intensity in each pixel
of the model images (see Benkevitch et al. 2010).  

Moschou presented  first results from
applications of her code to the study of radio-loud CMEs and their
relationship to other transient phenomena on the Sun.  The
resulting models capture 
the CME shock wave and allow watching
the development of an ensuing type~II radio burst. Future work by
Mochou's team will include adding the treatment of
polarization and magnetic fields in their code, as well as
application of the code to other stars. 

\subsubsection{Future Directions for Radio Studies of Active Stars} 
J. Linsky suggested several ways of obtaining crucial new physical
insights into active stars. He 
argued for further work in the mm domain, noting that
(sub)mm monitoring over several rotation periods might help  to reveal
periodic flux density modulations indicative of the presence of hotter regions or cooler starspots. Commensal
studies of flares in (sub)mm and hard X-rays are also expected to be fruitful
(see Krucker et al. 2013), since both bands show 
the same thermal structure in
flares, though soft X-rays show a more gradual development.
Linsky also advocated for expansion of the spectral types targeted for
flare studies. While typical radio observations of M
dwarfs routinely lead to flare detections within a few hours of
integration, active G stars with strong optical flaring ($\sim$100--1000 times
typical solar flare luminosities) have been discovered with {\it Kepler}
(e.g., Katsova \& Livshits 2015) and would be
interesting targets for future studies. In addition, Linsky stressed the
desirability of combining multiple radio instruments (e.g., the VLA and ALMA) to observe
flares from various types of stars
simultaneously from cm to mm wavelengths, as this could provide important
insights into the distribution of
the relativistic electron energies. He cited previous work by 
Raulin et al. (2004) that looked at how the frequency
of the peak in the gyrosynchrotron emission changes with
time in solar flares and noted that this has yet to be studied for other
stars. Finally,
Linsky raised the prospect of whether Doppler imaging might soon become possible for
active stars at radio wavelengths. As a step in this direction, 
A. Richards (Jodrell Bank Centre for Astrophysics/University of Manchester)
noted that  she and her colleagues have already produced a
spatially resolved velocity field for the red supergiant Betelgeuse using data from ALMA
(Kervella et al. 2018).

\section{Emission from Ultracool Dwarfs (UCDs)}
\protect\label{sec:ultracool}
\subsection{The Origin of the Radio Emission}
\protect\label{sec:coolorigins}
As described in the review by G. Hallinan (California Institute of
Technology), the past few years have seen significant progress
toward a comprehensive picture of activity on 
UCDs (i.e., stars with spectral types $\gsim$M7;
see reviews by Pineda, Hallinan, \& Kao 2017; Williams
2017). The ultra-wide bandwidths provided by the VLA and other 
current-generation radio facilities have been  instrumental in the
recent growth
of this  area, and not surprisingly, this topic received
considerable attention at RS-2.

Hallinan reminded us that the radio emission from  UCDs is characterized by both
``quiescent'' emission (including flares; Section~\ref{sec:UCD-quiet}), 
as well as periodic, pulsed emission that
is highly ($\sim$100\%) circularly polarized 
(e.g.,   Hallinan et al. 2006, 2007). Frequently, radio-detected UCDs are
rapid rotators, and the periodicity of the pulsed emission component is found to be
closely tied to the stellar rotation period (typically a few hours),
consistent with rotational beaming.  The high inferred brightness
temperature of the pulsed emission, coupled with its strong circular 
polarization, points to a coherent emission mechanism
that is now attributed to ECM. 

The atmospheres of UCDs are sufficiently cool to become
largely neutral, allowing the magnetic field to become decoupled from the
photospheric gas. Consequently, processes like magnetic reconnection  no longer
dissipate significant amounts of energy. This transition is found to correspond to 
a marked departure from the domain of
coronal/chromospheric magnetic activity to planet-like  auroral
magnetic activity. Importantly, in this regime currents are driven by a global
electrodynamic interaction in  a large-scale magnetic
field instead of local photospheric plasma motions  (see Pineda et
al. 2017).   

A recent radio and optical study of the M9 dwarf
LSR~J1835+3259 by Hallinan's team led to a simple, unifying model to 
explain the radio, H$\alpha$, and broadband optical emission through 
an electron beam powered by auroral currents (Hallinan et al. 2015). In this model, as the
electron beam hits the upper atmosphere,  a supply of free electrons
leads to the production of
H$^{-}$, whose opacity then dominates high  in the
atmosphere. This results in regions that are cooler than the surrounding
photosphere and produces a
largely blackbody spectrum (corresponding to $T_{\rm B}=$2180~K)---very different from  
a pristine photospheric spectrum, which is dominated by molecular
absorption.

Hallinan emphasized that the aurorae of relevance to
UCDs are quite different from those observed terrestrially. The latter are
dominated by a mechanism whereby the solar wind interacts with the
Earth's magnetosphere, producing
acceleration of electrons along magnetic field lines (Pineda et
al. 2017 and references therein). These events 
are a byproduct of local and impulsive current
systems on the Sun, driven by reconnection
events or shock fronts (see Section~\ref{sec:sun-as-star}).
In contrast,  UCDs have large-scale
current systems that are {\em global} and {\em quasi-stable} and that are
thought to power aurorae through energy dissipation (see also Section~\ref{sec:UCD-Bfields}).
Hallinan stressed that it remains an open question 
as to whether a similar type of  large-scale
magnetospheric currents may also occur in other classes of stars that
display pulsed emission and evidence for large-scale magnetic fields
(see Sections~\ref{sec:active} and \ref{sec:mag-OB}).

One lingering puzzle pointed out by Hallinan
is why only a  small fraction of brown dwarfs  ($\sim$7-10\%)  are
detected in  sensitive radio searches. 
One potential clue is a dependence on rotation
velocity, with a higher rate of detection for the rapid rotators
(Pineda et al. 2017), raising the possibility that there exists a critical
rotational velocity to achieve co-rotation breakdown. 
The low detection statistics have until recently thwarted attempts to better
characterize the magnetic field strengths and underlying electrodynamic engine
of brown dwarfs (see below).  However, as discussed by M. Kao (Arizona
State University),
the auroral nature of the brown dwarf radio emission  results in a
correlation between
characteristic time-varying emission in the H$\alpha$ line and the
IR continuum
(Hallinan et
al. 2015). These latter tracers have now been shown to serve as an effective proxy for the
identification of radio-bright brown dwarfs (Kao et al. 2016; Section~\ref{sec:UCD-Bfields}).  

\subsection{Origin and Characterization of Magnetic Fields on UCDs} 
\protect\label{sec:UCD-Bfields}
One of the chief outstanding challenges for the auroral model 
of radio emission from UCDs is to explain the generation  
and maintenance of the persistent magnetic fields 
in these fully convective stars.  Mechanisms that may 
plausibly generate aurorae in isolated UCDs are: (1) occurrence of a co-rotation 
breakdown when the field further out becomes too weak to
drag along the plasma;  (2) the presence of an embedded 
object within the UCD magnetosphere (Trigilio et al.  2018; 
see also Section~\ref{sec:exoeffects}). Presently no conclusive evidence exists 
to favor either mechanism, but both require the presence of 
persistent UCD magnetic fields. UCD magnetic field measurements based
on radio data (e.g., Route \& Wolszczan 2012; Kao et al. 2018), along with magnetic field topological
modeling  and comparisons with radio 
dynamic spectra may help to provide new insights (e.g., Lynch, Mutel, \& Guedel 2015; 
Leto et al. 2016).

Zeeman Doppler measurements are traditionally used to characterize the
magnetic fields of low-mass, magnetically active stars 
(Semel 1989), but this technique breaks down for spectral types
cooler than $\sim$M9. However, ECM emission is produced at or close
to the fundamental cyclotron frequency ($\nu_{\rm
  MHz}\approx2.8\times B$ where $B$ is the magnetic field
strength in G; Treumann 2006), making it a unique and powerful tracer
of the local field strength in radio-bright UCDs  (Hallinan et al. 2008; Route \&
Wolszczan 2012), provided that the plasma densities are sufficiently
low (see Mutel et al. 2006).
Despite this, magnetic
field measurements for L, T, and Y dwarfs have until recently remained sorely lacking
owing to the challenge of identifying suitable radio-emitting targets.
However, M. Kao presented results for a new sample of six brown dwarfs
 observed with the VLA in the 4--8~GHz
band. The objects were selected based on their IR and H$\alpha$
variability (see
Section~\ref{sec:coolorigins}). Highly circularly
polarized emission was detected from five of the six, including
one previously observed T6.5 dwarf (Kao et
al. 2016). Follow-up observations of the sample in 8--12~GHz and
12--18~GHz bands showed tentative evidence that  the auroral radio emission at these higher
frequencies starts to become more
variable, but two IR-variable Y
dwarfs were not detected. 
The future radio detection of
Y dwarfs would be helpful in probing a
temperature range where there currently exists no constraints on
magnetic fields, although
such detections may be beyond the sensitivity limits of current instruments.

Kao pointed to previous work by Christensen, Holzwarth, \& Reiners
(2009) showing that a
scaling of the convection-driven dynamo found in planets such as the
Earth and Jupiter appears to account for the magnetic fields
of low-mass stars, suggesting a single, unified mechanism 
governing magnetic activity in all rapidly rotating, fully convective
objects over more than three orders of magnitude  in
field strength.  
However, Kao's latest results appear to deviate from the Christensen et
al. relation. Accounting for age- and temperature-related effects,
including the decay or magnetic fields with time, does not remove this
discrepancy (Kao et al. 2018). 
J. Linsky pointed out that very young
brown dwarfs may still be burning deuterium, leading to very different
interior structures compared with older brown dwarfs.  

\subsection{Quiescent Radio Emission from UCDs}
\protect\label{sec:UCD-quiet}
Another outstanding puzzle concerning UCDs is the nature of their
quiescent  radio emission seen at GHz frequencies. Such emission is
ubiquitous in UCDs with pulsed radio emission and is observed
to show variations over a wide range of timescales, including
flare-like activity (e.g., Route 2017).  Hallinan noted that it is presently unknown whether the
same current system powers both the pulsed and quiescent emission, and
P. Williams
(CfA) highlighted the
shortcomings of existing models for describing recently observed radio
spectra.  For example,
while the quiescent emission is generally thought to be
gyrosynchrotron in nature, recent ALMA observations of the UCD
NLTT~33370~AB  at $\sim$100~GHz by Williams (Williams et al., in prep.)
show a surprisingly high
level of circular polarization ($\sim$30\%), posing a possible challenge for
this model (see Williams et al. 2015; but cf. Section~\ref{sec:firstthermal}).

Williams proposed a ``van
Allen belt'' model for the quiescent emission from UCDs, whereby energetic particles become
trapped in a stable, dipolar magnetosphere very different from the
magnetospheres of solar-type stars. He noted that this would go hand-in-hand with
the auroral model discussed by Hallinan, which also requires a stable
magnetosphere. 

The van Allen belt
model lends itself to a physically elegant description of charged
particle motion, and Williams has begun developing 
sophisticated  codes for the purpose of  simulating radio
(synchrotron) emission from van
Allen belt-type structures
under an open source project he calls
``vernon''.\footnote{\url{https://github.com/pkgw/rimphony}}
Crucial pieces include solving
the steady-state Fokker-Planck equation to get particle distributions
in a 6D phase space (that includes energy and pitch angle of the particles)
and then performing radiative transfer calculations to simulate the resulting
emission distribution.  Williams has built on the ``symphony'' code of Pandya et al. (2016)
by incorporating ingredients  necessary
for the proper treatment of the
planetary radiation belt problem, including anisotropy
in the radiative transfer and routines to calculate Faraday
coefficients.
Williams presented initial results from his code as applied to the
Jovian system, demonstrating that the results appear quite promising.

Recent 
efforts to detect flare-like emission from UCDs at moderately low
frequencies were described
by A. Zic (University of Sydney). 
Zic's ongoing study focused on two well-studied objects for which he
obtained 0.6 and 1.0~GHz observations with
the GMRT:
the late M dwarf TVLM~513-46, and the M7 binary LSPM~J1314+1320. 
Assuming that the quiescent radio emission (including flares) from
these two stars is gyrosynchrotron in nature (McLean et al. 2011; Lynch et al. 2015), 
detection of a low-frequency turnover beyond
which the radio emission becomes optically thick is expected to
break the degeneracy between magnetic field
strength and number density in the flare-emitting region. 
Zic presented first results based on the
1~GHz data.
TVLM~513-46 was undetected, but LSPM~J1314+1320 exhibited both
quiescent emission ($\sim$1~mJy) and short-duration ($<$1~minute)
flaring that peaked
at $\sim$5~mJy. These represent the lowest frequency and shortest
duration burst detected from this system to date.  

\subsection{Mass Measurements of Brown Dwarfs}
Direct mass measurements
are crucial for constraining evolutionary models of brown
dwarfs.
As highlighted by Hallinan, VLBI observations of the radio emission
from binary brown dwarfs are playing a crucial role in providing dynamical masses
(Dupuy et al. 2016; Forbrich et al. 2016a). These measurements are
unique, as most of these sources are  too faint to be observed by
{\it Gaia}, and adaptive optics measurements in the IR can yield
on only {\em relative}
motions, leading to a degeneracy in the masses of the individual
components.

\section{Radio Emission and Exoplanets and their Host Stars}
\protect\label{sec:exoplanets}
\subsection{Direct Radio Detections of Exoplanets}
While by design the RS-2 meeting was focused primarily on the study of stellar radio emission
rather than exoplanet searches and characterization,
G. Hallinan remarked that radio searches for exoplanets are expected to be
one of the next major
frontiers in astrophysics. The many low-frequency  ($<$300~MHz) arrays coming on line are
poised to aid in these quests, although detection of Jupiter-like objects will
likely remain challenging prior to the availability of the 
Square Kilometer Array (SKA). Nonetheless, detections are possible, since as noted by C. Lynch,
models predicting the radio fluxes are still poorly
constrained.  Furthermore, the first (free-floating) exoplanet may
have already been detected in the radio.  M. Kao and
collaborators  (Kao et al. 2016, 2018; Section~\ref{sec:ultracool}) 
recently detected a UCD whose derived mass of $\sim13M_{\rm
  Jupiter}$ (Gagn\'e et al. 2017)
places it near the boundary between brown dwarfs and
exoplanets. 
M. Anderson pointed out that ultimately the best place to
detect exo-Jupiters will be from space,  where it is possible to access
frequencies  below the ionospheric cutoff ($\lsim$10~MHz)---the best 
realm for detecting magnetic fields analogous to
those of solar system objects. 

\subsection{Extrasolar Space Weather}
\protect\label{sec:weather}
The recent discoveries of rocky planets surrounding cool and ultracool
stars (e.g., Dressing \& Charbonneau 2015; 
Gillon et al. 2017) naturally leads to speculation about the habitability of these
bodies. This question has captivated not
only astrophysicists, but also the public, and 
is one where radio astronomy is poised to play a key role. Indeed, 
R. Osten noted that this has been a major force
in reinvigorating interest in studies of stellar magnetic
activity at radio wavelengths (although they are
often reframed in terms of understanding how a star might influence its
environment). This topic is commonly referred to as ``extrasolar
space weather'', although J. Linsky pointed out that  extrasolar space weather often refers to
the sum total of
events that affect planets over their lifetimes, making
it more akin to ``climatology'' than ``weather''.

As noted by Linsky, Osten, and other speakers,  
the question of exoplanet habitability goes far beyond the
 presence of liquid water, to matters such as the
importance of CMEs  (which may compress planetary magnetospheres,
exposing their atmospheres to erosion) and whether various energetic processes (flares, ionized
winds, pick-up
processes, UV heating,
sputtering)  might destroy ozone and otherwise impact the atmospheric chemistry of
planets (e.g., Osten \& Crosley 2017; 
Airapetian et al. 2018). And importantly, 
not only do many late-type stars have significantly higher
activity levels than the Sun (e.g., Shibayama et al. 2013; Karmakar et
al. 2017), 
but as J. Villadsen reminded us, they may stay highly active for up to
several Gyr, in contrast with the Sun where
activity levels fall steeply within a few hundred Myr past the zero
age main sequence (see Guinan \& Engle 2008). 

In her review of extrasolar space weather,
Osten noted that if one makes
the assumption that CMEs accompany flares, an implication is that
low-mass stars must incur a very large mass-loss rate due to CMEs. 
This would dramatically impact habitability, since not only are
potentially habitable M dwarf planets closer to their host stars than
the Earth is to the Sun, but they also will likely 
be tidally locked and may possess a weak magnetic moment, 
providing little magnetospheric protection from CMEs 
(Lammer et al. 2007).  However, 
as discussed in Section~\ref{sec:active}, there have so far been no
unambiguous detections of CMEs (or their associated high-energy
particles) from other stars to support this hypothesis. 
Osten further pointed out that none of the 
hard X-ray flares that have been detected from other stars to date have been
convincingly shown to be {\em nonthermal} in nature (Osten et
al. 2016). 
 
As a consequence of this present lack of direct CME measurements,
researchers commonly resort to extrapolating properties of
the Sun's flares  to the low stellar mass
regime  in order
to infer information on space weather and planet habitability. However, Osten warned that there 
is no compelling evidence that such scalings are valid (e.g., 
Osten \& Wolk 2015). For example, previous work by
Segura et al. (2010) showed that a typical UV flare will
have only a $\sim$1\% effect on the ozone layer of an earth-like
planet. On the other hand, the same authors showed that 
the associated energetic particles (as scaled from
solar events) would be sufficient to completely destroy the atmosphere
of an Earth-like planet in the habitable zone of an M dwarf. 

Osten further noted that while
both solar and stellar flares have similar radiative energy partitions
(i.e, similar ratios of radiative energy of flares to the total
blackbody emission of the star; Osten \& Wolk 2015), earlier work of
G\"udel et al. (1996) showed
that stellar flares tend to produce stronger radio amplitudes relative to
their X-ray emission than do solar flares. 
Subsequent studies have also found that solar eruptive events tend to have
similar amounts of  energy in radiation and energetic
particles (as measured through X-rays; Emslie et al. 2012), while M dwarf flares tend to have a
larger fraction of energy in nonthermal particles (Smith, G\"udel, \& Audard
2005).
One positive step forward was recently made by Osten and her collaborators
(Crosley, Osten, \& Norman 2017), who
in effect observed the Sun as an unresolved star in order to develop
a framework for the interpretation of stellar events and the
derivation of parameters such as CME velocities, masses, and kinetic
energies. 

In a parallel effort,
M. Anderson (California Institute of Technology) described work underway with the Owens
Valley Radio Observatory Long Wavelength Array (OVRO-LWA; Hallinan \& Anderson 2017) to search
for signatures of extrasolar space weather phenomena at frequencies  where
emission from type~II burst-like emission and
ECM from exoplanets are expected to
peak ($\nu<100$~MHz). 
The all-sky coverage of the OVRO-LWA is uniquely powerful in allowing
it  to simultaneously observe many
objects and potentially catch rare events. Imaging in
Stokes V also capitalizes on the
the circularly polarized nature of the expected emission to improve the
signal-to-noise ratio (SNR) at these low 
frequencies where confusion dominates the background (see also Section~\ref{sec:activity}). 

The survey by Anderson's team is targeting a volume-limited sample of more than 4000 nearby
($d\lsim$25~pc) stars spanning a range in spectral type to look for
signatures of CMEs and radio emission from extrasolar planets
(Anderson \& Hallinan 2017). The sample includes $>$1000 L, T, and Y
UCDs, 1300 M dwarfs, and other higher mass stars. A primary goal of this work
it to gain   a better understanding of how
CMEs (and their impact on exoplanet habitability) 
scale with flare energy and frequency. Anderson's team also seeks to
further test the suitability of  scaling relations between
flare flux with CME mass based on the Sun (see also above) by comparing with 
results inferred by assuming  energy equipartition
between the bolometric flare energy and the kinetic energy. 

Observations by Anderson's team have been ongoing  since November
2016, and simultaneous optical monitoring is expected to 
provide critical information for tying together flares with radio
emission linked to CMEs. When complete, the resulting survey will be
equivalent to $>$5000 hours of targeted observations with sufficient
frequency and time 
resolution to identify the types of frequency drifts that are
characteristic of CME-associated type~II bursts and  plasma propagating out through the corona. 

\subsection{Effects of Exoplanets on their Stellar Hosts}
\protect\label{sec:exoeffects}
A few speakers briefly touched on the possible influences
exoplanets may have on the radio emission of their host stars. For
example, J. Linsky highlighted recent mm observations of 
Fomalhaut's disk that exhibit the
hallmarks of planetary interactions (MacGregor et al. 2017), and he
cited the importance of ALMA for answering the question of at what
stage of star formation Jupiter-like planets form (e.g., Ansdell et
al. 2017). R. Osten also drew attention to the 
study by Bower et al. (2016), which was the first to detect radio
emission from a non-degenerate star 
that hosts an exoplanet (in this case, a T Tauri star). 
She noted that future observations of this type may be a
promising way of yielding insights into the magnetic interaction between stars and
their planets.

In the case of UCDs  (Section~\ref{sec:ultracool}), one explanation discussed by G. Hallinan to
account for the electrodynamic engine underlying the observed
auroral-type emissions is the presence of an
embedded planet within the dwarf's magnetosphere. However, compelling
observational evidence for this model is presently lacking, and Hallinan noted that
the brown dwarf TRAPPIST-1 (which has many associated rocky
planets; Gillon et al. 2017), does not fit this picture (Pineda
\& Hallinan 2018).

\subsection{Radio Stars as Beacons for the Study of Exoplanets}
\protect\label{sec:exoprobes}
Observations of radio signals from spacecraft as they pass near
bodies within our solar system have long been used as a
means to study the atmospheres and magnetospheres of other planets.
P.  Withers (Boston University) considered the question of whether
this technique could be extended to exoplanets---i.e., whether the
observation of stellar radio emission that had passed through the
atmosphere of an exoplanet could be used to probe
exoplanet properties.  Such an occurrence
is expected to affect the amplitude, frequency, and polarization of
the radio signal and could in principle be used to infer parameters
such as the neutral density in the atmosphere, plasma density in the
ionosphere and magnetosphere, and the magnetic field in the plasma
environment  (Withers \& Vogt 2017). Withers is presently researching 
the most promising candidates for detecting such effects. S. Wolk pointed out that
there may be applicable lessons from analogous investigations recently
done in the X-ray (e.g., Wolk, Pillitteri, \& Poppenhaeger 2017).

\section{The Sun as a Radio Star}
\protect\label{sec:sun-as-star}
Radio emission from the Sun  samples a wide
variety of emission processes from both thermal plasma and nonthermal
electrons. At least five different solar radio emission mechanisms
have been
observed, including
bremsstrahlung, gyroresonance, gyrosynchrotron, and coherent emission in
the form of ECM and plasma emission (e.g., Figure~4.1 of Gary \& Hurford
2004). 
As pointed out by S. White, all five of these emission mechanisms
produce circularly polarized emission, which provides additional
diagnostic information.  (In contrast, linear polarization is wiped out by Faraday rotation
on the Sun, rendering it unobservable).

Radio observations of the Sun are in general not SNR
limited. Until recently, they have been instead limited by other
technological constraints. For example, dynamic spectroscopy of solar
emission has been limited by the
bandwidths and sampling rates of the available instruments, most of
which lacked
imaging capabilities. However, 
as was amply illustrated at the meeting, solar radio astronomy
has entered an exciting new era thanks to the latest
generation of instruments capable of dynamic spectro-imaging with
high time and spectral resolution and over ultra-wide bands. 

\subsection{The Sun at Low Frequencies}
\protect\label{sec:lowfreqSun}
At frequencies below $\lsim$1~GHz, solar emission is dominated by a
combination of thermal bremsstrahlung and
plasma emission, the latter of which is seen both at the fundamental and the
second harmonic of the plasma frequency. As discussed by
S. White, a number of puzzles remain concerning the observed properties of this
low-frequency solar emission. For example, the fundamental mode of the
plasma emission is predicted to be 100\% circularly polarized, but
this is not always observed. Furthermore, absorption is
thought to be quite high near the fundamental mode, necessitating
some kind of duct to get the fundamental emission to emerge
rather than getting absorbed. 

Fortunately, the latest generation of
low-frequency observatories are revolutionizing our understanding of
the meter-wave emission from the Sun, and
C. Lonsdale (MIT Haystack Observatory) described the exquisitely
detailed imaging  at frequencies of
80--300~MHz that are possible with the
MWA. The MWA is a powerful solar imaging telescope
owing to excellent instantaneous monochromatic $u$-$v$ coverage and
 high time and frequency
resolution  (e.g., Mohan \& Oberoi 2017).

To date, many petabytes of solar data have been gathered with the MWA, but
a hurdle to exploiting them has been the challenge of developing  
a robust and accurate reduction method  (e.g., Lonsdale et al. 2017).
As Lonsdale described,
significant progress has now been made, resulting in solar images that
reach a dynamic range of up to 75000:1, and efforts are underway to achieve
this on a routine basis.  
In comparison, the best that has
been achieved by other instruments at similar frequencies
is $\sim$300:1. Despite their high dynamic range, 
the MWA images are still not thermal noise-limited,
and the limiting factor is now thought to be ionospheric
microstructures tens to hundreds of meters in size that  produce
phase fluctuations of a few degrees on long baselines. 

Lonsdale reported that a handful of type~II solar bursts have been
captured by the MWA.   A hallmark of these bursts is a slow
drift to lower frequencies with time that is visible
in dynamic radio spectra, and these events 
are of considerable interest because of
their close link with CMEs (e.g., White 2007).
The analysis of a September 2014 event  
has provided the most detailed spatial information on a
type~II event ever achieved (Lonsdale et al. 2017; Oberoi et al.,
in prep.). During this burst, both the fundamental and second
harmonic of the plasma emission were recorded two minutes apart in 30~MHz bands 
centered at 90~MHz and 120~MHz, respectively.  Different frequency channels within
each band sample different plasma frequencies (and hence different
electron densities), revealing a significant amount of
complex, fine-scale structure. Analysis of these data also showed that the fundamental and
harmonic emission appear to originate from different spatial
locations, implying systematic outward motion from the Sun. The SNR was sufficient to enable
extraction of spatial
information on scales of a few percent of the MWA's resolution, and
the quality of the data enable high-fidelity imaging of every pixel in
a dynamic spectrum (with time and frequency
resolution of 0.5~s and 40~kHz, respectively).

For another type~II burst, MWA  data were analyzed
following the disappearance of the event, leading to the detection of
gyrosynchrotron emission from the associated, outwardly moving CME
material. Further analysis of these data will include attempting to measure
the linear polarization  to characterize the magnetic
field in the CME plasma (Kozarev et al. 2016 and in prep.).

Lonsdale also illustrated the power of the MWA for imaging other types
of solar events, including type~III bursts.  The latter give rise to
radio emission at the plasma frequency and its harmonics and are
characterized by rapid drifts in frequency with time (e.g., White
2007). Multi-frequency imaging
of one such type~III event by McCauley et al. (2017) showed that the intensity profiles of
the burst emission become
spatially more extended at lower frequencies and are double-peaked. The
explanation for this phenomenon is that the electron beams that are
traveling along the magnetic field lines are separating (over a time
span of $\sim$2~s) because of
travel along diverging field lines.

Crucial to mining the enormous volumes of solar data generated by the
MWA and other modern radio facilities used for dynamic spectro-imaging 
will be the assistance from automated feature identification algorithms.
Lonsdale drew attention to recent work by Suresh et al. (2017), who used a
wavelet-based approach to automatically identify and characterize the
numerous weak and short-lived emission features (undetectable with
previous instruments) that are found in typical MWA
solar data. Such features appear to be ubiquitous even during times of low
solar activity (see also Oberoi et al. 2011). At times of medium
activity on the Sun, up to half of the total solar flux
density within the MWA bands comes from the aforementioned
``mini-flare'' events  (Sharma, Oberoi, \&
Arjunwadkar 2018) and these are suspected to be of importance for coronal and
chromospheric heating.

Another low-frequency solar imaging instrument that is under
construction is the Radio Array of Portable Interferometric Detectors
(RAPID), a project led by MIT Haystack Observatory  (see Lonsdale
et al. 2017 and references therein). RAPID will comprise an array of
50--70 small, portable, reconfigurable, 
fully independent stations with self-contained power and data
storage. Its wide instantaneous frequency coverage and reconfigurable
layout will make it ideal for imaging different types of solar emissions,
surpassing even the  MWA. 

\subsection{The Sun at Centimeter Wavelengths}
\protect\label{sec:vlaSun}
B. Chen provided an overview of
the Sun at $\nu\gsim$1~GHz, where bremsstrahlung ($\nu\sim$1--3~GHz) and gyromagnetic
radiation processes ($\nu\gsim$3~GHz) dominate. As 
at lower frequencies (Section~\ref{sec:lowfreqSun}), solar work
in this domain has been revolutionized by the expansion in
bandwidth and sampling rates of the available instrumentation and the
ability to perform exquisitely detailed dynamic spectro-imaging.

Chen drew attention to the recent commissioning of the only
dedicated solar radio observatory currently operating in the United
States, the Expanded Owens Valley Solar Array (EOVSA).
It offers a usable frequency range of $\sim$2.5-18~GHz (Gary et
al. 2017) and it sweeps this entire range once per
second. While its relatively small number of antennas (14) do not
provide outstanding snapshot imaging capabilities, its broad frequency
coverage and dedicated solar coverage are key advantages. 

\begin{figure*}  
\centering
\scalebox{0.27}{\rotatebox{0}{\includegraphics{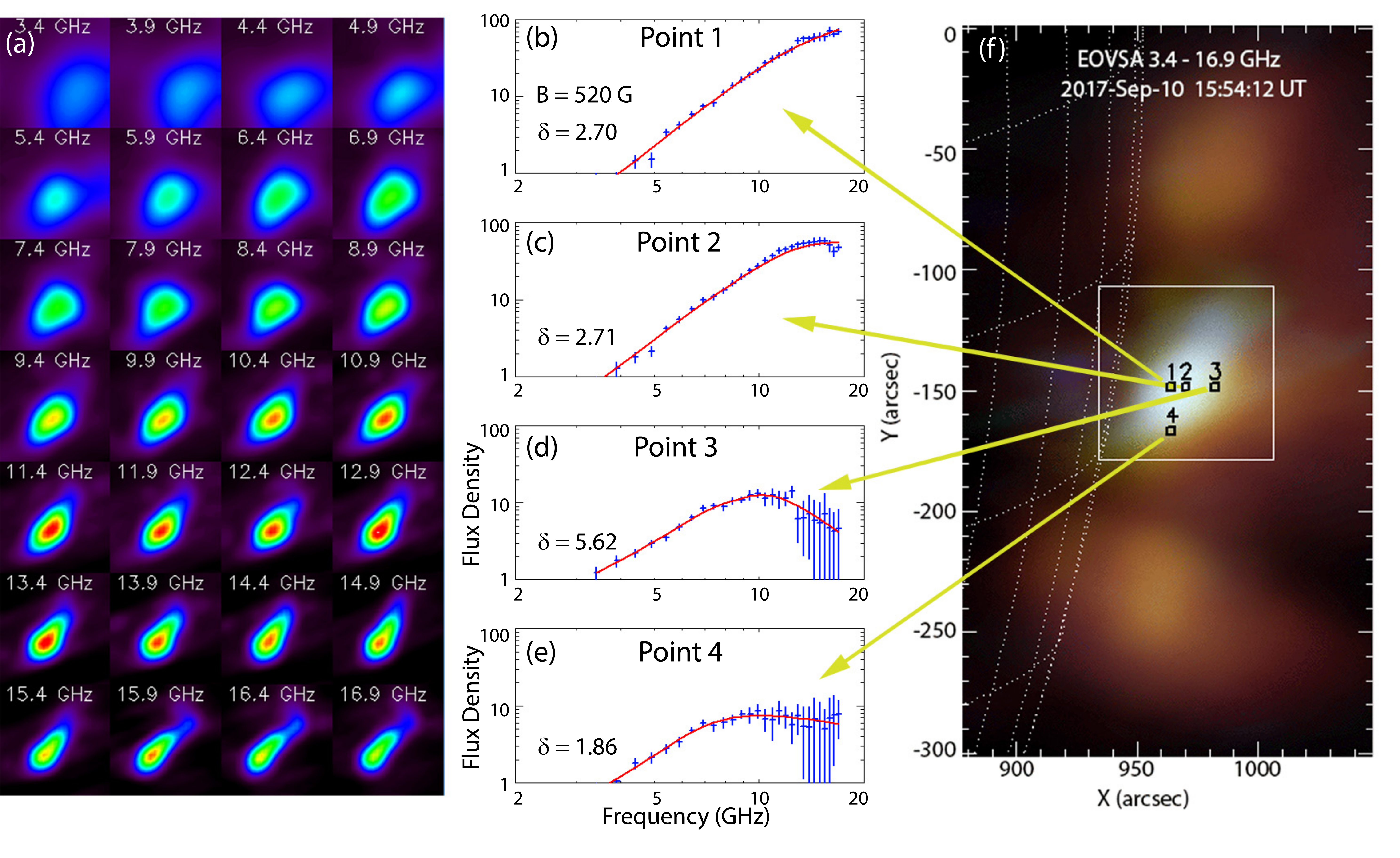}}}
\caption{Single-frequency images and spatially resolved gyrosynchrotron
  spectra of a solar limb flare observed with the EOVSA  on 2017
  September 10. The field-of-view of each image corresponds to the white box
  in the right panel. The spectra are extracted from single image
  pixels and have a scale in solar flux units per
  pixel, where the pixel area is $2''\times2''$. From Gary et
  al. (2018); reproduced by permission of the AAS. }
\label{fig:chen-solar}
\end{figure*}

Chen described recent work using EOVSA to observe the Sun
during the impulsive energy release phase of a
2017 X8.1 X-ray flare (Gary et al. 2018). The resolving power of the array was sufficient to
reveal a reservoir of hot electrons 
lying on top of the coronal loops imaged in the extreme UV, and the broad
spectral response confirms gyrosynchrotron emission spanning from 3--18~GHz. The
EOVSA data enabled studying the time-varying SED of this
emission (Figure~\ref{fig:chen-solar}), and fits to
the spectrum to yield the magnetic field strength and electron
density. The peak frequency of the emission decreases as material moves from low
altitude to high altitude owing to the corresponding drop in the
magnetic field, and the data provide information on the electron energy
spectrum as a function of position, and therefore clues on where the 
electrons are accelerated. 
Chen also described how the wide bandwidths provided by EOVSA
enable a new kind of study of sunspots; since
the frequency of the emission observed is proportional to the harmonic number times
the gyrofrequency, which is in turn a function of the magnetic field,
it is possible to effectively map the magnetic
field in 3D. 

Another key instrument for solar work in recent years is the upgraded VLA,
which has been 
available for solar observing 
in the 1--8~GHz range since 2011 (with  plans underway to commission
additional higher and lower frequency bands). 
The VLA can provide time and frequency resolution of up to 5~ms  and
1~MHz, respectively, with up to $\sim$10k channels,
implying that it is in theory possible to create millions of images per minute
of observation. Chen highlighted several recent results from VLA
solar observations.

A longstanding puzzle has been
the mechanism responsible for the conversion of magnetic
energy into the kinetic energy required for the acceleration of
relativistic particles during solar flares. A leading candidate has been termination shocks, which can
occur when a high-speed jet speed exceeds the local Alfv\'en speed. Such shocks are
predicted by numerical simulations, but their existence had yet to be
convincingly established observationally. As described by
Chen, this has changed thanks to
recent high cadence imaging spectroscopy of a solar flare with the
VLA, in combination with UV and X-ray observations
(Chen et al. 2015). With the VLA, Chen's team found short-lived bursts of
radio emission coincident with the predicted location of the
termination shock.  
Density fluctuations along such a termination shock map to different
frequencies, so by taking a dynamic spectrum it becomes possible to map the evolution
of the shock surface. The VLA dynamic spectra revealed that
upon arrival of the reconnection
outflow, the shock fragmented. In tandem, hard X-ray data showed
that the electron energy
distribution became softer. 
Thus Chen et al. concluded that
termination shock must be at least partially responsible for
electron acceleration in the flare. 

Another topic addressed by Chen is the
recent use of the VLA to trace fast electron beams associated with
type~III solar bursts. Electron beams travel up and down along
magnetic field lines, encountering different densities along the way, 
resulting in corresponding changes in the frequency of the
emitted radio waves. With the ability to make images at multiple
frequencies within a dynamic spectrum, it becomes possible to map the
electron trajectories in the solar corona. New data obtained by Chen
and collaborators show that during a
type~III burst, all of the electron beams originate from a single
point in the corona believed be the reconnection
site. In these types of events it therefore appears that the reconnection event is
responsible for accelerating electrons.

Finally, while  bremsstrahlung  and gyromagnetic
radiation dominate the solar emission at cm wavelengths, ECM emission
can also operate at these frequencies  and is believed to have powered
a 2006
spike burst event whose strength ($10^{10}$~Jy over 1--2~GHz) was sufficient to knock out global
positioning systems (Gary et al. 2007). The implications of such
events are clearly important to Earth, but as S. White pointed out,
the origin of this emission is is still not well understood and
the auroral  model for ECM that works for chemically
peculiar stars and UCDs (Section~\ref{sec:ultracool}) cannot be applied.

\subsection{The Sun at Millimeter Wavelengths}
\protect\label{sec:mmSun}
Solar observing with ALMA  using its Band~3 (100~GHz) and 6
(230~GHz) receivers became possible for the first time 
in late 2016 (e.g., White et al. 2017).
ALMA is now able to probe the solar chromosphere on scales from $\sim1-1.003R_{\odot}$
with unprecedented sub-arcsecond angular resolution, providing data
free from the complications of the non-LTE effects that impact traditional UV
emission line studies.
Commissioning of additional ALMA bands for solar work, as well solar polarization
capabilities, remains ongoing. 

Because ALMA's solar capability is so new, many of the first data are still
being analyzed, but RS-2 participants saw some exciting previews. 
For example, B. Chen showed how images of sunspot oscillations
obtained with ALMA can be compared
with those at other wavelengths (extreme UV, H$\alpha$) 
to constrain the height where the oscillation
is occurring. J. Linsky and S. White also briefly touched on the
potential of ALMA for measuring chromospheric magnetic
fields via circular polarization measurements (see also Section~\ref{sec:chromospheres}). 
However, such work will be challenging, as theoretical work by White and
collaborators found that
the expected degree of predicted polarization in the quiet Sun (excluding
sunspots) is quite small ($\lsim$1\%; Loukitcheva et al. 2017). 

\section{Stellar Winds across the H-R Diagram}
\protect\label{sec:winds}
K. Gayley began his review of stellar winds
with the observation that it is far
easier to explain why stars should {\em not} have winds
than it is to explain why they do.
He clarified by pointing out that powering a wind requires moving heat
from the stellar interior to its surface, and in general, the ratio of the surface
temperature to the interior temperature will be too small to
allow escape of material from the surface (e.g., a dwarf will
typically have $\sim10^{-3}$ times the energy at the surface needed
for escape). 
One way to overcome this is to deposit heat onto the surface in
places where the density is low, raising the outside temperature to a
level comparable to the inside temperature, and consequently pushing the
thermal velocity to close to the escape velocity.  Another option is to
mechanically push the surface out. This can be achieved with radiation pressure on
dust (as in cool asymptotic giant branch (AGB) stars; Section~\ref{sec:AGBwinds}) or line opacity 
(as occurs for high-luminosity stars; Section~\ref{sec:hotwinds});
magnetohydrodynamic
waves or field stresses (for rapid rotators; e.g., Section~\ref{sec:mag-OB}); or through gas pressure
heated by waves or reconnection events (as in coronal-type winds; Section~\ref{sec:coolwinds}).

Gayley pointed out that in dwarfs (Section~\ref{sec:coolwinds}),
coronal winds must overcome enormous and efficient radiative line
cooling that become increasingly important as temperatures approach
$\sim$100,000~K (or equivalently, $v_{\rm esc}\sim$40~\kms). Getting
over this radiative cooling
``barrier'' necessitates having very low-density gas, implying that it is
not possible to have a high-density coronal wind. On the other
hand, in giants, where the escape velocity is lower, achieving a high
thermal speed is no longer necessary ($T\sim20,000$~K), and somewhat
denser ``chromospheric winds'' can be achieved. 
Gayley proposed that this distinction likely accounts for the so-called ``coronal
graveyard''\footnote{The term ``coronal graveyard'' has been used to mark the dividing line
beyond which giants are not found to display coronal signatures in the X-ray and UV; in general,
coronae are not seen in giants redward of spectral type
K1~III (Ayres 2018 and references therein).} 
first identified by Linsky \& Haisch (1979), although more
work is needed (see also Suzuki 2007). He also
suggested that this type of driving mechanism may be responsible
for the ``superwinds'' of AGB (Section~\ref{sec:AGBwinds}) or 
red supergiant (RSG) stars (Section~\ref{sec:RSGs}), although additional radiative driving
would probably be required to power these fast, dense winds. 
For high-luminosity stars, radiation pressure can drive winds, given a
source of opacity in the form of dust [as in AGB stars
(Section~\ref{sec:AGBwinds}) and possibly
RSGs (Section~\ref{sec:RSGs})],
continuum radiation (as in luminous blue variables), or UV resonance
lines [as in Wolf-Rayet (W-R)
stars, blue supergiants, main sequence OB stars   (Section~\ref{sec:hotwinds}), and the central stars
of planetary nebulae (PNe; Section~\ref{sec:PNe})]. 

Gayley concluded by listing several unsolved problems concerning stellar winds
where radio observations are likely to provide new insights. These included:

\noindent {\it (1) How is mass loss affected by rapid rotation or strong magnetic
  fields?} While rotation helps to remove some of the gravitational barrier,
  it does not increase mass-loss at the
  equator. Instead, the
  star will tend to bulge at the equator (become oblate), causing the
  radiative flux to emerge from the poles. An exception is if the star
  is spun up to critical rotation, in which case a disk will form that
  is unrelated to the wind (as in the Be stars; Section~\ref{sec:Be}).

\noindent{\it  (2) How can we
gauge the past and future properties of winds? Are they continuous or
episodic? Do they increase or decrease with
  age? } Gayley noted that 21~cm line observations of AGB stars
(Section~\ref{sec:AGBwinds}) are 
of interest in this regard, since they probe  the
entire history of the mass loss, including distances beyond which
traditionally used molecular tracers of the circumstellar material
become dissociated by the interstellar radiation field. In
addition, the presence of knots and other features in the \HI\ gas
provide clues on whether the mass loss is episodic (see Matthews
et al. 2013).

\section{Radio Emission from Hot Stars}
\subsection{Hot Star Winds}
\protect\label{sec:hotwinds}
The most massive stars have strong winds that produce significant
rates of mass loss ($\sim10^{-6}-10^{-4}~M_{\odot}$ yr$^{-1}$),
with important consequences for the evolution of the star
 and  for the surrounding interstellar medium (ISM).  
The basics of hot star
winds are now relatively well-understood. These winds are
radiatively-driven, with line absorption providing the dominant opacity,
and are typically described by the so-called `CAK' theory (Castor,
Abbott, \& Klein 1975). However, some important questions
remain, and the review talks by 
I. Stevens (University of Birmingham) and K. Gayley touched on how radio
observations can be uniquely exploited in the quest to answer them.

Radio
emission mechanisms that are relevant for studying the winds of
massive stars include thermal free-free
emission from the ionized winds of single stars and the nonthermal (synchrotron)
emission 
that can arise from colliding winds in binaries
(Section~\ref{sec:colliding}). 
Other possible, though less explored radio
emission mechanisms include magnetic reconnection events in
interacting binaries (leading to
particle acceleration) and ECM emission (see Section~\ref{sec:mag-OB})---
but with different 
plasma conditions compared to the low-mass star case
(cf. Section~\ref{sec:ultracool}). 
The winds of runaway massive stars can also impact
the ambient medium, creating a bow shock that emits both
thermal and nonthermal radio
emission (Benaglia et al. 2010; Brookes et al. 2016a).

The winds of massive stars are dynamically unstable since the radiative driving is
very non-linear, leading to shock formation and clumping. Because free-free emission depends on the
density squared,  clumping will enhance the radio
emission and produce variations in the spectral index as a function of
frequency (e.g., Daley-Yates, Stevens, \& Crossland 2016). Radio
observations therefore provide unique insights into the nature of this
clumping 
(Section~\ref{sec:clumps}). Compared with 
X-ray and H$\alpha$ emission in massive star winds, which arise from within
a few stellar radii, different radio wavelengths also have the
advantage that they are able to probe from  tens
of stellar radii (sub-mm) to hundreds of stellar radii (1~GHz). 

While there is general agreement  that radio observations supply
one of the best means of deriving mass-loss rates for hot stars
(relying on fewer assumptions than other methods), 
Gayley  emphasized  that current factors of 2--3 uncertainty in
typical ${\dot M}$ values can have important consequences for the
impact of the mass loss on the evolution of a
given star.  Further refinement of these values through better 
observations and modeling therefore remains important.

Stevens noted that
higher radio frequencies (mm/submm) begin to probe the little-understood
acceleration regions of hot star winds. He pointed out that 
the effect of seeing into the acceleration region of the wind
at these wavelengths can produce a flux excess compared with a
canonical Wright \& Barlow (1975) model, leading to a
non-linear spectral index and causing errors in derived mass-loss
rates if  not accounted for (Daley-Yates et
al. 2016). The effects of clumping (in particular, radius-dependent
clumping; see also below) can further complicate interpretation of the mm/submm part of
the spectrum, leading to degeneracies in wind
models between the velocity law and the amount of clumping.  An
additional 
effect may be produced by He recombination in different regions of the
wind, leading to more electrons than expected in the
inner wind compared with the outer wind. Stevens noted that the
acquisition of continuous radio
spectrum over three orders of magnitude in frequency might be able to
disentangle these various effects, but we presently lack such
data, and indeed, most hot star radio
``spectra'' contain just a few points. 

\subsubsection{The Effects of Clumping}
\protect\label{sec:clumps}
R. Ignace (East Tennessee State University) discussed clumping and
other effects that must be properly treated in order to accurately
interpret radio observations of the ionized winds of 
hot stars.
He noted that with ``microclumping'' (effectively a modified smooth
wind), the clumps are optically thin, hence their shape
is unimportant. However, this type of clumping affects the observed radio
flux density and must be accounted for to 
accurately derive
mass-loss rates;
ignoring the clumping leads to
overestimates of ${\dot M}$.

In the case of ``macroclumping'', the  wind is composed of discrete density
structures with a range of optical depths. The resulting effects have been previously studied
in the X-ray and UV, but little explored until now in the
radio bands (Ignace 2016). Ignace showed that because the clumps are
optically thick, 
different clump geometries can in principle be expected
to have observable 
effects on both the radio fluxes and the slope of the SED. However, in
two sample geometries that he considered---fragments (``pancakes'')
and spheres---the effects are not significantly different from the
microclumping case, except when the spherical clumps have
a  very substantial volume filling factor. 

Ignace also revisited the earlier work of White (1985) that 
reported evidence of synchrotron
emission from apparently single OB stars. Some preliminary model calculations
performed by Ignace raise the question of whether the presence of synchrotron emission
might in some cases be misinterpreted as wind
clumping gradients (e.g., in the case of $\zeta$~Pup; cf. Ignace
2016).

\subsubsection{Colliding Wind Binaries}
\protect\label{sec:colliding}
In close massive binaries, wind
interactions can produce shock regions that are
believed to be sites of particle acceleration and the production of
nonthermal radio emission  (De~Becker \& Raucq 2013; De~Becker et
al. 2017). 
However, there are some close OB binaries with
no such radio emission. I. Stevens suggested that one possible
explanation is that the 
emission is dependent on the degree of magnetic field
alignment.

As described by Stevens, 
low-frequency (meter wave) observations are useful for identifying the turnover
frequency for colliding wind binaries, and he described such observations recently
undertaken for one such binary, WR~147, using the GMRT at 235 and 610~MHz. The system
was undetected at 235~MHz, indicating a steep downturn in the SED at
low frequencies. A lingering puzzle is that the nonthermal emission defies fitting
using a simple  model comprising synchrotron plus free absorption (Brookes, Stevens, \& Pittard
2016). Clumping may play a role, but
more data  points along the radio spectrum are needed to distinguish
between possible explanations.

\subsubsection{Massive Stars in Clusters}
\begin{figure} 
\centering
\vspace{-1.0cm}
\scalebox{0.4}{\rotatebox{0}{\includegraphics{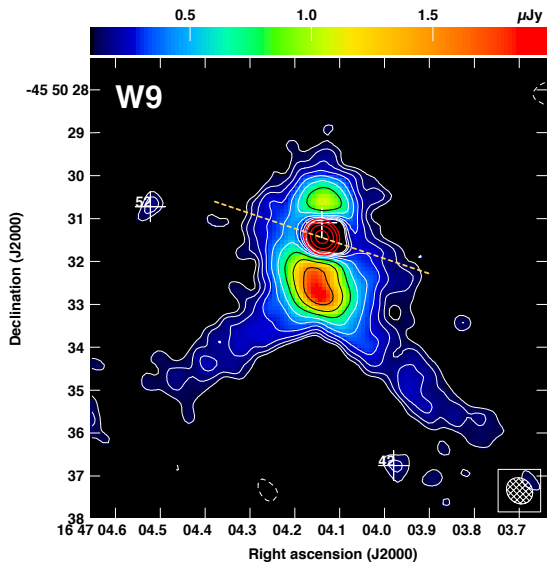}}}
\vspace{-2.0cm}
\caption{ALMA 100~GHz image (contours and color scale) of the sgB[e] star Wd1-9, a member of the
  5-Myr-old cluster Wd~1.  The extended emission
  contours (white) are ($-1$, 1, 1.4...16)$\times$99~$\mu$Jy
  beam$^{-1}$; for the compact source (red) they are (0.1,
  0.14, 0.2, 0.28)$\times$416~mJy
  beam$^{-1}$.  The dashed line indicates the orientation of an
  apparent bipolar outflow. Wd1-9 exhibits similarities to the enigmatic
  radio star MWC~349A (Section~\ref{sec:MWC349A}).  From Fenech et
  al. (2018); reproduced with permission from A\&A, \copyright ESO. }
\label{fig:Wd1-9}
\end{figure}

\protect\label{sec:hotclusters}
Work presented by 
D. Fenech (University College London) highlighted the unique
power of ALMA for studying the mass-loss properties of massive stars
and their interactions with their environments. She 
presented results from a comprehensive analysis of the
full Westerlund~1 (Wd~1) cluster based on 3~mm ALMA data (Figure~\ref{fig:Wd1-9}). 
Her team identified 50
radio-emitting stars within this 5~Myr-old cluster, spanning a wide range of
stages of post-main-sequence evolution for massive stars, including 21
W-R stars
and a number of cool and warm supergiants/hypergiants. They used the mm
data to derive mass loss rates for the sample (Fenech et
al. 2018). Unexpectedly,
they also uncovered emission nebulae surrounding a number of OB and W-R stars in
Wd~1, which they suggest result from the interaction between
the stellar ejecta and the intracluster medium. The presence or
absence of such a nebula around a massive star of a given type is
expected to have
important implications during the subsequent supernova explosions of
these stars, as it may affect the light curve, spectrum, and other
phenomenology of the evolving ejecta. 
Another surprise was that more than half of the stars appear to be
spatially resolved. It is unlikely that the observations are
resolving the stellar winds; the inferred sizes are $\sim$20-100
times the expected size of the radio photospheres (which are predicted
to range from $\sim$0.1 to 3.8~mas; Fenech et al. 2018), and the presence of dust around these
hot stars also seems unlikely. The
explanation for this finding is therefore unclear, but it may be related
to the interaction between the winds and the cluster
environment. 

Observations of Wd~1 were also performed recently using the
Australian Telescope Compact Array (ATCA) at 5.5 and 9~GHz by 
H. Andrews (University College London) and collaborators. Andrews 
presented preliminary results on some of the discovered radio sources.
The observations uncovered evidence
for asymmetric ejecta surrounding 3 M-type supergiants and 3 cool
hypergiants in the cluster. These detections provide the first ever
evidence of cometary-like
nebulae  associated with yellow hypergiants (Andrews et al. 2018). The alignment of
the observed asymmetries relative to the cluster center shows clear
evidence of a cluster outflow. F. Yusef-Zadeh noted that a similar
phenomenon has also been seen for stars near the Galactic Center. 

A second presentation by Fenech reported on the ``COBRaS'' project
(Morford, Prinja, \& Fenech 2017),  a 
 survey of the core of the young massive cluster Cyg~OB2  at 1.6 and 5~GHz 
using the optical fiber-linked Multi-Element Radio Linked
  Interferometer Network (e-MERLIN). When completed, the survey will comprise
$\sim$300 hours  of observations spanning multiple pointings  and covering
approximately 20 arcmin$^{2}$. Importantly, the combination of $\mu$Jy
sensitivity and excellent angular resolution ($\sim$40~mas at 5~GHz)
will  suppress contamination from diffuse emission and enable
the detection of free-free emission from
the winds of individual massive stars, allowing the determination of 
mass-loss rates. The 1.6~GHz observations have now been completed (Morford
et al. 2016) and one
surprising finding is that the only non-binary massive
stars detected are the luminous blue variables. To date, about half of the identified radio
sources show no counterparts at other wavelengths. Robust
upper limits on mass-loss rates were obtained for a number of OB giants and supergiants
that in turn provide useful constraints on the wind geometry
and clumping.
Four transients have also been identified in this sample. 

\subsubsection{Magnetic OB Stars}
\protect\label{sec:mag-OB}
Another class of hot star that emits nonthermal radio emission is the rapidly rotating Magnetic
Chemically Peculiar (MCP) stars. Stevens
highlighted recent work on
the MCP star CU~Vir where two radio peaks per rotation
period are seen, both with 100\% circular polarization. Trigilio et
al. (2011) attribute this
to magnetospheric ECM emission. However, there is a discrepancy between the
optical and radio light curves of this star that is not
understood.
Stevens stressed the importance of multi-frequency radio studies of
other MCP stars 
across their entire rotation period (including high-frequency
observations that will permit peering deeper into the magnetospheres) to determine whether CU~Vir is unique.

\subsection{Be Star Disks}
\protect\label{sec:Be}
Classical Be stars
are surrounded with self-ejected, rotationally-supported, slowly outflowing
gaseous disks that are ionized by the
hot, rapidly rotating central star. The 
disk physics are dominated by viscous processes, which provide the outward
transfer of angular momentum. Because the disks are ionized and emit free-free
emission, the cool outer regions can be uniquely probed by cm-wave
observations.  Radio observations can then be combined with data at
other wavelengths (extending shortward to the UV) 
that sample the inner disk and the star itself to
provide important tests of models.

As described in the presentation by
R. Klement (Center for High Angular Resolution Astronomy, Georgia
State University), the inner parts of Be disks
($r\lsim R_{\star}$) appear to be
well-explained by the so-called viscous decretion disk model
(Lee, Saio, \& Osaki 1991), 
but relatively little is known about the outer parts of the disks,
including the effects that may be produced by a binary
companion, and the location of the critical radius (transonic
boundary). The latter marks the point where the radial velocity
exceeds the  sound speed and viscosity can no longer 
transport angular momentum, leading to disk dissipation. 

Until recently, only 8 Be stars had been detected at radio wavelengths
(Waters et al. 1991; Dougherty \& Taylor 1992), but
Klement described the results of a recent study of six Be stars using
new and archival radio observations in combination with multiwavelength
data and radiative transfer modeling (Klement et al. 2017).  
Klement et al. found that in all cases, the cm-wavelength 
flux densities lie below the predictions of
non-truncated disk models. But at the same time, the SED shapes rule
out a {\em sharp} disk truncation. This suggests the presence of disk material
beyond the truncation radius, and a plausible explanation is that all of
these stars are binaries with circumbinary disks. Indeed, binarity is already
confirmed for two of the stars. One of these is
$\beta$~CMi, where the companion search was motivated by the radio results (Dulaney et al. 2017). 
Work is ongoing by Klement and collaborators to observe additional Be
stars at cm and mm wavelengths.

\clearpage
\section{AGB Stars}
For stars with initial masses in the range
0.8---8.0~$M_{\odot}$, the
AGB marks the second ascent of the red giant branch, at which point
post-He burning stars reach luminosities of up to $\sim10^{4}L_{\odot}$ and
their temperatures cool to 2000-3000~K. These cool temperatures are
conducive to dust and molecule formation, and a key property of AGB stars is their intense
stellar winds (${\dot M}\sim10^{-8}$ to $10^{-4}~M_{\odot}$ yr$^{-1}$)
which contribute significantly to the dust and heavy element
enrichment of the ISM (see H\"ofner \& Olofsson
2018). The continuing importance of radio wavelength observations for understanding
AGB stars was reviewed by H. Olofsson (Chalmers University).

\subsection{Radio Photospheres}
\protect\label{sec:photospheres}
The imaging of the radio photospheres of AGB stars was pioneered by
Reid \& Menten (1997).  Olofsson drew attention to 
recent developments on this topic, including the improved
ability of
current-generation radio interferometers to resolve details on the 
surfaces of nearby AGB stars (Matthews et
al. 2015, 2018; Vlemmings et al. 2017), and the possibility to perform
comparisons  of these results with the  predictions of state-of-the-art
3D models, which now include convection, pulsation, and dust
condensation 
(Freytag, Liljegren, \& H\"ofner 2017). 

Radio photospheres lie above the optical photosphere, but inside the
dust-condensation radius, making them an excellent probe of shocks and
thermal structure within the regions of the AGB star atmosphere where the stellar wind is
accelerated and launched.    Because of the
frequency-dependence of the free-free opacity, different radio frequencies
are expected to probe different depths in the atmosphere (Reid \& Menten 1997); 
this has now been confirmed observationally 
in one AGB star (Matthews et al. 2015) and one RSG (Lim et al. 1998; O'Gorman et al. 2015). 

The wide bandwidths available with the VLA and ALMA now permit the simultaneous imaging of
molecular lines along with the radio photosphere, and these lines can be
used as kinematic probes of the atmosphere.  One example is
the recent work of Olofsson and his collaborators that used ALMA to image W~Hya in the
338~GHz continuum, as well as in the CO $v$=1, $J=3-2$ line (Vlemmings
et al. 2017). 
These authors reported evidence for a ``hot spot'' with
$T_{B}>5\times10^{4}$~K that they interpreted as a chromosphere with a
low volume filling factor. Against
the stellar disk, the aforementioned CO line is seen in both emission and absorption
due to the presence of infalling and outflowing gas, suggestive of both cool ($T\approx$900~K)
and warm ($T\approx$2900~K) molecular layers and the presence of
shocks with velocities $\sim$20~\kms. However, Olofsson emphasized
that current models cannot fully explain these results, hence 
follow-up, including additional epochs of observations and
observations of additional stars, are needed.

\subsection{Circumstellar Masers in AGB Stars}
\protect\label{sec:AGBmasers}
L. Sjouwerman (NRAO) described results from the analysis of a large
sample of SiO maser-emitting evolved stars that have been observed as
part of the Bulge Asymmetries an Dynamical Evolution (BAaDE) project
(Trapp et al. 2018; Section~\ref{sec:denizens}). An important piece of
this project is the characterization of how biases and selection effects may impact
the results. As described by Sjouwerman, such investigations 
also have interesting implications for understanding the underlying stellar
astrophysics. As part of this effort, his team collected
near-simultaneous (same-day) measurements of the both 43~GHz and
86~GHz SiO $\nu$=1 maser lines for nearly 100 stars using the ATCA 
(Stroh et al. 2018). A systematic comparison of the line parameters
for the two transitions  revealed that the mean flux density ratio 
is consistent with unity. They also found that the 43~GHz
transition gets successively weaker in AGB stars with thinner
circumstellar envelopes (CSEs). But in contrast to both theoretical
predictions (Humphreys et al. 2002) and previous empirical findings based
on smaller samples (e.g., Pardo et al. 1998), they find no evidence for a turnover in the
NIR colors of Mira-type CSEs beyond which the 86~GHz line is
consistently brighter.
This has implications for the pumping mechanism of the masers and the
physical conditions under which the respective transitions can arise.

\subsection{Mass Loss from AGB Stars}
\protect\label{sec:AGBwinds}
Olofsson underscored that understanding AGB mass loss is fundamental to
understanding AGB stars, since their evolutionary tracks are strongly
dependent on how much mass these stars lose at various times along the
AGB.
However, this is a highly complex problem,
since mass loss depends on many stellar parameters, including mass,
luminosity, temperature, and metallicity, all of which
evolve with time. Indeed, Olofsson pointed to understanding the {\em temporal evolution} of AGB
mass-loss as a key unsolved problem.

 Current evidence points to increasing mass-loss
rates toward the end of the AGB, leading up to a
superwind phase. Initial-final mass relations 
(e.g., Cummings et al. 2016) necessitate that stars lose up to 80\%
of their initial mass
on the AGB. But the cause of the superwind remains unknown, and
observations  paint a confusing picture of the stages leading up to it.  
Olofsson drew attention to the puzzling findings of Justtanont et al. (2013),
who looked at a sample of extreme OH/IR stars thought to be in the
very late stages of the evolution on the AGB. Based on a dynamical
model, these authors inferred mass-loss rates as high as ${\dot
  M}\approx(2-10)\times10^{-4}~M_{\odot}$ yr$^{-1}$. However, values
inferred from  observations of low-$J$ CO lines are far lower ($\approx
3\times10^{-6}~M_{\odot}$ yr$^{-1}$).
A further conclusion is that the superwind phase lasts only 200-600
years---too short to produce the required amount of mass loss. 
One possibility is that there are several superwind
phases, and Olofsson proposed searching for possible signatures by using ALMA to
map out the distribution of multiple CO lines.

A prescription for mass-loss rate based on stellar parameters is also
an important ingredient for modeling the stellar populations and
chemical yields in galaxies, including the contributions of AGB stars to
integrated galactic light, dust production, and element synthesis. Currently the best 
mass-loss rate prescription we have is the one based on pulsation
period (Vassiliadis \& Wood (1993); Fig. 19 of H\"ofner \& Olofsson
2018). Indeed, A. Zijlstra noted that the very existence of a correlation between
period and ${\dot M}$ may indicate that pulsation is the primary
driver of mass loss during much of the AGB. However, 
Olofsson
stressed that there has been little progress on this topic for nearly 30
years. He proposed attempting to observe AGB stars 
during some well-defined process (e.g., a thermal pulse) and assess how that affects the
mass-loss rate (e.g., Kerschbaum et
al. 2017). 

To explore the effects of metallicity on mass loss, it is crucial to
expand the observations of AGB stars beyond the Milky Way. 
This has now become
possible with ALMA, and Olofsson was part of a team that  recently
obtained CO detections of four carbon stars in the
Large Magellanic Cloud (Groenewegen et
al. 2016). However, Olofsson noted that
even with ALMA, it will likely be difficult to perform
such observations beyond the Magellanic Clouds. 

Olofsson provided a reminder that the \HI\ 21~cm line can serve as  a
valuable tracer of the atomic component of the circumstellar
environments of AGB stars, particularly for warmer AGB stars. The weakness of the circumstellar \HI\ line,
coupled with the ubiquity of strong interstellar \HI\ emission along
the line-of-sight make such observations challenging, but
the number of AGB stars surveyed in the \HI\ line has
grown to well over 100 (G\'erard \& Le~Bertre 2006; G\'erard et
al. 2011 and in prep.; Matthews et al. 2013). Detection rates are high 
for irregular and semi-regular variables with moderate
mass-loss rates ($\lsim10^{-7}~M_{\odot}$ yr$^{-1}$), but detections
of Mira variables and higher mass-loss-rate stars are rare. Modeling 
by Hoai et al. (2015) has shown that this cannot be readily
attributed to radiation transfer effects, and most likely implies that
the circumstellar material in these cases is largely molecular. 

\begin{figure}
\centering
\vspace{-2.0cm}
\scalebox{0.4}{\rotatebox{0}{\includegraphics{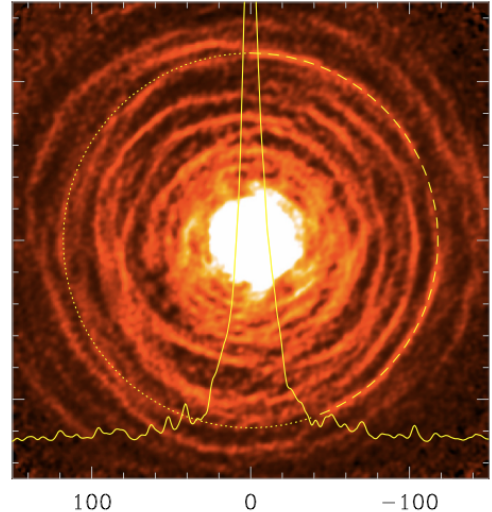}}}
\vspace{-2.0cm}
\caption{Circumstellar CO(2$-$1) emission  at the stellar systemic
  velocity ($-$26.5~\kms) surrounding the AGB star
  IRC+10216. The map  combines
interferometric data from the SMA and short-spacing data from
  the IRAM 30~m telescope. The scale is in arcseconds. The yellow line
  shows an intensity profile along an east-west strip through the
  stellar position. From Gu\'elin
  et al. (2018); reproduced with permission from A\&A, \copyright ESO.}
\label{fig:guelin-rings}
\end{figure}

A program  to study the CSE of an AGB star in
``3D'' 
was described by M. Gu\'elin (Institut de Radioastronomie
Millim\'etrique; hereafter IRAM). His team recently used a combination
of the Submillimeter Array (SMA), the IRAM 30~m telescope, and ALMA to
study the CSE of the nearby
carbon star IRC+10216 in a variety of molecular
lines and with various spatial resolutions. They then used the combined
results to constrain 3D models of the CSE velocity
structure, in addition to characterizing the star's mass-loss
history and the physical conditions of the CSE a function of radius (Gu\'elin et
al. 2018).
IRC+10216 is well known to have a massive, dusty envelope with a relatively
symmetric shape---albeit with considerable fine structure. The new
observations of Gu\'elin et al. show
a quasi-regular pattern of multiple thin rings of various radii
within the CSE, as
traced by molecules such as CO(2-1) (Figure~\ref{fig:guelin-rings}). Although roughly centered on the
star, these various rings are not concentric. However, Gu\'elin's team has shown that
these features are true shells (not part of a spiral pattern), 
with densities that are $\sim3\times$ higher than
the inter-shell regions. In the
outer envelope, the shell spacing of $\sim16''$ ($\sim$2000~AU) corresponds to
timescales of $\sim$700~yr, while in the inner CSE ($r<40''$) the
shell spacing is smaller. Gu\'elin's team has proposed that these
properties of the IRC+10216 CSE are the result of mass loss
modulated by a low-mass companion with an orbit in the plane of the
sky.
They also examined the question of
whether the mass loss from IRC+10216 is isotropic by
applying newly developed algorithms
to reconstruct their 3D position-velocity
data. Based on  the results of this analysis they concluded that
IRC+10216's  shells are spherical in shape, hence the mass
loss is isotropic (or nearly so).

\subsection{Circumstellar Chemistry}
H. Olofsson noted that $\sim$90 molecules have now been detected in
the CSEs of AGB stars, but numerous unidentified lines
are also present in the  mm and sub-mm bands.  He
emphasized the importance of unbiased wideband
spectral scans with single-dish telescopes for the discovery of additional
circumstellar species and for
advancing our understanding
circumstellar chemistry  (e.g., Cernicharo et al. 2010; Zhang et
al. 2009; Chau et al. 2012; Velilla Prieto et al. 2017; De~Beck \&
Olofsson 2018).  
As the number of such studies has recently grown, it has become possible
not only to compare in detail C- and O-rich stars, but also 
to discern more subtle differences among individual stars of the same
chemistry. 

Single-dish studies are a crucial complement to 
interferometric imaging, since the latter
often resolves out a significant fraction of the flux in
circumstellar lines
(e.g., Cernicharo et al. 2013). Further, with today's wide bandwidth systems, 
Olofsson pointed out that data from single-dish telescopes
can serve another vital role in situations where multiple
transitions of a given molecule with very different energy
requirements can be observed and interpreted in concert (e.g., Van de
Sande et al. 2018); in this case, excitation energy rather than spatial
resolution can serve as a radial discriminator. When used in
conjunction with radiative transfer models, such data help in understanding the
dynamical and chemical structure of AGB winds.

Isotope ratios are very important tracers of stellar
nucleosynthesis, and as an example, the $^{12}$C/$^{13}$C ratio in the
ISM is believed to be predominantly set by AGB stars. It
therefore serves as a marker of the stellar mass function and
star-formation history. Ramstedt \& Olofsson (2014) recently undertook a
comprehensive analysis of this
ratio for a large sample of AGB stars and found that while it varies
for the three different chemical types (M, S, and C), it is
surprisingly independent of mass-loss rate. 
Interestingly, their results also suggest that PNe (see
also Section~\ref{sec:PNe}) are
preferentially drawn from stars with low $^{12}$C/$^{13}$C.
Another isotope ratio of interest is $^{17}$O/$^{18}$O, which is set during
the red giant branch and does not evolve during the AGB. Olofsson and his colleagues have
shown that this ratio can serve as an initial mass estimator for stars
with $M_{\star}\lsim4M_{\odot}$ (de Nutte et al. 2017). Their work also
underscored an important limit of $M_{\star}\lsim1.5M_{\odot}$ below
which carbon stars do not form. 

\subsection{Disks around AGB Stars}
\protect\label{sec:AGBdisks}
The long-standing question of how nearly spherical AGB stars evolve into the
complex morphological varieties of PNe is still not fully solved (see also Section~\ref{sec:PNe}). However,
the formation of disks owing to the influence of a binary companion is
suspected of 
playing a key role.
L$_{2}$~Pup now represents the first example of a disk being directly
detected around a  bona fide AGB star (Lykou et al. 2015; Kervella et
al. 2016; Homan et al. 2017).  As described by Olofsson, high-resolution  ($\sim$15~mas)
observations of SiO lines using 
ALMA reveal that at least a portion of
the disk appears to be in Keplerian rotation, implying a mass for the
central star of 0.66~$M_{\odot}$. The estimated initial mass is
$\sim1~M_{\odot}$, implying that L$_{2}$~Pup is a solar-like star that has
lost $\sim$40\% of its mass (Kervella et al. 2016). In addition, the ALMA data
reveal evidence of a companion visible in both the continuum
and the CO(3-2) line. Its estimated mass ($\sim 12M_{J}$) lies at the border between
brown dwarfs and planets. However, as pointed out
by A. Zijlstra, such a low mass for the companion makes it difficult
to account for the disk's angular momentum.

\subsection{Magnetic Field Measurements of AGB Stars}
Olofsson reminded us that it has been notoriously difficult to
determine the properties of magnetic fields associated with AGB stars,
but this is of critical importance for better understanding the role
of B-fields in determining the mass-loss
characteristics of stars during the AGB and post-AGB phases. 
He pointed to the enormous potential that radio interferometry has  to
contribute to this topic through  polarization studies of both 
thermal and maser lines, including measurements of
the Zeeman effect in species such as SiO, H$_{2}$O, OH, and CN and 
the Goldreich-Kylafis effect in CO and other lines.
Olofsson pointed out that to date, there is only one measurement of a
magnetic field on an AGB star (the S-type star $\chi$~Cyg; L\`ebre et
al. 2014); in all other case the circumstellar
magnetic field is extrapolated inwards (e.g., Vlemmings 2012; Duthu et
al. 2017).

\section{Radio Emission from Supergiants}
\subsection{Cepheids}
\protect\label{sec:cepheids}
Studies of Cepheid variables provide crucial insights into the
evolution of intermediate mass stars, and understanding their physics 
is also of significant interest given their long-standing role
in establishing the extragalactic distance scale. Cepheids have 
received relatively little attention
at radio wavelengths, and with the exception of two Cepheids that appear to be
associated with 21~cm line-emitting circumstellar nebulae (Matthews et
al. 2012, 2016), no radio detections of Cepheid have ever been
reported  (Welch \& Duric 1988; Matthews et al. in prep.). However, 
recent work by N. Evans (CfA)
and collaborators
provides a new motivation for re-examining these stars in the radio.
Using {\it Chandra} and {\it X-Ray Multi-Mirror Mission (XMM-Newton)} 
observations of Cepheids, they have discovered surprising bursts of X-rays
at a pulsation phase just after maximum radius (Engle et al. 2014, 2017;
Evans et al. 2016). One possible explanation
is that these bursts are caused by shocks related to the
pulsation process; another is that they are a
flare-like phenomenon triggered by the collapse of the atmosphere. Evans noted that detections of
a radio counterpart to these events could provide additional
clues on their origins. The predicted radio flux densities are
highly uncertain, but present estimates suggest they are
likely to be detectable with the VLA during particular phases of the
pulsation cycle. Discovery of this phenomena is intriguing in light of
the long-standing question of whether Cepheids are undergoing
significant pulsationally-driven mass loss (e.g., Neilson 2014).   

\subsection{The Enigmatic Radio Star MWC~349A} 
\protect\label{sec:MWC349A}
V. Strelnitski (Maria Mitchell Observatory; MMO) described recent work on the
brightest stellar radio continuum
source in the Galaxy, MWC~349A. MWC~349A is surrounded by a dense \HII\ region, and to date is the only
star known to
exhibit high-gain atomic maser and  laser emission in H
RRLs. These masers appear to trace a disk and
photoevaporating wind
(Zhang et al. 2017). 
MMO has monitored MWC~349A in the optical ($VRI$, H$\alpha$)
and
radio (maser lines) for over 20 years, amassing a unique data set on
this star (Thomashow et al. 2018). 

MWC~349A appears to be a massive object ($>10M_{\odot}$), but its
evolutionary status has remained a matter of dispute. Indeed, part of the problem, as noted by
Strelnitski, is that 
evolution from the PMS to an evolved (supergiant) star spans only
$\sim$5~Myr for stars of this mass. 
Gvaramadze \& Menten (2012) argued that  MWC~349A is an
evolved massive star (luminous blue variable) based on its photometric
and spectroscopic variability, as well as evidence for the probable
ejection of MWC~349A and the neighboring B0III star MWC~349B
from the Cyg~OB2 association several Myr earlier.  However, new results from
Strelnitski and collaborators (Drew et al. 2017) find a radial
velocity difference of $\sim$35~\kms\ between MWC~349A and MWC~349B,
indicating that they are unlikely to have been bound. Based on this, along
with evidence of a possible physical association between MWC~349A and
a compact molecular cloud (Strelnitski et al. 2013), Strelnitski argued that MWC~349A may
instead be a
YSO. 

D. Fenech disputed the latter conclusion,
pointing to recent  ALMA 
observations of the B[e] supergiant Wd1-9 that show mm properties
extremely similar to MWC~349A (Fenech et al. 2017, 2018; Figure~\ref{fig:Wd1-9}; see also
Section~\ref{sec:hotclusters}). 
Fenech therefore argued that the latter too is
likely to be an evolved star. 
However, one difference noted by I. Stevens is that the RRL profiles of Wd1-9 are
Gaussian-like (Fenech et al. 2017), in contrast to the double-peaked
lines seen in MWC~349A (e.g., B\'aez-Rubio et al. 2013).
M. Reid 
went further to offer the provocative suggestion that MWC~349A might be a pre-planetary
nebula. Ultimately no consensus was reached at the meeting
regarding the evolutionary status of this
object.

\subsection{Red Supergiants}
\protect\label{sec:RSGs}
A. Richards described some recent advances in the study of RSG  winds
and outflows based
on the detailed analysis of circumstellar masers.
A wide variety of  mm/sub-mm
maser transitions lie within bands accessible with ALMA, spanning 
a considerable range of excitation conditions. When observed in combination
at high spatial resolution, these lines
supply powerful constraints on the physical
conditions (density, temperature) and kinematics over a wide range of radii within
the extended atmospheres of cool evolved
stars (see Gray et al. 2016). Importantly, such information can supply
constraints on models for the shaping and
acceleration of RSG winds. 

As an illustration, Richards
presented recent results for several mm/sub-mm H$_{2}$O transitions observed
with  ALMA toward VY~CMa (Richards et al. 2014). Richards's team found that the
183~GHz H$_{2}$O masers are far more extended than any other
previously observed H$_{2}$O transitions and appear to follow the
distribution of the extended, dust-scattered light  seen in {\it Hubble
  Space Telescope} images (Humphreys, Helton, \& Jones 
2007), implying a link with small, cool dust grains and
low density gas (Richards et al. 2018). Other high-frequency H$_{2}$O maser transitions
at 325, 321, and 658~GHz were seen to lie in thick, irregular
shells and trace, respectively, material
progressively closer to the central star. The surprisingly extended distribution of the
658~GHz emission poses a challenge for current maser pumping models. 

Richards noted that one surprise from the analysis of these new H$_{2}$O maser data
is that if it is assumed that dust that forms at
5--10~$R_{\star}$ (as current models predict), this should
produce fairly rapid wind acceleration---in contrast to the far more
gradual acceleration that is observed. Richards postulated that this
implies either that the dust forms more slowly than previously
thought, or that it undergoes some kind of annealing or
``reorganization'' within the wind.  She reported seeing  a similar
trend in other stars as well.
A. Zijlstra suggested an alternative possibility, namely that 
asphericity, together with velocity variations related to the time-variable mass loss,
may imprint similar signatures on the outflow. 

On a related point, Zijlstra
questioned the strength of current
evidence for dust-driven winds in RSGs. 
K. Gayley pointed out that confoundingly, temperatures are only cool enough for dust
to form at relatively large radii in RSGs---seemingly too
far out to drive the wind. He
suggested that  perhaps molecular opacity from molecular layer or ``MOLsphere''
(Tsuji 2000, 2001)  could play a role.  V. Strelnitski
drew attention to previous work by his team that showed this may be plausible
for C-rich giants and for some O-rich giants (Shmeld et
al. 1992; see also Jorgensen \& Johnson 1992; Helling, Winters, \&
Sedlmayr 2000). However, the precise role of molecular opacity remains
an open question, and the topic appears to have received relatively
little recent attention.
Overall there does not appear to be a present
consensus  on the driving mechanism for RSG winds (see also
Section~\ref{sec:winds}).

Other results from the analysis of Richards's team included evidence
favoring a 
time-variable mass-loss rate, and the need for gas clumps $\sim$50 times denser than
previously suspected in order to explain the observed
patterns of spatial overlap between various H$_{2}$O maser lines.
Richards noted that in future 22~GHz H$_{2}$O maser observations from
e-MERLIN it will be possible  to detect the stellar continuum
along with the H$_{2}$O maser emission and thus astrometrically align the
two. Such precise spatial registration between the masers and the star
will enable placing
even more
stringent constraints on pumping and dynamical models (e.g., Gray et
al. 2016).

High-resolution studies of the continuum (radio photosphere) have
already been performed for another RSG, Betelgeuse (Richards et
al. 2013; O'Gorman et al. 2017). The recent
ALMA study of O'Gorman et al. with 14~mas resolution made it
possible to discern thermal structure, including a region
$\sim$1000~K brighter than the mean disk, indicating localized heating. 

\section{Studies of Stars at Ultra-High Angular Resolution}
\subsection{Very Long Baseline Interferometry}
\protect\label{sec:VLBI}
A. Mioduszewski (NRAO) reviewed the
many aspects of stellar astrophysics that can be advanced by the ultra-high angular resolution measurements achievable with VLBI. 
A key advantage of this technique comes from the relevant physical scales that
can be probed. For example, for a fiducial 5000-mile
(8600~km) baseline, the achievable angular resolution at 20~cm and
7~mm is $\sim$5~mas and 
0.17~mas, respectively. (Note that 1~mas corresponds to $\sim$1~AU at
a distance of 1~kpc and to $\sim10R_{\odot}$ at
$d\approx$100~pc). In addition, astrometric
precision of $\lsim$50~$\mu$as in VLBI positional measurements  is now
routine (e.g., Forbrich et al. 2016; Ortiz-L\'eon et al. 2017). 
On the other hand, a drawback of VLBI is that the brightness
temperature sensitivity scales with baseline length $B_{\rm max}$ as:
$T_{B}=0.32\sigma_{N}\left(\frac{\rm mJy}{\rm beam}\right)B_{\rm max}({\rm
  km})^{2}$, and for present
VLBI arrays, the achievable noise level ($\sigma_{N}$) typically limits this technique
to sources with $T_{B}\gg 10^{4}$~K. Thus VLBI is sensitive only to
compact, nonthermal emission (e.g., masers,
synchrotron, ECM) or thermal sources seen in
absorption against nonthermal backgrounds. 

Mioduszewski summarized ongoing work by her team 
to study the radio emission from the weak-lined T~Tauri binary
V773~Tau (see also Section~\ref{sec:YSOs}). They find the radio emission
to be extremely variable (by factors of 5 or more), with significant brightening occurring near
periastron, providing further evidence that
PMS stars that produce nonthermal continuum emission are
interacting. Additionally, in her team's Gould's Belt VLBA distance survey of
young stars (Ortiz-L\'eon et
al. 2017 and references therein), over 50\% of those PMS stars with nonthermal
emission are confirmed close binaries, suggesting that close pairings 
help to create conditions conducive to the production
of such radio emission. However, the exact mechanism through which the
non-thermal emission is unhanced is unclear.

Position shifts of the two V773~Tau components are seen near periastron,
suggesting that when the stars are far apart, their radio centroids
correspond to the stars themselves, but when they are close together, the
emission arises from a region in between. Recent follow-up observations
with the High Sensitivity Array by Mioduszewski and her group
show evidence for a bright
connecting feature between the stars {\em after} they begin
brightening near periastron. The changes are very rapid, consistent
with a possible magnetic reconnection-type event.

Mioduszewski concluded her talk with a look forward to the future
of VLBI. In the near-term, the VLBA is returning to the NRAO's
portfolio, 
and options are being explored to increase VLBA bandwidths to 2~GHz, improving 
sensitivity by a factor of three.  
 Discussions are also underway to
include in plans for the ngVLA an upgrade of
existing VLBA sites with ngVLA technology, along with the possible introduction
of additional intermediate baselines to bridge ngVLA and
(current) VLBA baselines  (Murphy et al. 2018; Reid, Loinard, \&
Maccarone 2018).  
Such a plan could result in more than a
factor of 100 improvement in sensitivity over the current VLBA and
improve sensitivities to the point that brightness temperatures of
$T_{B}\sim$1000~K (i.e., thermal sources) would be within reach. This would
provide resolution of $<0.3R_{\odot}$ at 50~GHz, making it possible to
image the disks of the 100 closest stars. Elsewhere, cm-wave VLBI stations are being deployed in Africa to
form part of an African VLBI network (e.g., Copley et al. 2016) that 
is expected to play a crucial role in enabling VLBI with the
SKA.

\subsection{New Image Reconstruction Methods}
K. Akiyama (MIT Haystack Observatory) 
showcased a powerful new image reconstruction technique for radio
interferometry applications known as ``sparse
modeling'' (Honma et al. 2014). This technique has been shown to achieve
image resolution up to $\sim0.3\times$  the
nominal diffraction limit, but with substantially better
fidelity than traditional CLEAN deconvolution. While recent
sparse
modeling developments and applications have focused primarily on the
needs of mm VLBI experiments  (Akiyama et al. 2017a, b), Akiyama
showed that sparse modeling has a variety of other potential
applications where maximum angular resolution and high image fidelity
are important, 
including the imaging of circumstellar disks and
stellar surfaces.   As one illustration, he presented a sparse model
reconstruction of the radio photosphere of the
AGB star Mira based on ALMA data from Matthews et al. (2015;
see also Matthews et
al. 2018; Section~\ref{sec:photospheres}).
The resulting image reveals low-contrast 
surface features that cannot be discerned directly from images made using
CLEAN deconvolution.

\section{Evolution Beyond the AGB}
In his review talk, A. Zijlstra issued a reminder that we still lack
a robust formalism for describing how low-to-intermediate mass stars
evolve from the  AGB to the post-AGB and
PN stages. Still unsolved questions include how the shell of the
star gets ejected, what are
the maximum mass-loss rates achieved during this process, and what are
the timescales for the various processes involved. Underscoring this
latter question, Zijlstra
pointed to the recent
models of Miller Bertolami (2016) whose refined evolutionary timescales
differ by a factor of three or more compared with most earlier models.

\subsection{The Post-AGB and Pre-PN Stages}
\protect\label{sec:postAGB} 
Because stars transitioning from the post-AGB into PNe 
evolve considerably over timescales of only a few thousand
years, observable variations in the radio emission linked with source
evolution are expected. 
This has been observed in the case of the very young PN NGC~7027
by Zijlstra, van Hoof, \& Perley (2008),
who were able to use multi-frequency radio
observations, coupled with model fits, to derive an extremely precise
mass estimate for the central star. On the other hand,
Zijlstra cited other cases where the observed radio 
variability in similar 
objects is not straightforward to interpret. For example,
Cerrigone et
al. (2017) identified a few sources evolving toward PNe that showed either increases
or decreases in the radio flux over the course of $\sim$10--20
years, but it is unclear if these mirror the ionization changes
expected from source evolution or
simply variability of the central star. 
For the pre-PN CRL~618 (see also below), 
S\'anchez-Contreras et al. (2017) recently found evidence that the long-term variability
may be periodic.

Another important effect that occurs during the AGB$\rightarrow$PN
transition is that ionizing photons from the central star begin to
alter the chemistry of the circumstellar material.   Shocks linked
with the bipolar outflows that may appear during this time can also
impact the chemistry. N. Patel (CfA) described his
team's recent program to study the astrochemistry of objects spanning
the transition from the post-AGB to PNe in order to better
characterize these changes and test theoretical predictions (e.g., Bachiller et
al. 1997). Their work takes advantage of the new wide
bandwidth SWARM correlator on the SMA (Primiani et al. 2016), which provides up to
32~GHz instantaneous bandwidth with 140~kHz spectral resolution. 

\begin{figure*} 
\centering
\scalebox{0.9}{\rotatebox{0}{\includegraphics{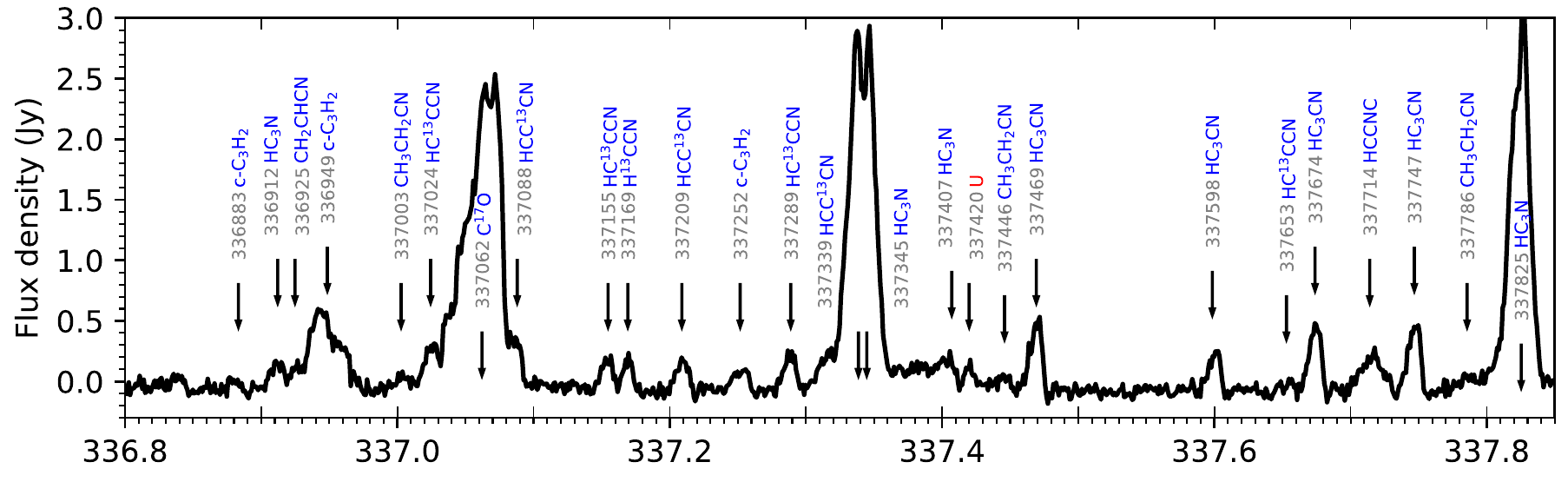}}}
\caption{Sample portion of the SMA spectrum of the pre-PN CRL~618
  obtained by Patel et al. (2018 and in prep.). The frequency axis is in GHz. Lines
labeled with a 'U' are unidentified.}
\label{fig:SMAspectrum}
\end{figure*}

As Patel explained, surveying these wide bandwidths is crucial, since
many molecules
(e.g., H$_{2}$CO, CH$_{3}$CN) emit lines spread out over a wide swath
of frequency. In cases where the lines are optically thin, 
obtaining simultaneous measurements of multiple
lines enables  construction of population diagrams  (e.g.,
Lee et al. 2013),
which in turn yield measurements of both the excitation temperature and column
density. As Patel showed, this method is also useful to confirm line identifications
and identify lines that may be blended (a significant
problem at sub-mm wavelengths).

As an example of his group's recent work, Patel presented spectra of
CRL~618 obtained with the SMA from 282--359~GHz
(Patel et al. 2018; Figure~\ref{fig:SMAspectrum}). This source has outflows spanning tens of
arcseconds, but
Patel's team focused on the inner few arcseconds, where they were able
to detect a total of 1075 lines, 373 of
which are presently unidentified. The detections included a number of 
H RRLs that were 
spatially unresolved by the \as{0}{5} beam. With the exception of SiO(8-7), no
Si-bearing molecules were detected. In comparison, Patel noted that only 440 lines
were detected toward the highly evolved C-rich AGB star 
IRC+10216 over a similar frequency
range (Patel et al. 2011). However, while the number of lines detected toward CRL~618
is larger, the number of distinct molecules is far smaller than in
IRC+10216. Indeed, the majority of the CRL~618 lines arise from just
two molecules, HC$_{3}$N and c-C$_{3}$H$_{2}$. 
Patel also reported that 
line survey data have been obtained for 
NGC~7027 over a $\sim$60~GHz band, but in this case, only 26 lines were
detected. 

Patel's group is currently
working with laboratory spectroscopists to reconcile
many of the unidentified lines toward their target sources. 
Preparations are also underway to perform a wide bandwidth survey the young PN
CRL~2688 with the SMA in 2018.  Patel noted that future work at lower (cm-wave)
frequencies, which are sensitive toward larger molecules and longer
carbon chains, would be an
important complement to his team's SMA studies of young and pre-PNe. 

\subsection{Planetary Nebulae (PNe)}
\protect\label{sec:PNe}
At cm wavelengths, typical flux densities for Galactic PNe range from
$\sim$1-1000~mJy, and over the last few decades there have been many radio
surveys of these objects (e.g., Zijlstra, Pottasch, \& Bignell 1989; Chhetri et
al. 2013). A recent survey of the Large Magellanic Cloud has also led
to the first radio detections of PNe
outside the Milky way (31 in total; Leverenz et al. 2017). 
A key advantage of radio wavelength studies of PNe is the ability to
overcome the the obscuring effects of
dust. However, as 
Zijlstra reminded us, a number of puzzles persist concerning the radio
properties of these object. 

\subsubsection{The Origin of the Shapes of PNe}
As summarized by Zijlstra, it has become clear that much of the complex structure observed in
PNe must originate from the effects of binary companions. Wide binaries can
produce modest levels of gravitational focusing, leading to
elliptically shaped nebulae; closer companions can produce more active
shaping and lead to the formation of disks and jets; and finally, a very
close-in companion (as occurs in $\sim$5\% of cases) can lead to common envelope ejection---the least
understood of the binary scenarios. 
Zijlstra noted that presumed ``single-star'' PNe (that are roughly
spherical) do exist, but they are rare, comprising $\sim$20-30\% of
PNe. These are typically more difficult to find, as they
are on average fainter owing to the absence of binary-induced
density enhancements in their ejecta, but Zijlstra predicted that more will be
found in deeper surveys.

\subsubsection{Synchrotron Emission from PNe}
Since both radio
emission and H$\beta$ emission arise from free-free processes, the
ratio of these two tracers should provide a straightforward measure of the
extinction. Nonetheless, there remains a persistent disagreement in the
extinction so inferred as compared with estimates from the
ratio of H$\alpha$/H$\beta$: the radio fluxes are consistently
too low. For years this was
attributed to the adoption of a standard Galactic ratio of total-to-selective
extinction ($R_{V}=$3.1) toward the Bulge,  which may not be
valid (Ruffle et al. 2004). However, this has been disputed by Pottasch \&
Bernard-Salas (2013), who instead argue that the 6~cm radio fluxes are
discrepant by up to a factor of two because of missing optically thick
emission (e.g., from clumpy material). This remains an ongoing
controversy.

Another open question is whether or not PNe emit synchrotron radiation  as a
result of shocks induced by the 
hot wind of the central star---an idea first suggested by Dgani \& Soker (1997). Zijlstra noted
that to
date, no such emission has ever been detected from within a PN,
although the expected signal ($\sim$1~mJy for a source at 1~kpc and
$\nu$=1~GHz) would be difficult to discern when superposed on the much brighter
free-free emission. Bains et al. (2009) 
detected variable nonthermal emission from 3 out of 28 post-AGB stars
surveyed, hinting that such emission may only occur {\em before} the nebula
is ionized (see also P\'erez-S\'anchez et al. 2013, 2017). Yet another
related object is V1018~Sco, which appears to be an OH/IR star with
associated synchrotron emission and PN emission (Cohen
et al. 2006) and Zijlstra
urged further studies of this enigmatic object. He also noted that
future
surveys capable of measuring flux variability at
the $\mu$Jy level could in turn allow measurements of the
expansion of many PNe, as well as  the derivation of their central star
masses, providing a
key missing ingredient in models of AGB evolution.

\subsubsection{Enigmatic X-ray Point Sources}
One more outstanding puzzle concerns the X-ray point
sources seen in approximately half of the PNe that are detected in
X-rays. Of these, $\sim$15\% show hard X-rays. Zijlstra and his collaborators find
that at least some of these cases may be attributed to an (old) main
sequence companion that has been rejuvenated by the AGB
wind  (Montez et al. 2015). However,  efforts to detect the
predicted weak radio emission
from these sources (expected to be $\sim 1~\mu$Jy) have been
unsuccessful, mostly likely due to sensitivity limitations. So far
there are no obvious trends seen in the ages or  nebular morphologies of
the PNe containing the hard X-ray sources.

\clearpage
\subsubsection{Radio Recombination Lines (RRLs) in PNe}
Zijlstra noted that RRLs in PNe are
relatively little observed because they tend to be weak and relatively
broad ($\sim$15~\kms\ thermal motions), making it difficult to use these lines for
dynamical measurements. Nonetheless, he emphasized that they are
excellent tracers of obscured regions. He also noted that there are a
few cases known where masing is seen near the H30 lines, which
appear to be linked with rotating disks (S\'anchez-Contreras et
al. 2017; Aleman et al. 2018). Angular momentum arguments then
dictate that a binary companion must be present. 
Zijlstra predicted that studies of  RRLs in PNe will be
facilitated by the increased collecting area of future telescopes such as the SKA. 

\subsubsection{Molecules in PNe}
Zijlstra pointed out that most molecules are typically found only in young
PNe, although CO tends to be found in PNe over a wide range of
ages. This includes cases where it is seen in an outer shell (e.g.,
NGC~7027) or in cometary globules which can shield it (e.g., the
Helix). ALMA has recently enabled the mapping of CO in PNe in
unprecedented detail, leading to a number of surprises. 
In the young PN M2-9, Castro-Carrizo et al. (2017) recently identified two CO rings
separated by a gap, most likely as a result of  mass-loss
events separated by $\sim$500~yr. The cause of these discrete
events is not entirely clear, although a binary companion has been implicated. 
In the case of the 
Boomerang nebula (which is so cold that it has been observed in
absorption against the cosmic background radiation)  CO is seen to trace both polar lobes and an
expanding torus (Sahai, Vlemmings,
\& Nyman 2017). This may be the result of a common envelope system,
although Zijlstra questioned whether such systems can drive jets, as
observed in this source. Finally, in the PN NGC~6302,
Santander-Garc\'\i a et al. (2017) were able to uncover a second inner
ring, inclined to a previously discovered expanding molecular 
torus. The former appear to have age ($\sim$2000~yr) similar to ionized lobes that are
also seen. This again suggests a double ejection event took place. 

\subsubsection{$^{3}$He$^{+}$}
$^{3}$He$^{+}$ is a product of the $p$-$p$ cycle within low-mass
stars,  
and is only dispersed when
these stars reach the end of their lives, making it the last addition to the
baryonic universe.
But how much $^{3}$He$^{+}$  solar-mass stars contribute
depends very sensitively on stellar interior processes during
the red giant branch and AGB phases (e.g., how much mixing occurs and how
much $^{3}$He$^{+}$ gets destroyed inside the star). 

$^{3}$He$^{+}$ can be
studied through a rather weak spin-flip transition at 8.6~GHz, and
so far it has been detected in three PNe (Balser et al. 1997,
2006; Guzman-Ramirez et al. 2016). All show a
double-peaked line profile, in contrast to the centrally peaked 
RRLs seen in these same objects.  The abundances are also higher
than expected, and the $^{3}$He$^{+}$ line appears significantly brighter
than predicted. This can be accounted for neither by
cosmological models nor stellar nucleosynthesis. 
A possible explanation recently proposed by Zijlstra and
his collaborators is that the $^{3}$He$^{+}$ line is {\em masing}, possibly as a
result of pumping by $^{4}$He$^{++}$ recombination (Guzman-Ramirez et
al. 2016). If true, this will make results derived from the $^{3}$He$^{+}$
line much more difficult to interpret.

\section{White Dwarf Binaries}
\subsection{Cataclysmic Variables (CVs)}
\protect\label{sec:CVs}
Cataclysmic variables (CVs) are binary systems
containing a white dwarf accreting mass from a
low-mass, main-sequence companion of M or K spectral type. CVs are further
divided into magnetic ($B\gsim10^{6}$~G) and
non-magnetic systems ($B\lsim10^{6}$~G) based on the field strength of the white
dwarf. 
A review talk by D. Coppejans (Northwestern University) described how
the increased sensitivity of modern radio interferometers has enabled
a steady stream of discoveries related to CVs within the past few years.

Coppejans noted that as of the 1980s, only 4\% of non-magnetic CVs (three systems in total)
had been detected in the radio. This led to the surprising inference
that CVs do not launch jets (Soker \& Lasota 2004),
which would make them a rarity
among accreting sources. However, this problem
was revisited by K\"ording et al. (2008), who looked at the dwarf nova SS~Cyg
during its (optical) outburst phase and discovered radio emission
attributed to synchrotron radiation from a transient jet. 
In a more recent study by Mooley et al. (2017) with the Arcminute
Microkelvin Imager Large Array,
data obtained over the course of one week not only confirmed this
behavior, but  showed the spectral index of SS~Cyg changing from
positive to negative as the jet evolves, plus the first ever detection of
a luminous flare near the end of the outburst.

Following the most recent VLA upgrade, Coppejans led a campaign to re-observe four
nova-like CVs that were previously
undetected in the radio. Three of the four  were detected
at 6~GHz during multiple epochs (Coppejans et al. 2015). Their emission was found to be
variable and  circularly polarized, 
indicating that it could not be
purely thermal in nature. 
For one target, TT~Ari, the
radiation was nearly 100\% circularly polarized, implying a coherent
emission component that is suspected of  being ECM. 
Coppejans noted that the emission from this source is reminiscent of magnetic
CVs (cf. Barrett et al. 2017). However, the interpretation is not
straightforward, since the secondary in this system is an M3.5 dwarf, which may
be prone to flaring.  TT~Ari's radio luminosity is nearly two orders of magnitude brighter than 
typical  flares for field dwarfs, but the secondary is
tidally locked into an orbit with a period of 3.5 hours, implying a
rotation velocity of $\sim$160~\kms---considerably faster than field
stars---which may lead to stronger flares.  
R. Osten also
pointed out  that in contrast to X-ray
emission, there is no saturation level for the radio luminosity of
flares, and at least one flare of comparable luminosity
to the TT~Ari flares ($\sim10^{16}$ erg s$^{-1}$ Hz$^{-1}$) has
already been detected from a field M dwarf in a wide binary (Osten et al. 2016).

\begin{figure}  
\centering
\scalebox{0.26}{\rotatebox{0}{\includegraphics{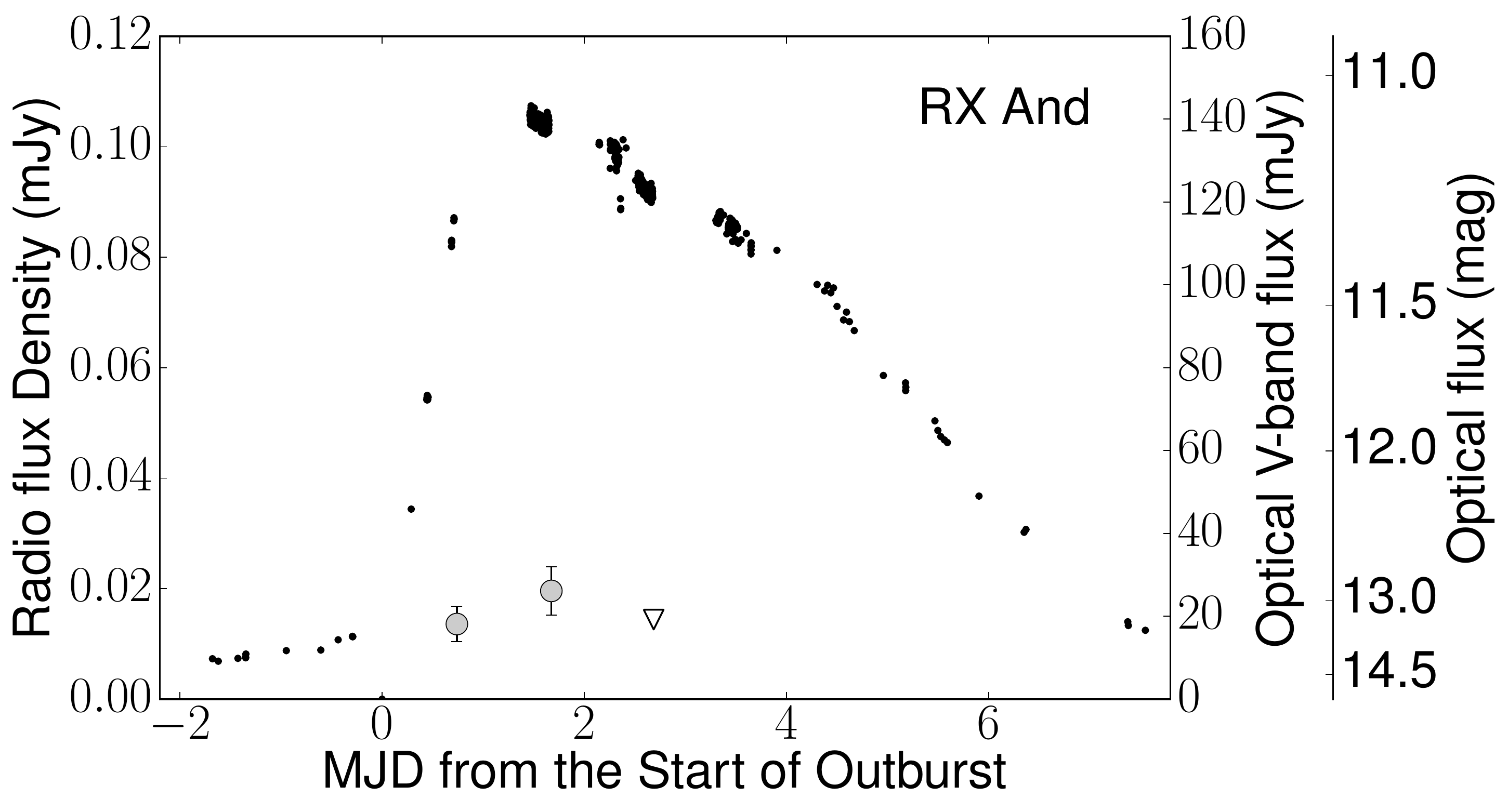}}}
\scalebox{0.26}{\rotatebox{0}{\includegraphics{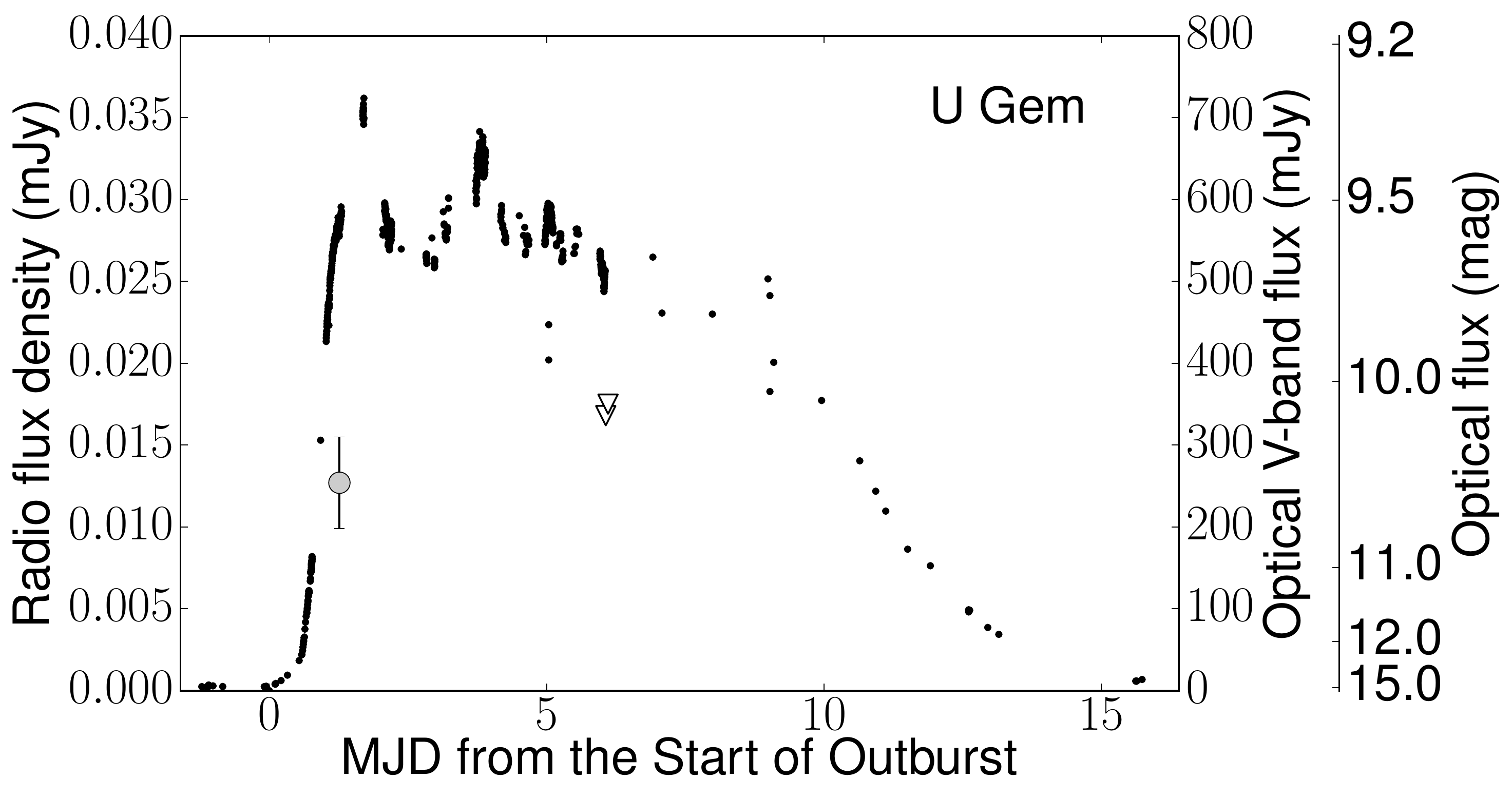}}}
\caption{Radio and optical light curves during the outbursts of two
  dwarf nova-type CVs. The 10~GHz radio data  
  obtained with the VLA are shown as grey symbols (triangles for
  upper limits). $V$-band optical data
  are shown as black dots. The $x$-axis indicates elapsed time in
  modified Julian date. Adapted from Figure~1 of Coppejans et al. (2016). }
\label{fig:CV-lightcurve}
\end{figure}

Another recent study led by Coppejans aimed to examine the behavior of
radio emission from CVs during 
optical outbursts. Such comparisons are challenging, since these
events occur unpredictably, and the radio emission is expected to peak within
24~hours of the outburst. To circumvent this, her team monitored a
sample of CVs to identify outbursts and enable rapid triggering of
radio wavelength follow-up with the VLA. In total, they obtained follow-up observations of five
outbursting CVs at 8--12~GHz. All were detected with specific radio luminosities in the range
$L_{\rm 10 GHz}\sim10^{14}-10^{16}$ erg s$^{-1}$ Hz$^{-1}$ and were found to exhibit variable
radio emission on timescales ranging from 200~s to several days 
(Coppejans et al. 2016; Figure~\ref{fig:CV-lightcurve}).  
No correlation was found between
the radio luminosity and any other properties of the CVs,
but Coppejans emphasized that this should be 
further explored with higher cadence radio monitoring (sampling times
of order hours, not days) since the
objects are so highly variable. 

Until recently, another unsolved question has been whether
the observed radio emission from CVs is dependent on their accretion
state. Coppejans reported that based on some recent  analysis by her team, the
answer is `yes', as none 
of the CVs they have observed with the VLA have been detected during 
quiescence. This seems to support
a jet model for the emission, but Coppejans cautioned that additional observations
are needed to establish the actual fraction of CV outbursts that are
radio-loud.

Another result highlighted by Coppejans was the recent discovery of the
first ``white dwarf pulsar'',  AR~Sco
(Marsh et al. 2016; Buckley et al. 2017). 
Marsh et al. monitored this  binary
over a wide range of wavelengths, including optical, IR, and 
9.0~GHz radio continuum with the ATCA, leading to the discovery of
pulsed emission  with a period of 1.97~min in all wavelength bands. 
The broadband  radio emission is thought to be synchrotron radiation 
produced by the interaction of the spinning white dwarf with  the
coronal loops of the M dwarf companion. Although AR~Sco is
presently a unique system, Coppejans suggested  it may 
represent a common evolutionary phase for CVs and may ultimately evolve
into a magnetic CV.  

Future radio studies of CVs are
expected to be enabled by 
ThunderKAT  (Fender et al. 2017), a transient survey slated to be
carried out commensally on the
MeerKAT array (Jonas 2009). In addition,
Coppejans is part of a team with 250 dedicated hours for CV work on
MeerKAT. She noted that a 1~m optical
telescope (MeerLICHT) will be slaved to MeerKAT to provide
simultaneous, multi-band optical
observations of every field studied. Coppejans urged consideration of
a similar facility in plans for the
ngVLA. 

\subsection{Classical Novae}
\protect\label{sec:novae}
Classical novae are accreting binary systems [either CVs (see Section~\ref{sec:CVs}) or
symbiotic systems] that exhibit
thermonuclear runaway on the surface of a white dwarf, leading to a non-recurring
ejection of large amounts of material. Approximately 35 classical 
novae  occur each year
in the Galaxy, out of which typically $\sim$8 are observed.
M. Rupen (National Research Council, Canada)  summarized the
rapid growth in discoveries related to these objects thanks to recent
radio observations.  

The long-prevailing paradigm for understanding the evolution of
classic novae is one where the ejecta are analogous to an \HII\ region
(powered by the hot white dwarf), and where the ejected material undergoes a homologous
(``Hubble flow'') expansion. 
In this model, the radio emission is
thermal bremsstrahlung, and the radio light curve is
characterized by  an optically thick rise with time $t$ ($t^{2}\propto\nu^{2}$) and an
optically thin decay ($t^{-3}\propto\nu^{-0.1}$).  Based on this
predicted behavior, it
is possible to directly estimate the mass of the
ejecta, although historically, such masses have been
systematically larger than theoretically predicted. 
Clumping in the ejecta has been invoked to reconcile this discrepancy
(e.g., Wendeln et al. 2017), but clumping models are poorly constrained.

While the aforementioned paradigm has been relatively successful, Rupen noted
that
cracks have begun to appear, thanks in large part to the results of
recent radio studies. He presented emerging evidence in favor of a revised picture
where shocks play a dominant role and where at
least two distinct outflows occur from a
single source. 

A significant problem for the ``\HII\ region'' model of
classical novae is
that free-free emission alone cannot fully account for
recent radio measurements.
In a study of the 2015 nova V1535~Sco by Rupen's
team (Linford et al. 2017), the optical data appear to adhere to
the classic model, while the radio data show a wide range of complex
behaviors. Radio light curves measured at various frequencies show
significant fluctuations rather than smooth evolution, with the
behavior differing between the frequency bands. Rupen suggested that this may be a
consequence of shocks between material from different ejection phases. The radio SEDs also
vary from day to day, and on several days exhibit slopes that can only
be explained by synchrotron radiation ($\alpha=-0.36$ to
$-0.88$). Furthermore, imaging of the source reveals
$T_{B}>6\times10^{5}$~K. To account for such high temperatures from thermal emission
alone would require far more emitting material than is inferred from concurrent
X-ray measurements. 
Another source studied by Rupen's team, V1723~Aql, showed an unexpectedly rapid
turn-on in the radio, an anomalous spectral index  
($\alpha\propto\nu^{1.3}$), a double-peaked radio light curve, and  
$T_{B}>10^{6}$~K (Weston et al. 2016b). A 
double-peaked radio light curve was also
seen in V1324~Sco (Finzell et al. 2018). 

Rupen reported that there are now at least 9 classical novae known to
emit $\gamma$-rays, and these statistics are consistent with all 
novae being $\gamma$-ray emitters. While an association
between synchrotron emission (shocks) and the emission of
$\gamma$-rays 
seems likely, an outstanding puzzle is why some sources such as
V1324~Sco show $\gamma$-rays but no X-rays, while others (e.g.,
V5589~Sgr) show hard X-rays (another shock tracer) but are undetected
in $\gamma$-rays (Weston et al. 2016a).  It is presently unknown if these
differences can be explained by variations in geometry/viewing angle. 

A first major clue came from the study of V959~Mon, a nova that was
detected in $\gamma$-rays and for which follow-up radio light curves
and VLBI imaging at multiple epochs were obtained (Chomiuk et
al. 2014). The model that emerged is one where 
there is a slow shell ejection early on. Interaction with the companion then
leads to enhanced mass loss perpendicular to the orbital plane. 
Later, fast ejection occurs in an orthogonal direction, shaped by the
initial torus, and shocks occur between the early and late winds. Subsequently,
the fast ejecta fade away, leaving behind the more slowly moving
material. However,
Rupen stressed that there is still considerable
work to be done to gain a more comprehensive understanding of
classical novae. He noted that to confirm or refute the new emerging paradigm
for understanding these objects, it will be crucial to study additional novae
with the type of high-quality detailed data that were obtained from
V959~Mon. The processes that might lead to a two-phase ejection are
still unclear, as is the reason for deviations from the expected rate
of rise of the radio light curves. Rupen concluded by pointing out that the
combined sensitivity and angular resolution of the
ngVLA would make it an ideal instrument for pursuing this topic.

\subsection{``Red'' Novae}
T. Kami\'nski (CfA) presenting recent findings concerning the enigmatic
nova-like object CK~Vul (Nova~1670)
based on mm and sub-mm observations (Kami\'nski et al. 2018). 
The appearance of this source was first recorded
in 1670, and in the two years following, it underwent three distinct episodes of
brightening
after
which it appeared extremely red  to the naked eye. 
CK~Vul's inner nebular remnant was first detected in the 1980s (Shara,
Moffat, \& Webbink 1985),
and a more extended ``hourglass'' nebula was later uncovered by Hajduk
et al. (2007). No stellar remnant has so far been
identified, although there is a weak radio source at the center of the
bipolar nebula. 

Recently Kami\'nski and his collaborators obtained 
spectroscopic observations of CK~Vul using the Atacama Pathfinder
EXperiment (APEX) and IRAM 30~m
telescopes, 
as well as interferometric imaging using the NOrthern Extended
Millimeter Array (NOEMA), the SMA, and ALMA.
These observations revealed a multitude of emission
lines, leading to the identification of over 320
unique spectral features  (Kami\'nski et al. 2017). The observed
linewidths are typically $\sim$300~\kms---far
smaller than in classic novae---and the interferometric images show
considerable kinematic complexity, including jets, rings, and
wide-angle outflows. Large quantities of  dust are also present, with
temperatures of a few tens of K. By analyzing the full SED,
it was possible to derive a luminosity of the central source
($\sim0.9L_{\odot}$) and a total mass for the remnant (gas+dust) of
$\sim0.1M_{\odot}$ (Kami\'nski et al. 2015). 

The collection of molecules detected in CK~Vul is highly unusual;  N
and F appear 
overabundant, and the species include some typically seen only in either
C-rich or O-rich environments. Perhaps most surprising is that for nearly all of the
molecules detected, rare isotopologues
are also seen, and all of the CNO isotopes are significantly enhanced
relative to solar composition. 

Having such rich information about the isotopologue ratios in CK~Vul
enables placing  new constraints on its progenitor.
Kami\'nski's team finds that the abundances can be explained
neither as the products of nucleosynthesis nor a classical nova event.
Instead, the object's origin is suspected of being a
stellar merger (Tylenda et al. 2013;
Kami\'nski et al. 2015), and based on recent findings, CK~Vul is now believed to be
being one of a small but growing group of objects dubbed
``red novae''. 

Another noteworthy feature of CK~Vul  is its
anomalously high abundance
of the radioactive nucleotide $^{26}$Al, which  is observed
as part of the isotopologue $^{26}$AlF.
Kami\'nski emphasized that this is also important in a
broader astrophysical context, since the decay of $^{26}$Al into Mg
produces a $\gamma$-ray line at 1.8~MeV. This line has long been
observed by $\gamma$-ray telescopes, pointing to an as-yet unidentified source of $^{26}$Al
in the Galaxy. CK~Vul 
represents the first astronomical object in which $^{26}$Al has been
directly identified. While  CK~Vul-like
objects alone are too rare to account for all of the $^{26}$Al
seen by $\gamma$-ray telescopes, Kami\'nski predicted that radio
telescopes may soon uncover additional sources of this isotope. 

\section{Radio Stars as Denizens of the Galaxy}
\protect\label{sec:denizens}
\subsection{Stellar Populations in the Galactic Center}
There has been long-standing interest in the stellar populations in
the neighborhood of Sgr~A*, the black hole at the Galactic Center (GC). 
Material shed by the winds of mass-losing, massive OB stars
near the GC are expected eventually feed onto
Sgr~A*. There has also been speculation as to whether star formation
can proceed {\it in situ} within the extreme environment close to the
black hole. 
As noted by M. Rupen, understanding the initial mass function in this
region is also relevant to understanding star formation in the extreme
environments present in submillimeter galaxies (e.g., Safarzadeh, Lu,
\&  Hayward 2017). However, as described by F. Yusef-Zadeh (Northwestern
University), only recently has radio emission been detected and
resolved from individual stars within the GC region (e.g., Yusef-Zadeh
et al. 2015a). He summarized results from several such
studies.

Taking advantage of the wide (8~GHz) continuum bandwidths now available with
the VLA at 34 and 44~GHz, Yusef-Zadeh's group has now detected over 300 compact radio
sources within 30$''$ of Sgr~A*, including 30 of the 90 massive stars with
previously known IR counterparts
(Yusef-Zadeh et al. 2015a). 
For those stars, mass-loss rates derived from radio and IR were
compared. After accounting for wind clumping, 
the radio measurements are systematically an order of 
magnitude lower (${\dot M}\sim10^{-7}M_{\odot}$ yr$^{-1}$),
implying a lower accretion rate onto
Sgr~A*  compared with previous estimates. 
Yusef-Zadeh argued that the radio estimates of ${\dot M}$ are likely
to be the most reliable because of the need for fewer assumptions (see
also Section~\ref{sec:hotwinds}).

The same VLA survey also uncovered $\sim$50 compact, partially resolved
radio sources with associated bow shocks, and some cases, cometary-like
tails that align toward the massive stars in the region. These
partially resolved sources do
not have IR counterparts,  and Yusef-Zadeh et
al. (2015b) have identified them as probable 
proplyds. The
corresponding size scales ($\sim$500~AU) are consistent with this
interpretation. Furthermore, arguments concerning both
the evaporation timescales and the stability against tidal disruption
imply that these objects must be self-gravitating and cannot be
merely clumps of ionized gas. A surprising implication is that low-mass
star formation is ongoing in the neighborhood of Sgr~A*. Follow-up ALMA
observations also uncovered dust from these objects (Yusef-Zadeh et
al. 2017a) and enabled mass determinations of the disk
components (0.03--0.06~$M_{\odot}$). The estimated ages are
$\sim10^{5}-10^{6}$~yr---far younger than the neighboring massive star
cluster, whose age is a few million years.

\begin{figure} 
\centering
\scalebox{0.3}{\rotatebox{0}{\includegraphics{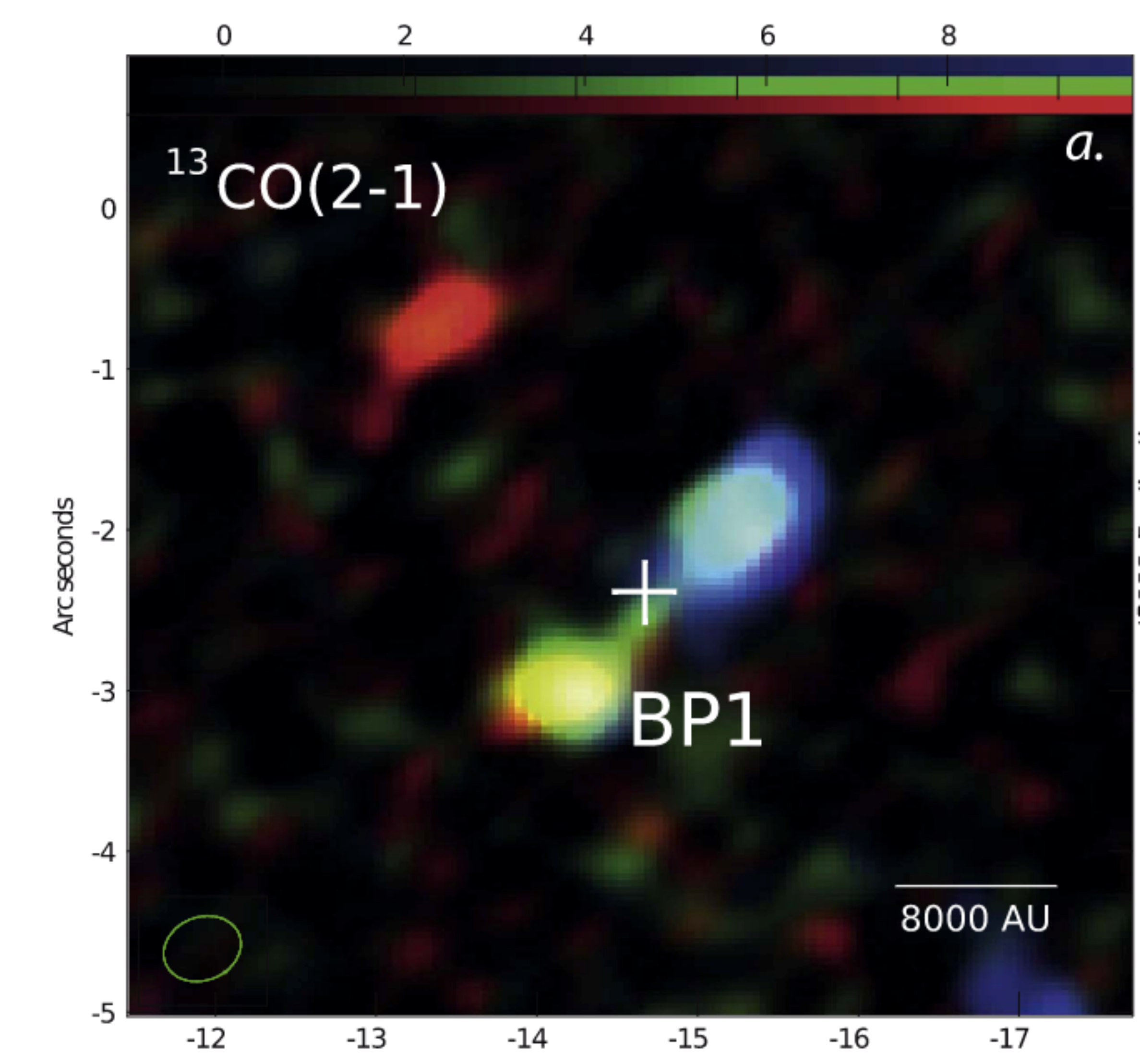}}}
\caption{Example of a YSO detected at a projected distance of
  $\sim$0.6~pc from Sgr~A* in deep ALMA observations. The two lobes of $^{13}$CO(2-1) emission are
  red- and blue-shifted by $\sim$2~\kms\ 
  and are connected by intermediate velocity gas. The intensity range
  is between $-1$ and 10~mJy beam$^{-1}$. The cross marks the
  position of a compact 226~GHz continuum source. From Yusef-Zadeh et
  al. (2017c); reproduced by permission of the AAS. }
\label{fig:GCbipolar}
\end{figure}

Another population of sources uncovered by Yusef-Zadeh's group
within a pc of Sgr~A* comprises 11 objects exhibiting
signatures of bipolar outflows (Yusef-Zadeh et al. 2017c). ALMA observations
in spectral lines including $^{13}$CO(2-1) show unambiguous signatures of red-
and blue-shifted gas lobes centered on a compact central source 
(Figure~\ref{fig:GCbipolar}). Inferred dynamical ages of
$\sim$6500~yr
again point to the ongoing formation of
low-mass stars in this extreme environment.  Yusef-Zadeh stressed that
the bipolar sources and proplyds found near the GC
 mark the first time that individual low-mass YSOs have been directly
detected at such large distances ($>$8~kpc) and
that this only possible at radio wavelengths. 

Lastly, Yusef-Zadeh described an enigmatic source in the 
GC known as IRS~3 (Yusef-Zadeh et al. 2017b).  
Whether this is dusty YSO or an AGB star is currently
unclear, although the latter classification is presently favored based
in part by detection of a shell-like structure of $^{13}$CO emission
surrounding the star. Numerical simulations show that
the non-spherical shape and extended size of the object
($\sim$8000~AU) can be explained by tidal distortion of
an extended radio photosphere. 

\subsection{Mapping the 3D Structure of the Galaxy}
Y. Pihlstr\"om (University of New Mexico) reviewed recent efforts to
use molecular maser emission from large samples of
evolved stars 
to provide new information on the structure and dynamics of the
Milky Way. She is a key team member of the BaADE
project (Trapp et al. 2018; see also Section~\ref{sec:RSGs}), which is 
aimed at making significant leaps in our
understanding of the bulge and bar regions of the Galaxy,  including the
question of whether the Milky Way's bulge was formed early in its history, or later
by buckling of the disk. 

As discussed by Pihlstr\"om, far-IR colors  based on measurements from the {\it Infrared
  Astronomical Satellite} ({\it IRAS}) provided the first efficient  means of
identifying large samples of low-mass, evolved stars within the
Galaxy, and follow-up
observations of many of these stars in maser lines (including OH and
SiO) subsequently provided crucial line-of-sight
velocity information for mapping Galactic structure (te Lintel
Hekkert, Dejonge, \& Habing
1991; Deguchi et
al. 2004a, b). However,  previous sample selection toward the
bulge and plane was severely restricted by confusion and high
extinction, leaving too few stars to derive 
detailed structural and dynamical information. 

A major advantage for BaADE is an improved method  to identify
larger numbers of SiO maser stars. Using mid-IR color data from the {\it Midcourse Space
eXperiment} ({\it MSX}), which had a beam of $\sim2''$ and which focused on
the Galactic Plane, Sjouwerman,
Capen, \& Claussen (2009)  found a high
success rate in detection of SiO masers ($\sim$50-90\%). 
In total,
28,000 candidate evolved stars were identified as targets for
follow-up SiO observations; $\sim$19,000 have now been observed at
43~GHz with the VLA (totaling over 400 hours),
and $\sim$9000 stars (those too far south for the VLA) are
currently being targeted at 86~GHz with ALMA.
 Multiple SiO transitions/isotopologues
are covered in the observing bands, along with a few lines from
C-bearing molecules that can be used to obtain velocities for
C-rich AGB stars lacking SiO lines.

\begin{figure*}
\centering
\scalebox{0.7}{\rotatebox{0}{\includegraphics{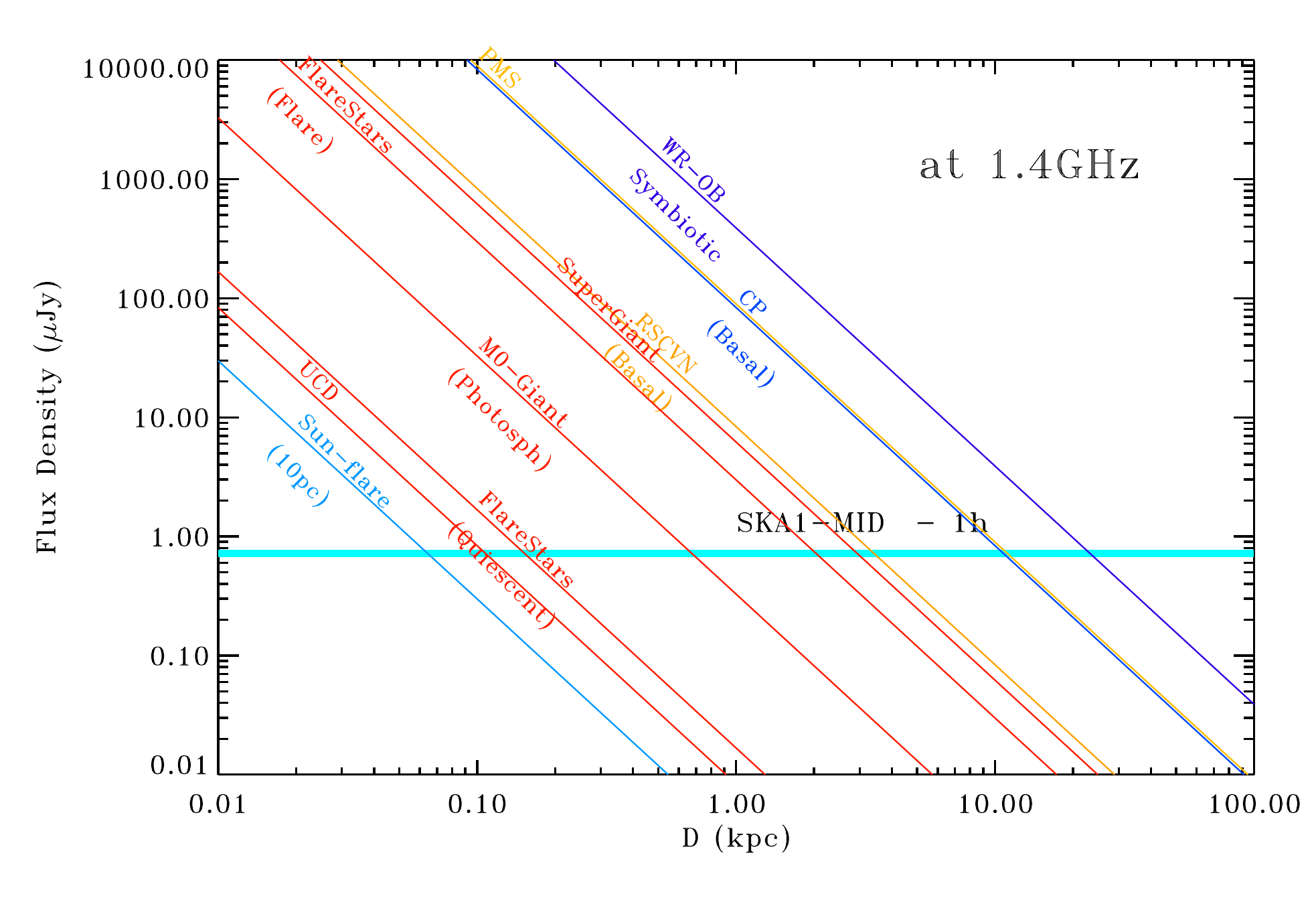}}}
\caption{Predicted 1.4~GHz flux densities (in $\mu$Jy) as a function of distance (in
  kpc) for various classes of stars. The horizontal line indicates
  the expected detection threshold for phase~1 of the SKA. Adapted by
  G. Umana from Umana et al. (2015b). }
\label{fig:radiosurveys}
\end{figure*}

Observations remain ongoing, but
the new data have already dramatically increased the sampling of the
Galactic potential compared with past surveys.
Measured dispersions allow separating the sample stars into disk and
bulge populations. Evidence for cylindrical rotation is clearly
seen. Future work will 
include follow-up VLBI astrometry for select stars to obtain
parallaxes and proper motions. The current database of 
over 20,000 SiO maser sources will also enable
statistical studies of maser pumping (through comparison of different
lines) and of the
correlations between the maser properties and other properties of
the stars themselves (see also Section~\ref{sec:AGBmasers}).  
Follow-up IR spectroscopy is planned to obtain metallicity information
for the sample.

Pihlstr\"om noted that BaADE complements the BeSSel Survey of Reid et al. (2014),
which used H$_{2}$O and CH$_{3}$OH maser
measurements of high-mass star-forming regions to determine the positions of spiral arms within the
Galaxy (see also Sanna et al. 2017) and to measure parameters
such as the Galactic rotation speed and distance to the GC. 
Radio measurements like these also provide an important complement
to optical astrometry for  understanding stellar populations,
Galactic structure,  and the dynamical evolution of
the inner Galaxy. L. H. Quiroga-Nu\~nez (Joint Institute
 for VLBI European Research Infrastructure Consortium/University of
 Leiden) described one such project, aimed at studying the population of
 SiO maser-emitting bulge stars from BaADE for which complementary measurements exist
 from {\it Gaia} and other optical/IR surveys. Quiroga-Nu\~nez's team
 plans to select a subset of these stars for astrometric VLBI
 follow-up to enable
 direct comparisons between astrometry obtained using radio
 and optical methods (Quiroga-N\~unez et
 al. 2018). The VLBI measurements are expected  to achieve 3D orbits and
 parallaxes with 50$\mu$s accuracy, comparable to  {\it Gaia}.
 In addition, his team plans to obtain VLBI astrometry for stars that
 are too optically bright
 ($g <9$) to permit {\it Gaia} astrometry.

\section{Radio Star Surveys - Current and Future}
\protect\label{sec:surveys}
G. Umana (Istituto Nazionale de Astrofisica/Observatatorio Astrofisico di Catania)
noted that we presently do not have a satisfactory answer to a
question frequently posed extragalactic astronomers,
namely 
``How many sub-mJy stellar sources are expected [above a certain
detection threshold] within one square
degree of sky?''.  She pointed out that there are
presently no good statistics on the detectability of radio emission
from various classes of stars, since existing survey data provide few
useful
constraints; they are typically 
too shallow to detect stellar sources, have poor angular resolution,
and/or concentrate only on high Galactic latitudes. Furthermore, to date, very few
stars with luminosities comparable to the Sun have been detected at
cm wavelengths, and those that have have required many hours of
integration (e.g., $\alpha$~Cen~A and B; Trigilio et al. 2018;
$\tau$~Cet, $\eta$~Cas~A, 40~Eri~A; Villadsen et al. 2014; Section~\ref{sec:cool}).   This
means that our current information comes  primarily from targeted
observations.

Umana predicted that the aforementioned limitations will finally be
overcome with the next generation of sensitive all-sky radio surveys
currently being planned.  This will include
the Evolutionary Map of the Universe (EMU) survey to be conducted with
the Australian Square Kilometre Array Pathfinder (ASKAP; Norris et
al. 2011), slated to map 75\% of the sky, including the Galactic
Plane, at 1400--1800~GHz to a depth
of 10$\mu$Jy beam$^{-1}$ at 10$''$ resolution. 
EMU is expected to detect UCDs to 20-30~pc, RS~CnV binaries to 500~pc, and
winds from W-R and OB stars to the GC. Eventually,
SKA-MID will be able to detect most classes of stars throughout the
Galaxy (Figure~\ref{fig:radiosurveys}), while SKA-2 would begin to detect many of these classes in
nearby galaxies (Umana et al. 2015a).

In the framework of EMU, Umana's team has recently begun a deep radio
survey of a $2^{\circ}\times2^{\circ}$  region of the Galactic plane
with $\sim$10$''$ resolution using the ATCA 
at 1.1--3.1~GHz. The
targeted flux limit is 40$\mu$Jy. This project,
known as SCORPIO [Stellar Continuum Originating from Radio Physics
in Our Galaxy (Umana et al. 2015b)], has two main
goals: (1) gauge the scientific potential of
deep radio surveys for stellar and Galactic
astronomy; (2) serve as a technical pathfinder for EMU and future SKA
surveys (including how to address issues such as source
identification, complexity, variability, and  separation 
from diffuse background emission). Data from a SCORPIO pilot experiment
have already been published (Umana et al. 2015b), and among the 614
sources detected, 10 were classified as stars
and many remain unidentified. To improve  source
classification, Cavallaro et
al. (2018) have developed automated pipelines to exploit spectral
indices, while Riggi et al. (2016)
have developed algorithms to improve automated detection of extended
sources. 
Another ongoing effort seeks to use a combination of radio and IR properties 
to improve automated classification (Ingallinera et al., in
prep.). 

I. Stevens also stressed the importance of contemporaneous, multi-frequency observations 
for understanding stars and urged the community to think more about how
to carry these out. He drew attention to the opportunity presented by the
many current and future
radio transient surveys and advocated for piggybacking  to
search for bursts of emission for various classes of stars, including
OB stars. Stevens further emphasized the need for wide-frequency observations
to identify spectral turnovers, 
coupled with sensitive time domain
studies capable of probing changes on timescales of relevance to studying winds and
other stellar processes ($\sim$1 hour or less).  Unfortunately, 
computational limitations presently preclude these kinds of 
commensal studies with many current interferometers.

While upcoming radio surveys hold exciting promise for stellar
astrophysics, S. White sounded a caution that while future survey 
instruments will be valuable for finding stars, it will be crucial to
retain instruments like   the VLA and ALMA to enable detailed follow-up studies
and to obtain the kinds of time and
multi-frequency coverage that are not possible with survey
instruments. The meeting concluded with a call to action to insure
that this needs gets
emphasized among the wider community.

\acknowledgments
The author gratefully
acknowledges the efforts of the RS-2 Local Organizing Committee
(H. Johnson, N. W. Kotary, M. Reynolds, J. SooHoo) and Scientific Organizing
Committee (G. Hallinan, R. Ignace, K. Menten, M. Rupen, L. Sjouwerman,
S. White). An anonymous referee provided useful comments on
this manuscript.
Financial support for the RS-2 workshop was provided by 
a grant from the National Science Foundation (AST-1743708). 

\vspace{0.5cm}

\references
Ainsworth, R. E., Scaife, A. M. M., Green, D. A., Coughlan, C. P., \&
Ray, T. P. 2016, MNRAS, 459, 1248

Ainsworth, R. E., Scaife, A. M. M., Ray, T. P., et al. 2014, ApJL,
792, L18

Airapetian, V. S., Adibekyan, V., Ansdell, M., et al. 2018, white
paper, arXiv:1803.03751

Akiyama, K., Ikeda, S., Pleau, M., et al. 2017a, AJ, 153, 159

Akiyama, K., Kuramochi, K., Ikeda, S., et al. 2017b, ApJ, 838, 1

Aleman, I., Exter, K., Ueta, T., et al. 2018, MNRAS, 477, 4499

ALMA Partnership, Brogan, C. L., P\'erez, L. M., Hunter, T. R., et al. 2015,
ApJL, 808, L3

Anderson, M. M. \& Hallinan, G. 2017, AAS, 49, id.401.02

Andrews, H., Fenech, D., Prinja, R. K., Clark, J. S., \& Hindson,
L. 2018, MNRAS, 477, L55

Anglada, G., Rodr\'\i guez, L. F., \& Carrasco-Gonz\'alez, C. 2018,
A\&ARev, 26, 3

Ansdell, M., Williams, J. P., Manara, C. F., et al. 2017, AJ, 153, 240

Ayres, T. R. 2018, preprint (arXiv:1808.06092)

Azulay, R., Guirado, J. C., Marcaide, J. M., et al. 2015, A\&A, 578, A16

Azulay, R., Guirado, J. C., Marcaide, J. M., et al. 2017a, A\&A, 607,
A10

Azulay, R., Guirado, J. C., Marcaide, J. M., et al. 2017b, A\&A, 602,
A57

Bachiller, R., Forveille, T., Huggins, P. J., \& Cox, P. 1997, A\&A,
324, 1123

B\'aez-Rubio, A., Mart\'\i n-Pintado, J., Thum, C., \& Planesas,
P. 2013, A\&A, 553, A45

Bains, I., Cohen, M., Chapman, J. M., Deacon, R. M., Redman,
M. P. 2009, MNRAS, 397, 1386

Balser, D. S., Bania, T. M., Rood, R. T., \& Wilson, T. L. 1997, ApJ,
483, 320

Balser, D. S., Goss, W. M., Bania, T. M., \& Rood, R. T. 2006, ApJ, 640, 360

Barrett, P. E., Dieck, C., Beasley, A. J., Singh, K. P., \& Mason,
P. A. 2017, AJ, 154, 252

Benaglia, P., Romero, G. E., Mart\'\i, J., Peri, C. S., \& Araudo,
A. T. 2010, A\&A, 517, L10

Benkevitch, L. V., Sokolov, I., Oberoi, D., \& Zurbuchen, T. 2001,
unpublished (arXiv:1006.5635)

Benz, A. O., Conway, J., \& G\"udel, M. 1998, A\&A, 331, 596

Benz, A. O. \& G\"udel, M. 1994, A\&A, 285, 621

Bower, G. C., Loinard, L., Dzib, S., Galli, P. A. B., Ortiz-L\'eon,
G. N., Moutou, C., \& Donati, J.-F. 2016, ApJ, 830, 107

Brookes, D. P., Stevens, I. M., Benaglia, P, Ishwara-Chandra, C. H.,
\& Mart\'\i , J. 2016a,  D. P. Brookes PhD dissertation, Chapter 6,
University of Birmingham

Brookes, D. P., Stevens, I. M., \& Pittard, J. M. 2016b, 
D. P. Brookes PhD dissertation, Chapter 4,
University of Birmingham

Buckley, D. A. H., Meintjes, P. J., Potter, S. B., Marsh, T. R., \&
G\"ansicke, B. T. 2017, NatAs, 1, 29

Carrasco-Gonz\'alez, C., Henning, T., Chandler, C. J., et al. 2016,
ApJL, 821, L16

Carrasco-Gonz\'alez, C., Rodr\'\i guez, L. F., Anglada, G., Mart\'\i,
J., Torrelles, J. M., \& Osario, M. 2010, Sci, 330, 1209

Castor, J. I., Abbott, D. C., \& Klein, R. I. 1975, ApJ, 195, 157

Castro-Carrizo, A., Bujarrabal, V., Neri, R., Alcolea, J., S\'anchez
Contreras, C., Santander-Garc\'\i a, M., \& Nyman, L.-\ang. 2017,
A\&A, 600, A4

Cavallaro, F., Trigilio, C., Umana, G., et al. 2018, MNRAS, 473, 1685

Cerrigone, L., Umana, G., Trigilio, C., Leto, P., Buemi, C. S.,
Ingallinera, A. 2017, MNRAS, 468, 3450

Chau, W., Zhang, Y., Nakashima, J, Deguchi, S., \& Kwok, S. 2012, ApJ,
760, 66

Chen, B., Bastian, T. S., Shen, C., Gary, D. E., Krucker, S., \&
Glesener, L. 2015, Sci, 350, 1238

Chhetri, R., Ekers, R. D., Jones, P. A., \& Ricci, R. 2013, MNRAS,
434, 956

Chomiuk, L., Linford, J. D., Yang, J., et al. 2014, Nature, 514, 339

Christensen, U. R., Holzwarth, V., \& Reiners, A. 2009, Nature, 457,
167

Cohen, M., Chapman, J. M.,  Deacon, R. M., Sault, R. J., Parker,
Q. A., \& Green, A. J. 2006, MNRAS, 369, 189

Copley, C. J., Thondikulam, V., Loots, A., et al. 2016,
unpublished (arXiv:1608.02187)

Coppejans, D. L., K\"ording, E. G., Miller-Jones, J. C. A., Rupen,
M. P., Knigge, C., Sivakoff, G. R., \& Groot, P. J. 2015, MNRAS, 451,
3801

Coppejans, D. L., K\"ording, E. G., Miller-Jones, J. C. A., Rupen,
M. P., Sivakoff, G. R., Knigge, C., Groot, P. J., Woudt, P. A.,
Waagen, E. O., \& Templeton, M. 2016, MNRAS, 463, 2229

Coughlan, C. P., Ainsworth, R. E., Eisl\"offel, J., et al. 2017, ApJ,
834, 206

Crosley, M. K. \& Osten, R. A. 2018, ApJ, 856, 39

Crosley, M. K., Osten, R. A., \& Norman, C. 2017, ApJ, 845, 67

Cummings, J. D., Kalirai, J. S., Tremblay, P.-E., \& Ramirez-Ruiz,
E. 2016, ApJ, 818, 84

Daley-Yates, S., Stevens, I. R., \& Crossland, T. D. 2016, MNRAS, 463,
2735

Davis, R. J., Lovell, B., Palmer, H. P., \& Spencer, R. E. 1978,
Nature, 273, 644

De Beck, E. \& Olofsson, H. 2018, A\&A, 615, 8

De Becker, M., Benaglia, P., Romero, G. E., \& Peri, C. S. 2017, A\&A,
600, A47

De Becker, M. \& Raucq, F. 2013, A\&A, 558, A28

Deguchi, S., Fujii, T., Glass, I. S., et al. 2014a, PASJ, 56, 765

Deguchi, S., Imai, H., Fujii, T., et al. 2014b, PASJ, 56, 261

de Nutte, R., Decin, L., Olofsson, H., et al. 2017, A\&A, 600, A71

Dressing, C. D. \& Charbonneau, D. 2015, ApJ, 807, 45

Drew, P., Strelnitski, V., Smith, H. A., Mink, J., Jorgenson, R. A.,
\& O'Meara, J. M. 2017, ApJ, 851, 136

Dulaney, N. A., Richardson, N. D., Gerhartz, C. J., et al. 2017, ApJ,
836, 112

Dulk, G. A. 1985, ARA\&A, 23, 169

Dupuy, T. J., Forbrich, J., Rizzuto, A., et al., 2016, ApJ, 827, 23

Duthu, A., Herpin, F., Wiesemeyer, H., Baudry, A., L\'ebre, A., \&
Paubert, G. 2017, A\&A, 604, 12

Emslie, A. G., Dennis, B. R., Shih, A. Y., et al. 2012, ApJ, 759, 71

Engle, S. G., Guinan, E. F., Harper, G. M., Cuntz, M., Evans, N. R.,
Neilson, H. R., \& Fawzy, D. E. 2017, ApJ, 838, 67

Engle, S. G., Guinan, E. F., Harper, G. M., Neilson, H. R., \& Evans,
N. R. 2014, ApJ, 794, 80

Evans, N. R., Pillitteri, I., Wolk, S., Karovska, M., Tingle, E.,
Guinan, E., Engle, S., Bond, H. E., Schaefer, G. H., \& Mason,
B. D. 2016, AJ, 151, 108

Feigelson, E. D. \& Montmerle, T. 1999, ARA\&A, 37, 363

Fender, R., Woudt, P. A., Armstrong, R., et al. 2017, to appear in
MeerKAT Science: On the Pathway to the SKA, PoS, (arXiv:1711.04132)

Fenech, D. M., Clark, J. S., Prinja, R. K., Dougherty, S., Najarro,
F., Negueruela, I., Richards, A., Ritchie, B. W., \& Andrews, H. 2018,
A\&A, 617, 137

Fenech, D. M., Clark, J. S., Prinja, R. K., Morford, J. C., Dougherty,
S., \& Blomme, R. 2017, MNRAS, 464, L75

Finzell, T., Chomiuk, L., Metzger, B. D., et al. 2018, ApJ, 852, 108

Forbrich, J., Dupuy, T. J., Reid, M. J., et al. 2016a, ApJ, 827, 22

Forbrich, J., Osten, R. A., \& Wolk, S. J. 2011, ApJ, 736, 25

Forbrich, J., Reid, M. J., Menten, K. M., Rivilla, V. M., Wolk, S. J.,
Rau, U., \& Chandler, C. J. 2017, ApJ, 844, 109

Forbrich, J., Rivilla, V. M., Menten, K. M., et al. 2016b, ApJ, 822, 93

Forbrich, J. \& Wolk, S. J. 2013, A\&A, 551, A56

Freytag, B., Liljegren, S., \& H\"ofner, S. 2017, A\&A, 600, A137

Gagn\'e, J., Faherty, J. K., Burgasser, A. J., et al. 2017, ApJL, 841, L1

Gary, D. E., Chen, B., Dennis, B. R., et al. 2018, ApJ, 863, 83

Gary, D. E., Chen, B., Nita, G. M., et al. 2017, AGU \#SH41A-2755

Gary, D. E. \& Hurford, G. J. 2004, in Solar and Space Weather
Radiophysics, ed. by D. E. Gary and C. U. Keller, (Kluwer: Dordrecht),
ASSL, 314, 71 

Gary, D. E., White, S. M., Hurford, G. J., Nita, G. M., \& Liu,
Z. 2007, AAS, 210.9329

G\'erard, E. \& Le Bertre, T. 2006, AJ, 132, 2566

G\'erard, E., Le Bertre, T., \& Libert, Y. 2011, in 
Proceedings of the Annual Meeting of the French Society of Astronomy
and Astrophysics, ed. 
G. Alecian et al. 419

Gillon, M., Triaud, A. H. M. J., Demory, B.-O., et al. 2017, Nature,
542, 456

Gray, M. D., Baudry, A., Richards, A. M. S., Humphreys, E. M. L.,
Sobolev, A. M., \& Yates, J. A. 2016, MNRAS, 456, 374

Groenewegen, M. A. T., Vlemmings, W. H. T., Marigo, P., et al., 2016,
A\&A, 596, 50 

G\"udel, M. 2002, ARA\&A, 40, 217

G\"udel, M. \& Benz, A. O. 1993, ApJ, 405, L63

G\"udel, M., Benz, A. O., Schmitt, J. H. M. M., \& Skinner,
S. L. 1996, ApJ, 471, 1002

Gu\'elin, M., Patel, N. A., Bremer, M., et al. 2018, A\&A, 610, A4

Guinan, E. F. \& Engle, S. G. 2008, in The Ages of Stars, IAU
Symposium No. 258, ed. E. E. Mamajek, D. R. Soderbolm, \&
R. F. G. Wyse, 395

Guirado, J. C., Azulay, R., Gauza, B., et al. 2018, A\&A, 610, A23

Guzman-Ramirez, L., Rizzo, J. R., Zijlstra, A. A., Garc\'\i a-Mir\'o,
C., Morisset, C., \& Gray, M. D. 2016, MNRAS, 460, L35

Gvaramadze, V. V. \& Menten, K. M. 2012, A\&A, 541, A7

Hajduk, M., Zijlstra, A. A., van Hoof, P. A. M., et al. 2007, MNRAS,
378, 1298

Hallinan, G. \& Anderson, M. M. 2017, BAAS, 49, id.401.01

Hallinan, G., Antonova, A., Doyle, J. G., Bourke, S., Brisken, W. F.,
\& Golden, A. 2006, ApJ, 653, 690

Hallinan, G., Antonova, A., Doyle, J. G., Bourke, S., Lane, C., \&
Golden, A. 2008, ApJ, 684, 644

Hallinan, G., Bourke, S., Lane, C., et al. 2007, ApJL, 663, L25

Hallinan, G., Littlefair, S. P., Cotter, G., et al. 2015, Nature, 523,
568

Helling, C., Winters, J. M., \& Sedlmayr, E. 2000, A\&A, 358, 651

Hjellming, R. M. 1988, in Galactic and
  Extragalactic Radio Astronomy, Second Edition, ed.
G. L. Verschuur and K. I. Kellermann, (Springer-Verlag: New York), 381

Hillenbrand, L. A. \& White, R. J. 2004, ApJ, 604, 741

Hoai, D. T., Nhung, P. T. G\'erard, E., Matthews, L. D., Villaver, E.,
\& Le~Bertre, T. 2015, MNRAS, 449, 2386

H\"ofner, S. \& Olofsson, H. 2018, A\&ARv, 26, 1

Homan, W., Richards, A., Decin, L., Kervella, P., de Koter, A.,
McDonald, I., \& Ohnaka, K. 2017, A\&A, 601, A5

Honma, M., Akiyama, K., Uemura, M., \& Ikeda, S. 2014, PASJ, 66, 95

Humphreys, E. M. L., Gray, M. D., Yates, J. A., Field, D., Bowen,
G. H., \& Diamond, P. J. 2002, A\&A, 386, 256

Humphreys, R. M., Helton, L. A., \& Jones, T. J. 2007, AJ, 133, 2716

Ignace, R. 2016, MNRAS, 457, 4123

Jonas, J. 2009, IEEEP, 97, 1522

Jorgensen, U. G. \& Johnson, H. R. 1992, A\&A, 265, 168

Justtanont, K., Teyssier, D., Barlow, M. J., Matsuura, M., Swinyard,
B., Waters, L. B. F. M., \& Yates, J. 2013, A\&A, 556, A101

Kami\'nski, T., Menten, K. M., Tylenda, R., Hajduk, M., Patel, N. A.,
\& Kraus, A.  2015, Nature, 520,
322

Kami\'nski, T., Menten, K. M., Tylenda, R., Karakas, A., Belloche, A.,
\& Patel, N. A. 2017, A\&A, 607, A78

Kami\'nski, T., Tylenda, R., Menten, K. M., et al. 2018, NatAs, 109, 

Kao, M. M., Hallinan, G., Pineda, J. S., Escala, I., Burgasser, A.,
Bourke, S., \& Stevenson, D. 2016, ApJ, 818, 24 

Kao, M. M., Hallinan, G., Pineda, J. S., Stevenson, D., \& Burgasser,
A. 2018, ApJS, 237, 25

Karmakar, S., Pandey, J. C., Airapetian, V. S., \& Misra, K. 2017,
ApJ, 840, 102

Katsova, M. M. \& Livshits, M. A. 2015, SoPh, 290, 3663

Kerschbaum, F., Maercker, M., Brunner, M. et al. 2017, A\&A, 605, A116

Kervella, P., Decin, L., Richards, A. M. S., et al. 2018, A\&A, 609, A67

Kervella, P., Homan, W., Richards A. M. S., Decin, L., McDonald, I.,
Montarg\`es, M., \& Ohnaka, K. 2016, A\&A, 596, 92

Klement, R., Carciofi, A. C., Rivinius, T., et al. 2017, A\&A, 601, A74

K\"ording, E., Rupen, M., Knigge, C., Fender, R., Dhawan, V.,
Templeton, M., \& Muxlow, T. 2008, Sci, 320, 1318

Kozarev, K. A., Oberoi, D., Morgan, J., Crowley, M., Benkevitch,
L. V., Lonsdale, C., Erickson, P. J., Winter, H. D. III, McCauley, P.,
\& Cairns, I. 2016,  American Geophysical Union Fall General Assembly,
id. SH22B-01

Krucker, S., Gim\'enez de Castro, C. G., Hudson, H. S., et al. 2013,
A\&AR, 21 58

Kundu, M. R., Pallavicini, R., White, S. M., \& Jackson, P. D. 1988,
A\&A, 195, 159

Lammer, H., Lichtenegger, H. I. M., Kulikov, Y. N., et al., 2007,
AsBio, 7, 185

L\`ebre, A., Auri\'ere, M., Fabas, N., Gillet, D., Herpin, F.,
Komstamtinova-Antova, R., \& Petit, P. 2014, A\&A, 561, A85

Lee, U., Saio, H., \& Osaki, Y. 1991, MNRAS, 250, 432

Lee, C.-F., Yang, C.-H., Sahai, R., \& S\'anchez Contreras, C. 2013,
ApJ, 770, 153

Leto, P., Trigilio, C., Buemi, C. S., Umana, G., Ingallinera, A., \&
Cerrigone, L. 2016, MNRAS, 459, 1159

Leverenz, H., Filpovi\'c, M. D., Vukoti\'c, B., Uro$\breve{\rm s}$evi\'c, D., \&
Grieve, K. 2017, MNRAS, 468, 1794

Lim, J., Carilli, C. L., White, S. M., Beasley, A. J., \& Marson,
R. G. 1998, Nature, 392, 575

Linford, J. D., Chomiuk, L., Nelson, T., et al. 2017, ApJ, 842, 73

Linsky, J. L. 1996, , in Radio Emission from the
  Stars and the Sun, ASP Conf. Series, Vol. 93, ed.
A. R. Taylor and J. M. Paredes, (ASP: San Francisco), 439

Linsky, J. L. \& Haisch, B. M. 1979, ApJ, ApJL, 229, L27

Lonsdale, C., Benkevitch, L., Cairns, I., et al. 2017, Proceedings of 8$^{\rm th}$ International
Workshop on Planetary, Solar, and Heliospheric Radio Emissions,
ed. G. Fischer, G. Mann, M. Panchenko, \& P. Zarka, (Austrian Academy
of Sciences Press: Vienna), 425

Loukitcheva, M., White, S. M., Solanki, S. K., Fleishman, G. D., \&
Carlsson, M. 2017, A\&A, 601, A43

Lykou, F., Klotz, D., Paladini, C., et al. 2015, A\&A, 576, A46

Lynch, C. R., Lenc, E., Kaplan, D. L., Murphy, T., \& Anderson,
G. E. 2017, ApJL, 836, L30

Lynch, C., Mutel, R. L., \& G\"udel, M. 2015, ApJ, 802, 106

MacGregor, M. A., Matr\`a, L., Kalas, P., et al. 2017, ApJ, 842, 8

Marsh, T. R., G\"ansicke, B. T., H\"ummerich, S.,  et al. 2016, Nature,
537, 374

Matthews, L. D. 2013, ``Radio Stars and Their Lives in the
Galaxy'', PASP, 125, 313

Matthews, L. D., Le Bertre, T., G\'erard, E., \& Johnson, M. C. 2013,
AJ, 145, 97

Matthews, L. D., Marengo, M., Evans, N. R., \& Bono, G. 2012, ApJ,
744, 53

Matthews, L. D., Marengo, M., \& Evans, N. R., 2016, AJ, 152, 200

Matthews, L. D., Reid, M. J., \& Menten, K. M. 2015, ApJ, 808, 36

Matthews, L. D., Reid, M. J., Menten, K. M., \& Akiyama, K. 2018,
AJ, 156, 15

McCauley, P. I., Cairns, I. H., Morgan, J., Gibson, S. E., Harding,
J. C., Lonsdale, C., \& Oberoi, D. 2017, ApJ, 851, 151 

McLean, M., Berger, E., Irwin, J., Forbrich, J., \& Reiners, A. 2011,
ApJ, 741, 27

Miller Bertolami, M. M. 2016, A\&A, 588, A25

Mohan, A. \& Oberoi, D. 2017, SoPh, 292, 168

Montez, R. Jr., Kastner, J. H., Balick, B., et al. 2015, ApJ, 800, 8

Mooley, K. P.,  Miller-Jones, J. C. A., Fender, R. P., et al. 2017,
ApJL, 467, L31

Morford, J. C., Fenech, D. M., Prinja, R. K., Blomme, R., \& Yates,
J. A. 2016, MNRAS, 463, 763

Morford, J., Prinja, R., \& Fenech, D. 2017, in Formation, evolution,
and survival of massive star clusters, IAU Symposium No. 316,
ed. C. Charbonnel and A. Nota, 169

Murphy, E. J. Bolotto, A., Chatterjee, S., et al. 2018, in Science
with a Next-Generation Very Large Array, ASP Conf. Series, Monograph
7, ed. E. J. Murphy, (ASP: San Francisco), in press

Mutel, R. L., Menietti, J. D., Christopher, I. W., Gurnett, D. A., \&
Cook, J. M. 2006, JGRA, 111, A011660

Neilson, H. R. 2014, in Precision Asteroseismology, IAU Symp. 301,
ed. J. A. Guzik, W. J. Chaplin, G. Handler, and A. Pigulski,
(Cambridge University Press: Cambridge), 205

Norris, R. P., Hopkins, A. M., Afonso, J., et al. 2011, PASA, 28, 215

Oberoi, D., Matthews, L. D., Cairns, I. H., et al. 2011, ApJL, 728, L27

O'Gorman, E., Harper, G. M., Brown, A., Guinan, E. F., Richards,
A. M. S., Vlemmings, W., \& Wasatonic, R. 2015, A\&A, 580, A101

O'Gorman, E. Kervella, P., Harper, G. M., et al. 2017, A\&A, 602, L10

Ortiz-L\'eon, G. N., Loinard, L., Kounkel, M. A., et al. 2017, ApJ,
834, 141

Osten, R. A. \& Crosley, M. K. 2017, Next Generation Very Large Array
Memo 31, {\url{http://library.nrao.edu/public/memos/ngvla/NGVLA\_31.pdf}}

Osten, R. A., Kowalski, A., Drake, S. A., et al. 2016, ApJ, 832, 174

Osten, R. A. \& Wolk, S. J. 2015, ApJ, 809, 79

Padovani, M.,  Hennebelle, P.,  Marcowith, A., \& Ferri\`ere, K. 2015,
A\&A, 582, L13

Pandya, A., Zhang, Z., Chandra, M., \& Gammie, C. F. 2016, APJ, 822,
34

Pardo, J. R., Cernicharo, J., Gonzalez-Alfonso, E., \& Bujarrabal,
V. 1998, A\&A, 329, 219

Paredes, J. M. 2005, EAS, 15, 187

Patel, N. A., Gottlieb, C., Young, K., et al. 2018, AAS, 231, id. 408.06

Patel, N., Young, K. H., Gottlieb, C. A., et al. 2011, ApJS, 193, 17

P\'erez-S\'anchez, A. F., Tafoya, D., Garc\'\i a L\'opez, R., Vlemmings,
W., \& Rodr\'\i guez, L. F. 2017, A\&A, 601, A68

P\'erez-S\'anchez, A. F., Vlemmings, W. H. T., Tafoya, D., \& Chapman,
J. M. 2013, MNRAS, 436, L79

Pineda, J. S. \& Hallinan, G. 2018, accepted to AAS journals (arXiv:1806.00480)

Pineda, J. S., Hallinan, G., \& Kao, M. M. 2017, ApJ, 846, 75

Pottasch, S. R. \& Bernard-Salas, J. 2013, A\&A, 550, A35

Primiani, R. A., Young, K. H., Young, A., et al. 2016, JAI, 5, 1641006

Quiroga-N\~unez, L. H., van Langenvelde, H. J., Pihlstr\"om, Y. M.,
Sjouwerman, L. O., \& Brown, A. G. A. 2018, in Astronomy and
Astrophysics in the {\it Gaia} Sky, IAU Symposium No. 330, edited by
A. Recio-Blanco, P. de Laverny, A. G. A. Brown, and T. Prusti,
(Cambridge University Press: Cambridge), 245

Ramstedt, S. \& Olofsson, H. 2014, A\&A, 566, A145

Raulin, J.-P., L\"uthi, T., Trottet, G., \& Correia, E. 2004, COSPAR,
35, 3118

Reid, M., Loinard, L., \& Maccarone, T. 2018 in Science
with a Next-Generation Very Large Array, ASP Conf. Series, Monograph
7, ed. E. J. Murphy, (ASP: San Francisco), in press

Reid, M. J. \& Menten, K. M. 1997, ApJ, 476, 327

Reid, M. J., Menten, K. M., Brunthaler, A., et al. 2014, ApJ, 745, 191

Richards, A. M. S., Davis, R. J., Decin, L., et al. 2013, MNRAS, 432,
L61

Richards, A. M. S., Gray, M. D., Baudry, A., et al.
E. M. L. 2018, in Astrophysical Masers: Unlocking the Mysteries of the
Universe, Proceedings of IAU Symp. 336, ed. A. Tarchi, M. J. Reid, \&
P. Castangia, 347

Richards, A. M. S., Impellizzeri, C. M. V., Humphreys, E. M., et
al. 2014, A\&A, 572, L9

Riggi, S., Ingallinera, A., Leto, P., et al. 2016, MNRAS, 460, 1468

Rodr\'\i guez-Kamenetzky, A., Carrasco-Gonz\'alez, C., Araudo, A., et
al. 2016, ApJ, 818, 27

Route, M. 2017, ApJ, 845, 66

Route, M. \& Wolszczan, A. 2012, ApJL, 747, L22

Rowlinson, A., Bell, M. E., Murphy, T., et al. 2016, MNRAS, 458, 3506

Ruffle, P. M. E., Zijlstra, A. A., Walsh, J. R., et al. 2004, MNRAS,
353, 796

Safarzadeh, M., Lu, Y.,  \& Hayward, C. C. 2017, MNRAS, 472, 2462

Sahai, R., Vlemmings, W. H. T., \& Nyman, L.\ang. 2017, ApJ, 841, 110

S\'anchez-Contreras, C., B\'aez-Rubio, A., Alcolea, J., Bujarrabal,
V., \& Mart\'\i n-Pintado, J. 2017, A\&A, 603, A67

Sanna, A., Reid, M. J., Dame, T. M., Menten, K. M., \& Brunthaler,
A. 2017, Sci, 358, 227

Santander-Garc\'\i a, M., Bujarrabal, V., Alcolea J., et al. 2017, A\&A,
597, A27

Segura, A., Walkowicz, L. M., Meadows, V., Kasting, J., \& Hawley,
S. 2010, AsBio, 10, 751

Semel, M. 1989, A\&A, 225, 456

Shara, M. M., Moffat, A. F. J., \& Webbink, R. F. 1985, ApJ, 294, 271

Sharma, R., Oberoi, D., \& Arjunwadkar, M. 2018, ApJ, 852, 69

Shmeld, I. K., Strelnitskij, V. S., Fedorova, A. V., \& Fedorova,
O. V. 1992, in Astrochemistry of Cosmic Phenomena, IAU Symposium 150,
ed. by P. D. Singh, 
(Kluwer Academic Publishers, Dordrecht), 413

Shibayama, T., Maehara, H., Natsu, S., et al. 2013, ApJS, 209, 5

Sjouwerman, L. O., Capen, S. M., \& Claussen, M. J. 2009, ApJ, 705, 1554

Smith, K., G\"udel, M., \& Audard, M. 2005, A\&A, 436, 241

Soker, N. \& Lasota, J.-P. 2004, A\&A, 422, 1039

Spangler, S. R. \& Moffett, T. J. 1976, ApJ, 203, 497

Strelnitski, V., Bieging, J. H., Hora, J., Smith, H. A., Armstrong, P,
Lagergren, K., \& Walker, G. 2013, ApJ, 777, 89

Stroh, M. C., Pihlstr\"om, Y. M., Sjouwerman, L. O., Claussen, M. J,
Morris, M. R., \& Rich, R. M. 2018, ApJ, submitted

Suresh, A., Sharma, R., Oberoi, D., et al. 2017, ApJ, 843, 19

Suzuki, T. K. 2007, ApJ, 59, 1592

te Lintel Hekkert, P., Dejonghe, H., \& Habing, H. J. 1991, PASAu, 9,
20

Thomashow, E., Jorgenson, R. A., Strelnitski, V., Walter, G., and
Maria Mitchell Observatory Research Experience for Undergraduates
Interns, 2018, AAS, 231, id. 136.10

Tingay, S. J., Hancock, P. J., Wayth, R. B., Intema, H., Jagannathan,
P., \& Mooley, K. 2016, AJ, 152, 82

Trapp, A. C., Rich, R. M., Morris, M. R., Sjouwerman, L. O.,
Pihlstr\"om, Y., Claussen, M., \& Stroh, M. 2018, ApJ, 861, 75

Treumann, R. A. 2006, A\&ARv, 13, 229

Trigilio, C., Umana, G., Buemi, C. S., \& Leone, F. 2011, ApJL, 739,
L10

Trigilio, C., Umana, G., Cavallaro, F., Agliozzo, C., Leto, P. Buemi,
C. S., Ingallinera, A., Bufano, G. F., \& Riggi, S. 2018, MNRAS, 481, 217

Tsuji, T. 2000, ApJ, 540, L99

Tsuji, T. 2001, in Galaxies and their Constituents at the Highest
Angular Resolutions, IAU Symposium 205, ed. R. T. Schilizzi, 316

Tylenda, R., Kami\'nski, T., Udalski, A., et al. 2013, A\&A, 555, A16

Umana, G., Trigilio, C., Cerrigone, L. 2015a, PoS, id. 118

Umana, G., Trigilio, C., Franzen, T. M. O., et al. 2015b, MNRAS, 454, 902

Van de Sande, M., Decin, L., Lombaert, R., Khouri, T., de Koter, A.,
Wyrowski, F., De Nutte, R., \& Homan, W. 2018, A\&A, 609, A63

Vassiliadis, E. \& Wood, P. R. 1993, ApJ, 413, 641

Velilla Prieto, L., S\'anchez Contreras, C., Cernicharo, J., et
al. 2017, A\&A, 597, A25

Villadsen, J. Hallinan, G., Bourke, S., G\"udel, M., \& Rupen, M. 2014,
ApJ, 788, 112

Villadsen, J. Hallinan, G., Bourke, S., \& Monroe, R. 2017, AASTCS5,
id. 400.04

Vlemmings, W. H. T. 2012, in Cosmic Masers - from OH to H$_{0}$, IAU
Symp. 287, ed. R. S. Booth, E. M. L. Humphreys, and W. H. T. Vlemmings, 31

Vlemmings, W., Khouri, T., O'Gorman, E., et al. 2017, NatAs, 1, 848

Wedemeyer, S., Bastian, T., Braj$\breve{\rm s}$a, R., et al. 2016,
SSRv, 200, 1

Welch, D. L. \& Duric, N. 1988, AJ, 95, 1794

Wendeln, C., Chomiuk, L., Finzell, T., Linford, J. D., \& Strader,
J. 2017, ApJ, 840, 110

Weston, J. H. S., Sokoloski, J. L., Chomiuk, L., et al. 2016a, MNRAS,
460, 2687

Weston, J. H. S., Sokoloski, J. L., Metzger, B. D., et al. 2016b,
MNRAS, 457, 887

White, R. L. 1985, ApJ, 289, 698

White, S. M. 2004, in Solar and Space Weather Radiophysics, ed.
D. E. Gary and C. U. Keller, (Kluwer: Dordrecht), 89

White, S. M. 2007, Asian J. Phys. 16

White, S. M. \& Franciosini, E. 1995, ApJ, 444, 342

White, S. M., Iwai, K., Phillips, N. M. 2017, SoPh, 292, 88

Williams, P. K. G. 2017, in Handbook of Exoplanets, ed. H. J. Deeg and
J. A. Belmonte, submitted (arXiv:1707.04264)

Williams, P. K. G., Casewell, S. L., Stark, C. R., Littlefair, S. P.,
Helling, C., \& Berger, E. 2015, ApJ, 815, 64

Winters, J. G., Henry, T. J., Lurie, J. C., et al. 2015, AJ, 149, 5

Withers, P. \& Vogt, M. F. 2017, ApJ, 836, 114

Wolk, S. J., Pillitteri, I., \& Poppenhaeger, K. 2017, in Living
Around Active Stars, IAU Symp. 328, 290

Wright, A. E. \& Barlow, M. J. 1975, MNRAS, 170, 41

Yusef-Zadeh, F., Bushouse, H., Sch\"odel, R., Wardle, M., Cotton, W.,
Roberts, D. A., Nogueras-Lara, F., \& Gallego-Cano, E. 2015a, ApJ,
809, 10

Yusef-Zadeh, F., Cotton, B.,  Wardle, M., Royster, M. J., Kunneriath,
D., Roberts, D. A., Wootten, A., \& Sch\"odel, R. 2017a, MNRAS, 467, 922

Yusef-Zadeh, F., Roberts, D. A., Wardle, M., Cotton, W., Sch\"odel,
R., \& Royster, M. J. 2015b, ApJL, 801, L26

Yusef-Zadeh, F., Wardle, M., Cotton, W., Sch\"odel, R., Royster,
M. J., Roberts, D. A., \& Kunneriath, D. 2017b, ApJ, 837, 93

Yusef-Zadeh, F., Wardle, M., Kunneriath, D., Royster, M., Wootten, A.,
\& Roberts, D. A. 2017c, ApJL, 850, L30

Zhang, Q., Claus, B., Watson, L., \& Moran, J. 2017, ApJ, 837, 53

Zijlstra, A. A., Pottasch, S. R., \& Bignell, C. 1989, A\&AS, 79, 329

Zijlstra, A. A., van Hoof, P. A. M., \& Perley, R. A. 2008, ApJ, 681, 1296

\end{document}